\documentclass[12pt,number,sort&compress,final]{elsarticle}

\usepackage{graphicx}
\usepackage{booktabs,multicol} 
\usepackage{subcaption} 
\usepackage{lmodern}
\usepackage[T1]{fontenc}
\usepackage{textcomp}
\usepackage{gensymb,amssymb}
\usepackage{ulem}

\usepackage{algorithm}
\usepackage[noend]{algpseudocode}

\usepackage[usenames,dvipsnames]{xcolor}

\usepackage{etoolbox}
\newtoggle{markup}
\togglefalse{markup}

\usepackage[xcolor]{changebar}

\usepackage{xspace}
\usepackage{relsize}
\newcommand{\smallunderscore}{\textscale{.5}{\textunderscore}\xspace}

\usepackage{catchfilebetweentags}

\usepackage{xr-hyper}

\usepackage[version=4]{mhchem} 
\usepackage{amsmath,eqparbox,xparse,mathtools}
\usepackage[binary-units]{siunitx}

\usepackage[textsize=footnotesize,textwidth=1.5\linewidth]{todonotes}

\usepackage{macros}

\usepackage[hyphens]{url}
\usepackage[breaklinks=true, linkcolor=blue, citecolor=blue, colorlinks=true]{hyperref}
\usepackage[english,capitalise]{cleveref}

\crefname{appendix}{}{}

\usepackage{placeins}

\DeclareSIUnit\doubles{doubles}

\graphicspath{{./figures/}}

\begin{document}

\begin{frontmatter}

\title{Using SIMD and SIMT vectorization to evaluate sparse chemical kinetic Jacobian matrices and thermochemical source terms}

\author[1]{Nicholas J.~Curtis\corref{corr}}
\ead{nicholas.curtis@uconn.edu}
\author[2]{Kyle E.~Niemeyer}
\author[1]{Chih-Jen Sung}

\address[1]{Department of Mechanical Engineering, University of Connecticut, Storrs, CT 06269, USA}
\cortext[corr]{Corresponding author}
\address[2]{School of Mechanical, Industrial, and Manufacturing Engineering, Oregon State University, Corvallis, OR 97331, USA}

\begin{abstract}
Accurately predicting key combustion phenomena in reactive-flow simulations, e.g., lean blow-out, extinction\slash ignition limits and pollutant formation, necessitates the use of detailed chemical kinetics.
The large size and high levels of numerical stiffness typically present in chemical kinetic models relevant to transportation\slash power-generation applications make the efficient evaluation\slash factorization of the chemical kinetic Jacobian and thermochemical source-terms critical to the performance of reactive-flow codes.
Here we investigate the performance of vectorized evaluation of constant-pressure/volume thermochemical source-term and sparse/dense chemical kinetic Jacobians using single-instruction, multiple-data (SIMD) and single-instruction, multiple thread (SIMT) paradigms. These are implemented in pyJac, an open-source, reproducible code generation platform.
Selected chemical kinetic models covering the range of sizes typically used in reactive-flow simulations were used for demonstration.
A new formulation of the chemical kinetic governing equations was derived and verified, resulting in Jacobian sparsities of \SIrange{28.6}{92.0}{$\percent$} for the tested models.
Speedups of \SIrange{3.40}{4.08}{$\times$} were found for shallow-vectorized OpenCL source-rate evaluation compared with a parallel OpenMP code on an \avx/ central processing unit (CPU), increasing to \SIrange{6.63}{9.44}{$\times$} and \SIrange{3.03}{4.23}{$\times$} for sparse and dense chemical kinetic Jacobian evaluation, respectively.
Furthermore, the effect of data-ordering was investigated and a storage pattern specifically formulated for vectorized evaluation was proposed; as well, the effect of the constant pressure\slash volume assumptions and varying vector widths were studied on source-term evaluation performance.
Speedups reached up to \SI{17.60}{$\times$} and \SI{45.13}{$\times$} for dense and sparse evaluation on the GPU, and up to \SI{55.11}{$\times$} and \SI{245.63}{$\times$} on the CPU over a first-order finite-difference  Jacobian approach.
Further, dense Jacobian evaluation was up to \SI{19.56}{$\times$} and \SI{2.84}{$\times$} times faster than a previous version of \texttt{pyJac} on a CPU and GPU, respectively.
Finally, future directions for vectorized chemical kinetic evaluation and sparse linear-algebra techniques were discussed.
\end{abstract}


\end{frontmatter}

\section{Introduction}

As the combustion and reactive-flows community has recognized the importance of detailed chemical kinetics for predictive reactive-flow simulations~\cite{LU2009192}, chemical kinetic models have grown in size and complexity to describe current and next-generation fuels relevant to transportation and power generation.
For example, a recent biodiesel model~\cite{WESTBROOK2011742} consists of \textasciitilde\num{3500} chemical species and over \num{17000} reactions.
Moreover, the cost of evaluating the chemical source-terms scales linearly with the size of the model, while evaluating and factorizing the chemical kinetic Jacobian respectively scale quadratically and cubically with the number of species in the model, if naively implemented via a finite-difference method~\cite{LU2009192}.
These factors often prohibit using detailed chemical kinetics in practice; e.g., in a direct numerical simulation using a 22-species model, evaluating reaction rates consumed around half of the total run time~\cite{Spafford:2010aa}.
In addition, most common implicit integration techniques need to evaluate and factorize the Jacobian matrix to deal with stiffness.
As a result, these operations are bottlenecks when using even moderately sized chemical models in realistic reactive-flow simulations, necessitating other cost-reduction strategies~\cite{LU2009192}.

A host of techniques have been developed to lessen the computational demand of chemical kinetic calculations while maintaining fidelity, falling broadly into three categories: removal of unimportant species and reactions~\cite{Lu:2006bb,Pepiot-Desjardins:2008,Hiremath:2010jw,Niemeyer:2010bt,Curtis:2015},
lumping of species with similar thermochemical properties~\cite{Lu:2007,Ahmed:2007fa,Pepiot:2008kq},
and time-scale methods that reduce numerical stiffness~\cite{Maas:1992ws,Lam:1994ws,Lu:2001ve,Gou:2010}.
We refer interested readers to the recent review by Tur\'anyi and Tomlin~\cite{turanyi2016analysis} for a comprehensive overview.

In addition to the previously mentioned cost reduction methods, effort has gone into improving the integration algorithms and codes that evaluate the chemical kinetics~\cite{Gou:2010,SCHWER2002270,Niemeyer:2016aa,GAO2015287}.
In particular, a carefully derived analytical formulation of the Jacobian matrix can greatly increase sparsity~\cite{SCHWER2002270} and drop the cost of Jacobian evaluation to linearly depend on the number of species in the model~\cite{LU2009192}; sparse-matrix techniques can then reduce the cost of Jacobian factorization~\cite{superlu99}.
In addition, studies have shown that Single-Instruction, Multiple-Data (SIMD) and the related Single-Instruction, Multiple-Thread (SIMT) processors can accelerate chemical kinetic simulations~\cite{Shi:2012aa,Niemeyer:2014aa,Sewerin20151375,Niemeyer:2016aa,CurtisGPU:2017,stone2018}.

SIMD and SIMT programming are two important vector-processing paradigms used increasingly in scientific computing.
Traditional multicore parallelism is now used to increase central processing unit (CPU) performance, as the exponential growth in processing power---colloquially known as Moore's law---has slowed~\cite{khan2018science}.
Recently, SIMD\slash SIMT processors, e.g., in the form of graphics processing units (GPUs), have gained recognition due to their increased floating operation throughput.
The parallel programming standard OpenCL~\cite{stone2010opencl} has further enabled adoption of vector processing in scientific computing by providing a common application program interface (API) for execution on heterogeneous systems, e.g., CPU, GPU, or Intel's Many Integrated Core (MIC) architecture.
Here we will largely use OpenCL terminology to describe these processing paradigms, as it provides a convenient way to classify otherwise disparate processor types (e.g., CPUs and GPUs).
However, the concepts discussed herein broadly apply to SIMD\slash SIMT processing.

\begin{figure}[htb]
  \centering
  \begin{subfigure}[t]{0.45\linewidth}
      \includegraphics[width=\textwidth]{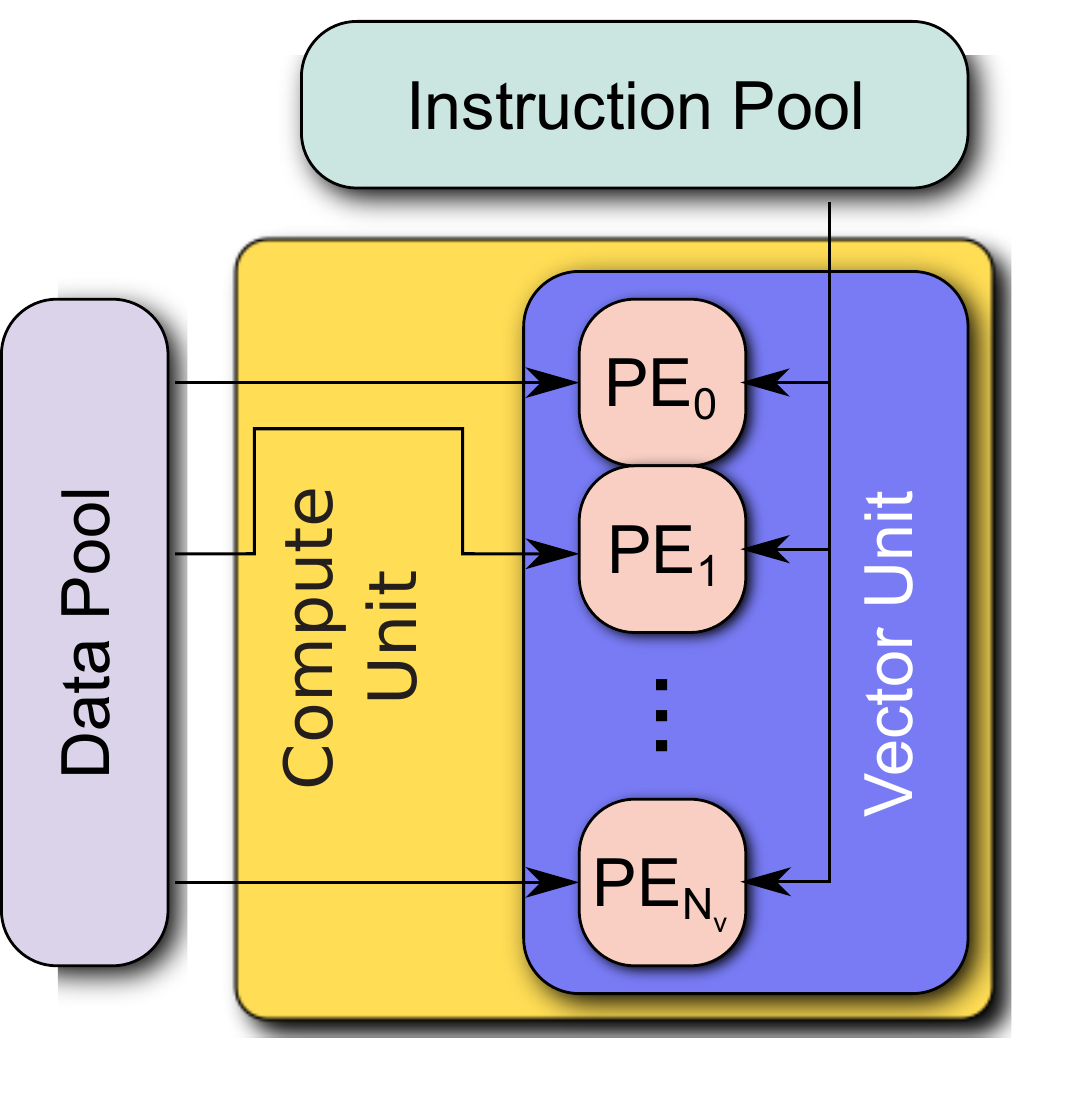}
      \caption{Schematic of SIMD processing.  A single compute unit (e.g., a CPU core) contains a vector unit with $N_v$ processing elements (PEs), together called a vector-lane.  The vector unit executes a single instruction concurrently on multiple data.}
      \label{F:SIMD}
  \end{subfigure}
  \hfill
  \begin{subfigure}[t]{0.45\linewidth}
      \includegraphics[width=\textwidth]{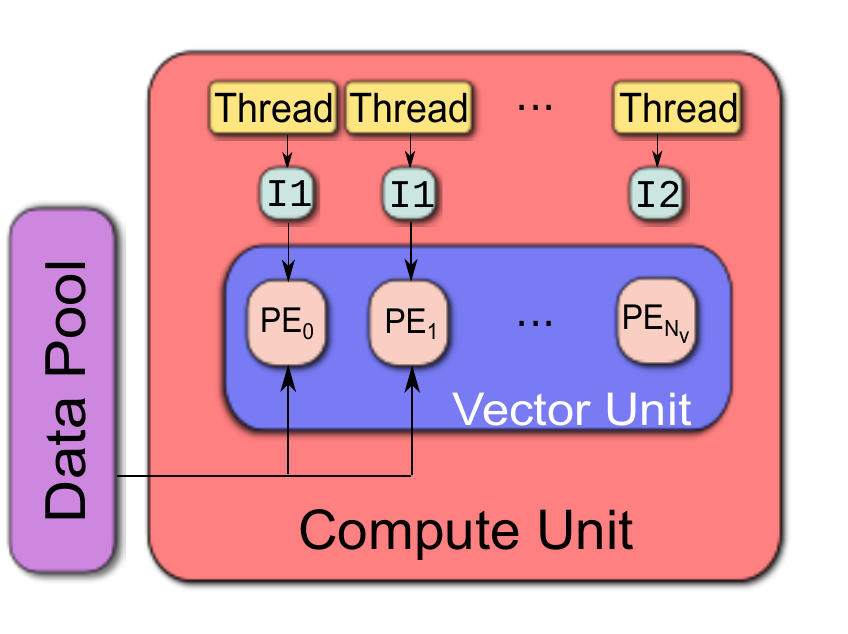}
      \caption{Schematic of SIMT processing. A single compute unit (e.g., a GPU streaming multiprocessor) contains many processing elements (PEs) and hosts many threads, each with an instruction to execute (I1, I2).  Threads with the same instruction execute concurrently on multiple data while the others must wait (leading to thread divergence).}
      \label{F:SIMT}
  \end{subfigure}
  \caption{Simple diagrams explaining the fundamentals of the SIMD and SIMT vector-processing paradigms.}
\end{figure}

A typical modern CPU contains multiple compute units (i.e., cores), each with specialized vector processing units capable of running SIMD instructions, as \cref{F:SIMD} depicts.
A SIMD instruction uses the vector processor to execute the same floating-point operation (e.g., multiplication, division) on different data concurrently.
The vector-width is the number of possible concurrent operations, typically around two to four in double precision.\footnote{OpenCL allows for use of vector-widths different from the actual hardware vector-width via implicit conversion, and may provide some performance benefit as Sec.~\ref{S:results} discusses.}
Specialized hardware accelerators have also been developed, like Intel's Xeon Phi co-processor (i.e., the MIC architecture), that have tens of cores with wide vector-widths (e.g., \numrange{4}{8} double-precision operations).
Cutting-edge and forthcoming Intel CPUs also include these wide vector-widths, like the Skylake Xeon and Cannon Lake architectures.

Modern GPUs rely on the related computing paradigm of SIMT processing, where a single compute element hosts large
numbers of threads (a streaming multiprocessor in Nvidia terminology)~\cite{lindholm2008Nvidia}.
\Cref{F:SIMT} depicts a SIMT compute unit, where a group of threads---typically \num{32}, known as a warp on Nvidia GPUs---execute the same SIMT instruction on multiple data concurrently.
If some threads must execute a different instruction, they are forced to wait and execute later; this may occur due to if\slash then branching or predication.
This phenomenon, known as thread-divergence, is a key consideration for SIMT processing and can cause serious performance degradation for complicated algorithms~\cite{CurtisGPU:2017}.

\subsection{Related work}
\label{S:related}

Recognizing the need to accelerate chemical-kinetic Jacobian evaluation and factorization, a number of recent works have been published on constructing analytical Jacobian matrices; although as will be discussed at the end of this section, here we offer several key improvements over past efforts.
Schwer et al.~\cite{SCHWER2002270} were among the first to recognize the critical importance of a sparse analytical Jacobian to accelerate chemical kinetic simulations.
Later, Safta et al.~\cite{Safta:2011vn} developed the \texttt{TChem} software package, which was one of the first developed that provides analytical Jacobian evaluation. However, \texttt{TChem} has several limitations, including incompatibility with modern reaction types---i.e., pressure-dependent Arrhenius (or P-Log) and Chebyshev reactions---and its lack of thread-safety to enable parallel execution~\cite{Curtis2017:tchem}.
Youssefi~\cite{Youssefi:2011tm} explored the importance of analytical Jacobian matrices for time-scale analysis techniques as well as their effect on computational efficiency in zero-dimensional homogeneous reactor simulations.
Bisetti~\cite{Bisetti:2012jw} developed an isothermal, isobaric analytical Jacobian code-generation utility;
this approach significantly increases Jacobian sparsity, although the chosen isothermal assumption is not typical in most combustion simulations.
In the same work Bisetti also provided a novel way to compute dense matrix-vector multiplications resulting from a change of system variables without storing the full Jacobian.
Perini et al.~\cite{Perini:2012gy} developed an analytical Jacobian code for constant-volume combustion, with additional options to increase sparsity (at the expense of strict correctness) and tabulate temperature-dependent properties; they reported an \SI{80}{\percent} speedup over a finite-difference-based Jacobian when used in a multidimensional reactive-flow simulation.
Gao et al.~\cite{GAO2015287} derived a sparse analytical Jacobian, but did not verify it outside the context of use with an implicit-integration technique.
In addition, since the Jacobian was based on an over-constrained system~\cite{HANSEN2018257}, the effect on strict conservation of mass\slash energy was not studied.

Recently, some groups have developed frameworks for constructing analytical Jacobians for evaluation on modern SIMD or SIMT processors.
Dijkmans et al.~\cite{Dijkmans:2014bb} developed a GPU-based analytical Jacobian code with optional tabulation of temperature-dependent properties, and showed speedups up to \SI{120}{$\times$} for zero-dimensional chemical kinetic integration with large chemical models (\textasciitilde\num{3000} species).
Bauer et al.~\cite{Bauer:2014} used warp-specialization to improve GPU-vectorization over a standard data-parallel vectorization approach; they achieved speedups of up to \SIrange{2.81}{3.75}{$\times$}, \SIrange{1.91}{2.58}{$\times$}, and \SIrange{1.4}{1.5}{$\times$} for evaluating viscosity, species diffusion, and chemical source terms, respectively.
Niemeyer et al.~\cite{Niemeyer:2016aa} created and verified the open-source analytical chemical kinetic Jacobian code-generator, \texttt{pyJac}, which supports parallel execution on CPUs and SIMT execution on GPUs; \texttt{pyJac} enables a speedup of \SIrange{3}{7.5}{$\times$} over a finite-difference Jacobian on the CPU.

Relevant to all of the aforementioned efforts, Hansen and Sutherland~\cite{HANSEN2018257} explored the choice of thermochemical state vectors and the resulting effect on consistency and errors in conserved properties such as mass and energy.
They also characterized how the choice of state vector affects implicit\slash linearly implicit integration algorithms and chemical mode analysis techniques.
Overall they found that while many literature Jacobian formulations are not strictly correct or over-specified, such flaws negligibly affect Newton--Krylov methods---perhaps because the incorrect Jacobian reasonably approximates the true Jacobian.
On the other hand, linearly implicit algorithms like Rosenbrock methods and analysis techniques like chemical explosive mode analysis~\cite{lu_yoo_chen_law_2010} need accurate and correct Jacobians.

A number of recent works have investigated using high-performance SIMT devices like GPUs to accelerate reactive-flow and chemical kinetics simulations.
Spafford et al.~\cite{Spafford:2010aa} coupled GPU-based chemical source-term evaluation with an explicit direct numerical simulation code, achieving an order of magnitude speedup compared to a CPU-based serial implementation.
Shi et al.~\cite{Shi:2011aa} combined GPU-based chemical kinetic source-term evaluation and Jacobian factorization with two implicit CPU solvers, achieving an order-of-magnitude speedup for homogeneous reactor simulations of large chemical models  over a serial CPU implementation.
Niemeyer et al.~\cite{Niemeyer:2011aa} implemented an explicit solver for non-stiff chemistry on a GPU, achieving a speedup of nearly two orders of magnitude over a sequential CPU code.
Shi et al.~\cite{Shi:2012aa} proposed a strategy for chemical-kinetic integration in three-dimensional reactive-flow simulations, where a traditional implicit integrator handled the stiffest computational cells on a CPU and a stabilized-explicit solver solved the less-stiff cells on a GPU; this hybrid solution technique performs \SIrange{11}{46}{$\times$} faster than the implicit CPU solver alone for simulation of a premixed diesel engine.
Le et al.~\cite{Le2013596} found a \SIrange{30}{50}{$\times$} speedup for a GPU-based shock-capturing reactive-flow code as compared with a sequential CPU version of the same.
Stone and Davis~\cite{Stone:2013aa} investigated a GPU-based version of a common implicit integrator (VODE~\cite{Brown:1989vl}), finding an order-of-magnitude speedup over a serial CPU implementation.
Niemeyer and Sung~\cite{Niemeyer:2014aa} developed a GPU-based stabilized explicit integrator for use with moderately-stiff chemical kinetics, achieving an order of magnitude speedup over a multithreaded VODE solver on a six-core CPU.
Sewerin and Rigopoulos~\cite{Sewerin20151375} studied a fifth-order implicit Runge--Kutta solver on both consumer-grade and high-end GPUs\slash CPUs; the high-end GPU solver was at best \SI{1.8}{$\times$} slower than the high-end CPU version running on \num{16} cores.
Yonkee and Sutherland~\cite{Yonkee2016} implemented accelerated evaluations of thermodynamic parameters, multicomponent transport properties, and species production rates on both the CPU and GPU, achieving speedups over serial evaluation between \SIrange{8}{13}{$\times$} on a 16-core CPU and \SIrange{20}{40}{$\times$} on the GPU.
In addition, \textasciitilde\SI{9}{$\times$} and \textasciitilde\SI{25}{$\times$} speedups were achieved for the simulation of a partially premixed methanol flame for solving partial differential equations (PDE) on 16 CPU cores and the GPU, respectively.
Curtis et al.~\cite{CurtisGPU:2017} implemented a fifth-order implicit Runge--Kutta method~\cite{wanner1991solving}, as well as two fourth-order exponential integration techniques~\cite{Hochbruck:1998,Hockbruck:2009} paired with an analytical Jacobian code~\cite{Niemeyer:2016aa} on the GPU and CPU.
The GPU-based implicit Runge--Kutta method performed equivalently to a standard implicit integrator~\cite{Hindmarsh:2005} running on \numrange{12}{38} CPU cores for two relatively small chemical models with an integration time step of $10^{-6}$ s.

In contrast, SIMD-based chemical kinetics evaluation\slash integration have been studied far less.
Linford et al.~\cite{Linford:2011} implemented a three-stage, second-order Rosenbrock integrator for atmospheric chemical kinetics on the CPU, GPU, and cell broadband engine (CBE)---a specially designed vector processor---and found speedups regularly exceeding~\SI{25}{$\times$} over a serial CPU implementation.
Kroshko and Spiteri~\cite{kroshko2013efficient} implemented a SIMD-vectorized third-order stiff Rosenbrock integrator for atmospheric chemistry on the CBE and found a speedup of \SI{1.89}{$\times$} (a parallel scaling efficiency of \SI{94}{$\percent$}) over a serial version of the same code.
Stone et al.~\cite{stone2018} implemented a linearly implicit fourth-order stiff Rosenbrock solver in the OpenCL for various platforms including CPUs, GPUs, and MICs.
They found that SIMD vectorization improves integrator performance over an OpenMP baseline vectorized by simple compiler hints (i.e., \texttt{\#pragmas}) by \SIrange{2.5}{2.8}{$\times$} on the CPU and \SIrange{4.7}{4.9}{$\times$} on the MIC, while the GPU performs only \SIrange{1.4}{1.6}{$\times$} faster than the OpenMP baseline due to thread divergence~\cite{stone2018}.

\subsection{Goals of this study}
\label{S:Goals}

In this article we
\begin{itemize}
 \item Derive and verify a new Jacobian formulation that greatly increases sparsity;
 \item Detail the implementation of cross-platform SIMD\slash SIMT vectorization for CPUs, GPUs, and other accelerators;
 \item Investigate the performance of SIMD\slash SIMT-vectorization for a wide range of chemical kinetic models, and compare with the previous version of our analytical chemical kinetic Jacobian code~\cite{Niemeyer:2016aa}; and finally
 \item Discuss future extensions to this work as well as several promising directions for SIMD\slash SIMT vectorization in reactive-flow simulations.
\end{itemize}
This work builds upon our previous analytical chemical kinetic Jacobian code, \texttt{pyJac}~\cite{Niemeyer:2016aa}, using the new formulation, \texttt{pyJac v2}, to achieve these goals.
To our knowledge is the first open-source, verified effort that vectorizes the evaluation of chemical-kinetic source terms and Jacobian matrices for any chemical model on a wide selection of platforms.

\section{Methodology}
\subsection{Data ordering and vectorization patterns}
\label{S:data}

\begin{figure}[htb]
  \centering
  \begin{minipage}{0.45\linewidth}
    \begin{subfigure}[t]{\textwidth}
      \includegraphics[width=\textwidth]{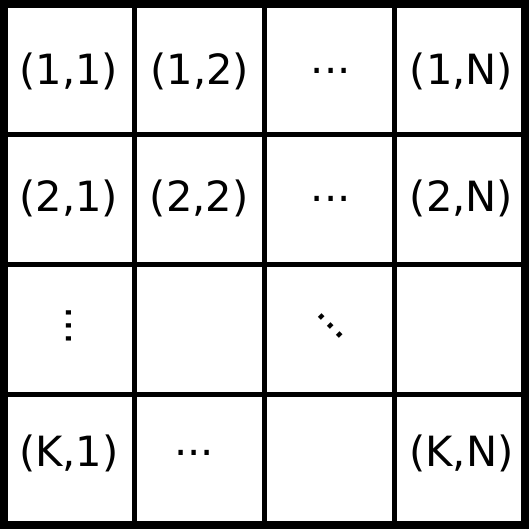}
      \caption{A simple 2-D data array with K rows and N columns.}
      \label{F:mem}
    \end{subfigure}
  \end{minipage}
  \hfil
  \begin{minipage}{0.45\linewidth}
    \begin{subfigure}[t]{\textwidth}
	\includegraphics[width=\textwidth]{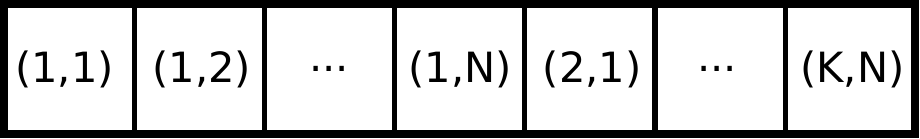}
	\caption{Row-major data ordering}
	\label{F:row_major}
    \end{subfigure}
    \\
    \\
    \\
    \\
    \begin{subfigure}[t]{\textwidth}
	\includegraphics[width=\textwidth]{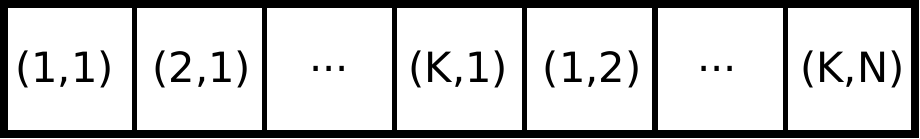}
	\caption{Column-major data ordering}
	\label{F:column_major}
    \end{subfigure}
  \end{minipage}
  \caption{Simple data-layout patterns for 2-D arrays}
\end{figure}

When storing arrays for a chemical kinetic model, the data-storage layout and vectorization patterns are critical to achieving high-performance code.
\Cref{F:mem} depicts an example data array with $K$ rows and $N$ columns where index ($i$, $j$) corresponds to the $i$th row and $j$th column.
For example, the concentration of species $j$ for the $i$th thermochemical state would be stored in $[C]_{i, j}$ with $1 \le i \le N_{\text{state}}$ (the number of thermochemical states considered for evaluation) and $1 \le j \le \ns/$ (the number of species in the model).
The stored concentrations would then have $K = N_{\text{state}}$ rows and $N = \ns/$ columns.

The ``C'' (C row-major) format stores the concentrations of all species for a single thermochemical condition $i$  sequentially in memory, i.e., with  $[C]_{1, 1}$ in index 1 (using one-based index notation), $[C]_{1, 2}$ in index 2, and so on, as shown in \cref{F:row_major}.
Conversely, in the ``F'' (Fortran column-major) format the concentrations of a single species $j$ over all thermochemical states lie adjacent in memory, corresponding to storing $[C]_{1, 1}$ in index 1, $[C]_{2, 1}$ in index 2, and so on, as shown in \cref{F:column_major}.
This ordering strongly affects the performance of SIMD\slash SIMT-vectorized algorithms, as does the device (CPU, GPU, etc.\@) and vectorization pattern in question.

In a \textit{shallow} SIMD\slash SIMT vectorization (also referred to as ``per-thread'' in previous works using GPUs~\cite{Stone:2013aa}), each SIMD lane or SIMT thread in a compute unit evaluates the source terms or Jacobian for a different thermochemical state.
If the data is stored in ``F''-order, the SIMD lanes\slash SIMT thread accessing $[C]_{1, j}\ldots[C]_{N_v, j}$ will load sequential locations in memory, where $[C]_{i, j}$ is the concentration of species $j$ for state $i$ and $N_v$ is the SIMD vector-width or the number of threads in a SIMT warp.
The first $(j+1)$th species concentration, $[C]_{1, j+1}$, will be $N_{\text{state}}$ memory locations away; this increases the likelihood of cache misses on the CPU~\cite{gray2000rules}, but conversely well matches the pattern of coalesced memory access on the GPU~\cite{Nvidia:2018}.

In a \textit{deep} SIMD\slash SIMT vectorization (also referred to as ``per-block'' in previous GPU works~\cite{Stone:2013aa,CurtisGPU:2017}), a compute unit uses its SIMD lanes\slash SIMT threads cooperatively to evaluate the thermochemical source terms for a single thermochemical state; thus SIMD lanes loading $[C]_{1, j} \ldots [C]_{1, j + N_v}$ will access sequential memory locations if the data is stored in ``C''-order.
Further, in ``C'' ordering any two species concentrations within the same thermochemical state lie at most $N_{\text{sp}}$ locations away, with $N_{\text{sp}} \ll N_{\text{state}}$ in most cases; this greatly improved data locality increases the chances of a cache hit on the CPU, but may lead to uncoalesced memory accesses on the GPU.
 Deep vectorization requires synchronization between SIMD lanes\slash SIMT threads via memory fences\slash barriers, a potentially expensive operation.
In addition, deep vectorization may result in SIMD waste or SIMT thread divergence caused by different lanes\slash threads executing different instructions (e.g., resulting from different if\slash then branches).
Shallow vectorization may also experience SIMD waste or SIMT thread divergence, e.g., in chemical kinetic integration due to varying internal solver time-step sizes~\cite{CurtisGPU:2017}.
However, in this work shallow vectorization is largely unaffected by this concern as the only major code paths that differ between vector lanes are high\slash low-temperature polynomial evaluations and differing pressures for P-Log reactions, which cause far fewer issues compared with differing internal ODE integration time-steps~\cite{CurtisGPU:2017}.

\begin{figure}[htb]
  \centering
  \begin{minipage}{0.6\linewidth}
    \begin{subfigure}[t]{\textwidth}
	\includegraphics[width=\textwidth]{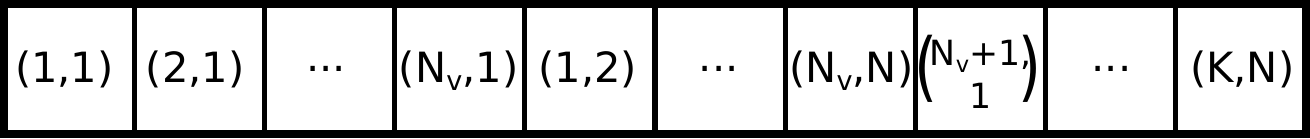}
	\caption{Row-major, shallow-vectorized data ordering}
	\label{F:row_major_split}
    \end{subfigure}
    \\
    \\
    \\
    \\
    \begin{subfigure}[t]{\textwidth}
	\includegraphics[width=\textwidth]{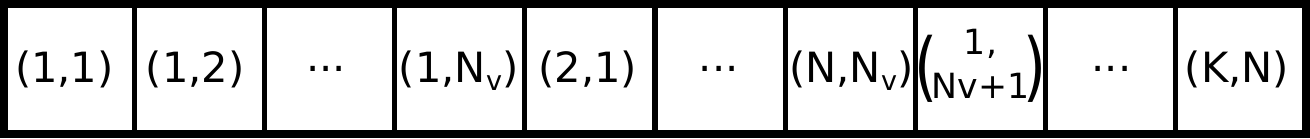}
	\caption{Column-major, deep-vectorized data ordering}
	\label{F:column_major_split}
    \end{subfigure}
  \end{minipage}
  \caption{Vectorized data-ordering patterns}
  \label{F:vector_data}
\end{figure}

Finally,~\cref{F:vector_data} shows a vectorized data-ordering that improves the caching patterns of a shallow, ``C''-ordered SIMD vectorization on the CPU (\cref{F:row_major_split}) and a deep, ``F''-ordered SIMT vectorization on the GPU (\cref{F:column_major_split}).
We accomplish this by splitting the slower-varying axis of the data array---columns for ``C''-ordering, and rows for ``F''-ordering---into chunks of size $N_v$ (the SIMD vector width or SIMT warp size) and laying these data out contiguously in memory.
For example, using the shallow-vectorized ``C''-ordering pictured in~\cref{F:row_major_split}, the concentrations of species $j$ for states $i$ to $i+N_v$ ($[C]_{i, j}, \ldots, [C]_{i + N_v, j}$) lie contiguously in memory and are followed by the concentrations of species $j + 1$ for the same states ($[C]_{i, j + 1}, \ldots, [C]_{i + N_v, j + 1}$).
This pattern ensures that any SIMD operation occurs on data contiguous in memory, which greatly improves caching and SIMD throughput;
it is also similar to OpenCL's native vector data-types, e.g., \texttt{double\num{8}} treats eight contiguous double-precision floating-point numbers as a single vector datum.
Conversely, the data-ordering in~\cref{F:column_major_split} enables coalesced memory accesses for ``F''-ordered, deep SIMT vectorization on the GPU.
We will discuss the effects of these various data-ordering and vectorization patterns on performance in~\cref{S:results}.

\subsection{Thermochemical source terms and Jacobian}
This new version of \texttt{pyJac} is capable of evaluating the thermochemical source-terms for using the constant-pressure (\conp/) or constant-volume (\conv/) assumption\footnote{Note: in this context, the ``constant-pressure'' and ``constant-volume'' assumptions refer to evaluation within a reaction sub-step in the operator splitting scheme, rather than a general constant-pressure or constant-volume reactive-flow simulation.}
In this section, we will outline a brief summary of the system evaluated by \texttt{pyJac}; the supplemental material contains the complete---and lengthy---derivations.

The thermochemical state vector consists of the temperature, a non-constant thermodynamic state parameter (volume or pressure for \conp/ and \conv/, respectively), and the number of moles of all species except the last species in the chemical model, typically taken to be the bath gas (e.g., \ce{N2}):
\loadeq{state}
where $T$ is the temperature, $V$ and $P$ the volume and pressure respectively, and $n_j$ the number of moles of the $j$th species in the model (containing $\ns/$ total species).

This state vector---inspired by Schwer et al.~\cite{SCHWER2002270}---has a number of beneficial features.
First, the state vector results in highly sparse chemical kinetic Jacobians, as will be detailed in~\cref{S:sparsity}.
Second, this formulation explicitly conserves mass, because the number of moles and rate of change of the final species are calculated from the ideal gas law and conservation of mass, respectively; see the supplemental material for the full details of the governing equations.
The system is not over-constrained~\cite{HANSEN2018257} and does not require use of a more-complicated differential algebraic equation solver (as compared to an ODE integrator) for integration.
Finally, the chemical kinetic Jacobian for this formulation changes relatively little between the \conp/ and \conv/ forms, making maintaining the codebase much simpler.
Although most current combustion codes do not use species moles as a state variable, conversion to\slash from the more-common mass\slash mole fractions and moles is straightforward, and the choice of variables no longer matters once inside the integration of an chemical kinetic initial-value problem (IVP).

The evolution of the thermochemical state vector is described by a set of chemical kinetic ordinary differential equations:
\loadeq{dstate}
For both \conp/ and \conv/, the molar source terms are~\cite{TurnsStephenR2012Aitc}:
\loadeq{dmolar}
where $\dot{\omega}_k$ is the $kth$ species' overall molar production rate:
\begin{equation}
 \dot{\omega}_{k} = \sum_{i=1}^{\nr/} \nu_{k, i} R_{i} c_{i} \;,
\end{equation}
$\nu_{k, i}$ is the net stoichiometric coefficient of species $k$ in reaction $i$, $N_{reac}$ is the total number of reactions, $R_{i}$ is the net rate of progress of reaction $i$, and $c_{i}$ is the pressure-dependent modification term, i.e., for third-body or falloff\slash chemically-activated reactions.
\texttt{pyJac} is capable of evaluating all modern reaction types, e.g., P-Log and Chebyshev reactions.

The temperature source-term~\cite{TurnsStephenR2012Aitc} is:
\loadeq{dtemp}
where $H_k$, $U_k$, $C_{p,k}$, and $C_{v, k}$ are the enthalpy, internal energy, constant-pressure specific heat, and constant-volume specific heat of species $k$ in molar units, respectively, while $[C]_{k}$ is the concentration, given by
\begin{equation}
 [C]_{k} = \frac{n_{k}}{V} \;.
\end{equation}
By differentiating the ideal gas law, given by
\begin{equation}
 PV = n\mathcal{R}T
\end{equation}
where $\mathcal{R}$ is the ideal-gas constant in molar units, we find the volume and pressure source terms (where $W_k$ and $W_{\ns/}$ are the molecular weights of species $k$ and $\ns/$, respectively):
\loadeq{dparam}

\texttt{pyJac} arranges the computed Jacobian entries such that entry ($i$, $j$) corresponds to the partial derivative of the $i$th source-term in~\cref{e:source_terms} by the $j$th state variable in~\cref{e:state}:
\loadeq{jacgeneral}

The supplemental material for this article contains the complete derivation of the Jacobian used by \texttt{pyJac}, for interested readers.

\subsection{Code generation and testing infrastructure}
\label{s:unittest}

The new version of \texttt{pyJac} uses the Python package \texttt{loo.py}~\cite{kloeckner_loopy_2014} for code generation, which translates pseudo-code and data to OpenCL\slash C code.
As the name implies, \texttt{loo.py} generates code using for loops; this differs from the previous version of \texttt{pyJac}~\cite{pyjac16} that generates static code---i.e., fully unrolled loops, with thermodynamic\slash reaction parameters written directly in code rather than stored in arrays.
In our previous work~\cite{Niemeyer:2016aa}, this static code generation caused some issues with large file sizes, long compilation times, and even occasionally broke the \texttt{gcc} and \texttt{nvcc} compilers (the latter issue necessitated splitting the Jacobian\slash source-term evaluations into separate files).
We will discuss the implications of this change in~\cref{S:jacobian_results}, where the performance of the new version of~\texttt{pyJac} will be compared with the previous version.

In addition, \texttt{loo.py} allows the user to more easily make changes to the structure of the generated program, e.g., the data ordering, vectorization, and threading patterns, as well as switch the target language for code generation (and more simply extend to additional languages, e.g., CUDA).
Further, \texttt{loo.py} can execute developed subroutines from Python (natively for C code, or via \texttt{PyOpenCL}~\cite{kloeckner_pycuda_2012} for OpenCL), enabling unit testing\slash verification for each component of the Jacobian or source terms; the unit testing suite also helps ensure that bugs are not present in less commonly used code-paths, or are introduced by future code changes.
The source terms and sub-components thereof (e.g., rates of progress, pressure-modification terms) are directly compared with Cantera~\cite{Cantera}, while the automatic differentiation code Adept~\cite{hogan2014fast,adept-v11} provides reference values for Jacobian sub-components.
We use the Portable OpenCL (POCL) implementation~\cite{poclIJPP} and OpenMP~\cite{dagum1998openmp} to perform OpenCL and C unit testing, respectively, on the continuous-integration framework Travis CI~\cite{travis:2018}.
We will discuss verification of the complete (as opposed to the sub-component testing discussed here) generated source-terms and Jacobian codes in detail in~\cref{s:verification}.

\section{Results and discussion}
\subsection{Testing platforms}
\label{s:test_platforms}

\begin{table}[htb]
\centering
\begin{tabular}{@{}l l l@{}}
\toprule
CPU Model        & Xeon X5650      & E5-2690 V3     \\
\midrule
Instruction Set  & SSE4.2 	   & AVX2 	    \\
Vector Width     & two doubles & four doubles \\
Cores            & $2 \times 6$    & $2 \times 12$  \\
Identifier       & \sse/ 	   & \avx/  	    \\
OpenCL Version   & \num{1.2}       & \num{1.2}      \\
\bottomrule
\end{tabular}
\caption{The Intel CPU configurations used in this study.
	 The vector widths are reported in (ideal) number of double operations per SIMD instruction, as this will be used in measuring SIMD efficiency; for reference, the vector widths of the \sse/ and \avx/ machines are \SI{128}{\bit} and \SI{256}{\bit}, respectively.
	 The identifier field will be used as a shorthand descriptor in the performance plots to quickly identify the CPU type.}
\label{t:cpus}
\end{table}

We ran the performance and verification studies for this work on a variety of CPU and GPU platforms.
\Cref{t:cpus} shows the number of cores, vector instruction set, and model of the CPUs used in this work; each CPU had both \texttt{v16.1.1} of the Intel OpenCL runtime~\cite{intelopencl:2018} and \texttt{v1.0} of the POCL~\cite{poclIJPP} runtime installed, both enabling OpenCL \texttt{v1.2} execution.
Additionally, \texttt{v5.0} of the LLVM\slash\texttt{clang}~\cite{Lattner:2004:LCF:977395.977673} compiler chain was installed on all machines to enable use of POCL.
\Cref{t:gpus} lists the model, number of CUDA cores, and Nvidia driver of each GPU we used.
Nvidia's OpenCL runtime is bundled with the Nvidia driver~\cite{Nvidia:2018}, hence the driver version is used to specify the OpenCL runtime version.

\begin{table}[htb]
\centering
\begin{tabular}{@{}l l l l l@{}}
\toprule
Nvidia Model   & Tesla C2075    & Tesla K40m    \\
\midrule
Driver Version & \num{384.81}   & \num{387.26}  \\
CUDA Cores     & \num{448}      & \num{2880}    \\
Identifier     & \gpuold/ 	& \gpunew/	\\
OpenCL Version & \num{1.1}	& \num{1.2}	\\
\addtocounter{footnote}{1}Memory\footnotemark[\thefootnote] & \SI{6}{\giga\byte} & \SI{12}{\giga\byte}  \\
\bottomrule
\end{tabular}
\caption{The Nvidia GPU configurations used in this study.  Nvidia's OpenCL runtime is provided with the graphics driver, rather than any specific version of CUDA.
The identifier field will be used as a shorthand descriptor during analysis of the performance results.
}
\label{t:gpus}
\end{table}
\footnotetext{A driver implementation issue limited total memory to \SI{4}{\giga\byte} and \SI{10}{\giga\byte} on the \gpuold/ and \gpunew/ GPUs, respectively~\cite{Nvidia_memory}.\addtocounter{footnote}{1}}

\Cref{t:platforms} lists the platforms and vectorization\slash execution patterns that they are capable of running.
The Intel and Nvidia OpenCL runtimes lack implementations of atomic operations on double-precision variables; \texttt{pyJac} currently needs these to run deep-vectorized code.
On the other hand, POCL is an open-source OpenCL runtime that works on all CPU types tested here, and does implement these atomic operations.
However, POCL's implicit vectorization module---which uses the LLVM compiler~\cite{Lattner:2004:LCF:977395.977673} to translate OpenCL code to vectorized machine code---typically fails to achieve much, if any, speedup.
Thus POCL is useful for verification but not necessarily for performance studies; it is noted that while POCL is currently used by \texttt{pyJac} for unit-testing purposes, it is not required to use \texttt{pyJac}.
We will expand upon this discussion in~\cref{S:future} to highlight future directions.

\begin{table}[htb]
\centering
\begin{tabular}{@{}l c c c@{}}
\toprule
Platform & Parallel & Shallow Vectorization & Deep Vectorization \\
\midrule
OpenMP & \checkmark & -- & -- \\
POCL OpenCL & \checkmark & \checkmark & \checkmark \\
Intel OpenCL & \checkmark & \checkmark & -- \\
Nvidia OpenCL & -- & \checkmark & -- \\
\bottomrule
\end{tabular}
\caption{The platforms used in this study and the execution \slash vectorization patterns that they are capable of running.}
\label{t:platforms}
\end{table}

Finally, \cref{t:models} displays the chemical kinetic models used in this work, as well as number of partially stirred reaction conditions (PaSR) used in the condition database for each.
Our previous works describe the creation of the PaSR databases in detail works~\cite{CurtisGPU:2017,Niemeyer:2016aa}.

\begin{table}[htb]
\centering
\begin{tabular}{@{}l c c @{}}
\toprule
Model &  Number of Conditions & Reference \\
\midrule
\ce{H2}\slash\ce{CO} & \num{900900} & \cite{Burke:2011fh} \\
GRI-Mech 3.0         & \num{450900} & \cite{smith_gri-mech_30} \\
USC-Mech II           & \num{91800}  & \cite{Wang:2007} \\
\ce{iC5H11OH}         & \num{450900} & \cite{Sarathy:2013jr} \\
\bottomrule
\end{tabular}
\caption{The chemical kinetic models used in this study and number of conditions in the partially stirred reactor database for each.}
\label{t:models}
\end{table}

\subsection{Source-term verification}
\label{s:verification}
We verified the reaction rates of progress (ROP), species production rates, and temperature rates in this study by comparing with values calculated using Cantera~\cite{Cantera}.
However, special care must be taken due to floating-point arithmetic issues.

For a direct comparison, a relative error norm of a quantity $X_{ij}$ over all states $j$ and reactions (or species) $i$ was computed using the $L^{\infty}$ norm:
\begin{equation}
E_{X} = \left\lVert \frac{\left\lvert X_{ij,\text{CT}} - X_{ij}\right\rvert}{\num{e-10} + \num{e-6} \times \left\lvert X_{ij,\text{CT}} \right\rvert} \right\rVert_{\infty} \;,
\label{e:rel_err}
\end{equation}
where the \text{CT} subscript indicates values from Cantera~\cite{Cantera}.

However, computing the net ROP of reaction $i$ for state $j$ from the forward and reverse ROP, i.e., $R_{ij} = R_{ij}^{\prime} - R_{ij}^{\prime\prime}$, can easily lose accuracy as the net ROP may be many orders of magnitude smaller than the forward and\slash or reverse rates---particularly near chemical equilibrium.
To quantify this phenomena, we first define the error in forward ROP as
\begin{equation}
\varepsilon^{\prime}_{ij} = \left\lvert R_{ij}^{\prime} - R_{ij,\text{CT}}^{\prime} \right\rvert \;,
\end{equation}
while the error in reverse ROP, $\varepsilon^{\prime\prime}_{ij}$, can be defined analogously.
Finally, for the reaction $i^{*}$ and the state $j^{*}$ that result in the largest error in net ROP, i.e., $E_{R}$, an estimate of the error attributable to floating-point error accumulation from the forward and reverse ROPs can be obtained using
\begin{equation}
E_{\varepsilon} = \frac{\max(\varepsilon^{\prime}_{i^{*}j^{*}}\text{, }\varepsilon^{\prime\prime}_{i^{*}j^{*}})}{\num{e-10} + \num{e-6} \times \left\lvert R_{i^{*}j^{*},\text{CT}} \right\rvert} \;.
\end{equation}
This estimate allows for directly comparing the error in forward or reverse ROPs with the value of the net ROP itself; the error in net ROP will be large if these are similar in magnitude.

\begin{table}[htbp]
\sisetup{retain-zero-exponent=true}
\centering
\begin{tabular}{@{}l S[table-format=1.2e1] S[table-format=1.2e1] S[table-format=1.2e1] S[table-format=1.2e1] @{}}
\toprule
{Model} & {\ce{H2}\slash\ce{CO}} & {GRI-Mech.~3.0} & {USC-Mech II} & {\ce{iC5H11OH}} \\
\midrule
$E_{R^{\prime}}$                    & 1.56e-8 & 2.95e-8 & 9.42e-8 & 4.86e-4 \\
$E_{R^{\prime\prime}}$              & 6.92e-8 & 6.53e-8 & 1.20e-7 & 5.07e-4 \\
$E_{R}$                             & 1.49e1  & 1.11e0  & 2.80e0  & 4.82e-1 \\
$E_{\varepsilon}$                   & 1.48e1  & 1.13e0  & 2.93e0  & 5.03e-1 \\
$E_{\frac{\text{d} n}{\text{d} t}}$ & 2.53e1  & 2.60e0  & 7.62e0  & 1.58e1 \\
$E_{\frac{\text{d} T}{\text{d} t}}$ & 3.94e5  & 3.35e8  & 3.95e6  & 7.11e7 \\
$E_{\frac{\text{d} S}{\text{d} t}}$ & 3.52e12 & 3.46e12 & 3.44e12 & 3.38e12 \\
\bottomrule
\end{tabular}
\caption{Summary of errors in rates of progress, species, temperature, and thermodynamic state-parameter rate compared with Cantera.
Error statistics are based on the infinity-norm of the relative error detailed in~\cref{e:rel_err} for each quantity.
The ``S'' in $E_{\frac{\text{d} S}{\text{d} t}}$ refers to the thermodynamic state parameter, either $V$ or $P$ for \conp/ and \conv/, respectively.
}
\label{T:source_error}
\end{table}

Table~\ref{T:source_error} compares \texttt{pyJac} v2's source-term evaluations with Cantera's~\cite{Cantera} using the data set of PaSR conditions (\cref{t:models}).
The forward and reverse ROPs agree closely for all models, though the error norm is \textasciitilde\numrange{3}{4} orders of magnitude larger for the isopentanol model.
This discrepancy results from differences in evaluation of P-Log reactions between \texttt{pyJac} and Cantera: \texttt{pyJac} computes the logarithm of the upper and lower reaction Arrhenius rates analytically (see supplemental material) while Cantera evaluates this term numerically.
If we neglect the errors from P-Log reactions in~\cref{e:rel_err}, the errors for the forward and reverse ROPs fall to \num{5.44e-08} and \num{1.59e-07}, respectively.
This discrepancy does not imply any actual error in either \texttt{pyJac} or Cantera---in fact, the error still lies well within the proscribed tolerances in~\cref{e:rel_err}---but merely emphasizes
how even small code changes can affect the accumulation of floating-point errors.

The error in the net ROP further underscores this point: it is
\textasciitilde\numrange{3}{9} orders of magnitude (or \numrange{7}{9} orders of magnitude when including P-Log reaction contributions) larger than the error in forward or reverse ROP.
\Cref{T:source_error} shows that the magnitudes of $E_{\varepsilon}$ and $E_R$ agree in all cases, indicating that the accumulation of floating-point error from the forward and reverse ROPs causes this large increase in error as previously discussed.
The magnitudes of the errors in molar species production rate and net ROP agree, but
thermodynamic properties amplify the error in net species production rates and lead to high discrepancies in temperature and state-parameter rates.
Again, these discrepancies in net ROP will not necessarily cause errors when integrating the chemical kinetics---either in \texttt{pyJac} or Cantera---as this loss of accuracy only occurs when the forward and reverse ROPs are nearly equal (i.e., near equilibrium).

\subsection{Jacobian verification}
\label{S:jac_valid}

As in our previous work~\cite{Niemeyer:2016aa}, we determined Jacobian matrix correctness by comparing with that obtained by automatic differentiation of the \texttt{pyJac}-generated source term, using the \texttt{Adept} software library~\cite{adept-v11,hogan2014fast}.
We previously explained this choice fully~\cite{Niemeyer:2016aa}, but broadly speaking automatic differentiation provides relatively fast, highly accurate Jacobian matrix evaluation with minimal additional programming effort.
(In contrast, it is challenging to obtain robust, accurate Jacobians using finite differences.)
The discrepancy between the analytical and automatic-differentiation Jacobians for thermochemical state $k$, denoted by $\mathcal{J}_k$ and $\hat{\mathcal{J}}_k$ respectively, is determined by the relative error Frobenius norm over all Jacobian indices $i, j$:
\begin{equation}
 \label{e:jac_error_base}
 E_{\text{rel}, k} = \left\lVert \frac{\hat{\mathcal{J}}_{ij,k} - \mathcal{J}_{ij,k}}{\hat{\mathcal{J}}_{ij,k}} \right\rVert_{F} \;.
\end{equation}

To avoid large relative discrepancies in small nonzero Jacobian elements due to accumulation of floating-point error, the Frobenius norm of the automatically differentiated Jacobian is calculated over all thermochemical states $k$:
\begin{equation}
 \label{e:thresh}
 \mathcal{T} = \left\lVert \mathcal{\hat{J}} \right\rVert_{F} \;.
\end{equation}
The error statistics reported in this section are then based only on matrix elements where $\mathcal{J}_{ijk} \ge \frac{\mathcal{T}}{\mathcal{C}}$, where $\mathcal{C}$ is a tunable threshold parameter; this filtered form of~\cref{e:jac_error_base} is denoted $E_{\mathcal{C},k}$.
Finally, the Frobenius norm is calculated over all the states $k$ in the PaSR thermochemical condition database:
\begin{equation}
 \label{e:thresholded_error}
 E_{\mathcal{C}} = \left\lVert E_{\mathcal{C},k} \right\rVert_{F} \;.
\end{equation}

This error norm is quite different from the relative error Frobenius norm suggested by Anderson et al.~\cite{Anderson:1999aa} for quantifying the error of matrices in LAPACK, e.g., over Jacobian indices $i, j$:
\begin{align}
 \label{e:error_lapack}
 \lhs{lapack}{E_{\mathcal{L}, k}} &=  \rhs{lapack}{\frac{\left\lVert \hat{\mathcal{J}}_{ijk} - \mathcal{J}_{ijk} \right\rVert_{F}}{\left\lVert \hat{\mathcal{J}}_{ijk} \right\rVert_{F}}} \nonumber \\
\intertext{and states $k$:}
 \lhs{lapack}{E_{\mathcal{L}}} &= \rhs{lapack}{\left\lVert  E_{\mathcal{L}, k} \right\rVert_{F}}[c] \;.
\end{align}
In our experience, the accuracy of larger elements in a Jacobian often dominates the LAPACK error norm, while the filtered error norm can identify errors in both large and small Jacobian entries.
Further, with the tunable threshold parameter $\mathcal{C}$, we can assess the error of different ranges of element sizes and isolate the effects of floating-point error.
For reference, both our error norm and the LAPACK error norm will be reported.

\begin{table}[htbp]
\sisetup{retain-zero-exponent=true}
\centering
\begin{tabular}{@{}l S[table-format=.0] S[table-format=1.3e1] S[table-format=1.3e1] S[table-format=1.3e1] S[table-format=1.3e1] @{}}
\toprule
Model                 & \multicolumn{1}{c}{$E_{\mathcal{L}}$} & \multicolumn{1}{c}{$\bar{\mathcal{T}}$} & \multicolumn{1}{c}{$E_{\mathcal{C} = 10^{20}}$}   & \multicolumn{1}{c}{$E_{\mathcal{C} = 10^{15}}$} \\
\midrule
\ce{H2}\slash \ce{CO} & \num{1.862e-14}      & \num{6.431e+18}      & \num{1.741e+0}  & \num{4.508e-5} \\
GRI-Mech 3.0          & \num{1.567e-14}      & \num{7.783e+19}      & \num{3.842e-7}  & \num{3.687e-7} \\
USC-Mech II           & \num{1.137e-14}      & \num{2.830e+21}      & \num{1.199e-2}  & \num{1.983e-7} \\
\ce{iC5H11OH}         & \num{1.227e-10}      & \num{2.733e+26}      & \num{1.363e-3}  & \num{2.764e-5} \\
\bottomrule
\end{tabular}
\caption{Summary of Jacobian matrix verification results.
The reported error statistics are the maximum filtered relative error $E_\mathcal{C}$ and LAPACK error $E_{\mathcal{L}}$ over all test platforms, vectorization patterns (\cref{t:platforms}),  \conp/\slash \conv/, and sparse\slash dense Jacobians.
The Frobenius norm described in~\cref{e:thresh} varies slightly between the \conp/ and \conv/ cases; the reported $\bar{\mathcal{T}}$ is the average of the two, with the appropriate value used during calculations of the error statistics.
}
\label{T:error}
\end{table}

\Cref{T:error} reports the maximum $E_{\mathcal{C}}$ and $E_{\mathcal{L}}$ values over all test platforms and vectorization patterns (see~\cref{t:platforms}), sparse and dense (see~\cref{S:sparsity}), as well as \conp/ and \conv/ formulations.
The most stringent filtered error norm ($\mathcal{C} = 10^{20}$) ranges from \numrange[retain-zero-exponent=true]{e-7}{e0}; the largest error is for the \ce{H2}\slash\ce{CO} model.
For this model, $\mathcal{T}$ is smaller than the tolerance of \num{e20}, and hence the error norm considers Jacobian entries smaller than $\mathcal{O}(1)$.
GRI-Mech 3.0 has a $\mathcal{T}$ roughly an order of magnitude larger and so the stringent error norm is significantly smaller: $\mathcal{O}(\num{e-7})$.
Given the intricacy of floating-point error evaluation, the use of different languages and OpenCL platforms (the effect of these differences will be explored in \cref{A:per_platform}), and the general complexity of \texttt{pyJac} it would be exceedingly difficult to pinpoint an exact cause for this phenomenon, as was done in~\cref{s:verification}.
To ensure no bugs or errors present in the Jacobians generated by \texttt{pyJac}, relaxed filtered error norms ($\mathcal{C} = 10^{15}$) are also presented for each model in~\cref{T:error}.
This relaxed norm is smaller by \numrange{2}{5} orders of magnitude for all models---except GRI-Mech 3.0, where the stringent case already has small error as previously discussed---which indicates that accuracy is higher when the smaller Jacobian entries are excluded.
This result suggests that floating-point error accumulation controls the stringent filtered error norm.

The relative LAPACK error norm---ranging from \textasciitilde\numrange{e-10}{e-14}---re-enforces this finding, as it indicates roughly \numrange{10}{14} digits of accuracy~\cite{Anderson:1999aa}.
The \ce{iC5H11OH} model has the largest LAPACK error norm, likely due to the presence of P-Log\slash Chebyshev reactions and the resulting complicated derivatives with many logarithms, exponetiations, and summations.
Further, the LAPACK error norm does not correlate well with the stringent filtered error norm, e.g., USC-Mech II has the smallest LAPACK error norm (\num{1.137e-14}) but the second-largest stringent filtered error norm (\num{1.119e-2}).
Conversely, the model with the largest LAPACK error norm, \ce{iC5H11OH} has the second smallest stringent filtered error norm: $E_{\mathcal{L}} = \num{1.227e-10}$ and $E_{\mathcal{C} = 10^{20}} = \num{1.363e-3}$, again suggesting that floating-point error accumulation influences the stringent error norm.
These findings, along with the individual unit-testing of all chemical source-terms and Jacobian sub-components described in~\cref{s:unittest}, gives high confidence in the correctness of \texttt{pyJac} v2.

\subsection{Sparsity patterns}
\label{S:sparsity}

In general, the Jacobian matrices generated by \texttt{pyJac} are largely sparse with non-zero entries corresponding to species that participate in the same reaction or non-default efficiency third-body species in a reaction, with dense rows\slash columns corresponding to temperature and the thermodynamic state parameter.
However, the explicit-mass conservation formulation of \texttt{pyJac} can introduce additional non-zero entries in two ways.
First, if the last species in the model (i.e., the bath gas) participates directly in any reaction, the derivative of its forward or reverse rate of progress is non-zero with respect to all other species in the model, regardless of whether the other species participate in that reaction or not.
Similarly, if the last species has a third-body efficiency not equal to the default (one), this will again create nonzero derivatives for the pressure-modification term with respect to all other species (see the supplemental material).
Either case will result in a fully dense Jacobian row for all species with a non-zero net stoichiometric coefficient in such a reaction.

However, \texttt{pyJac} v2 allows the user to ignore these derivatives (via a command-line switch) and avoid the adverse effects on Jacobian sparsity.\footnote{Alternatively, one may choose the last species as one that does not participate in any reactions.}
The rationale behind this choice is that many common implicit integration techniques (e.g., CVODE~\cite{Hindmarsh:2005}) used to solve chemical-kinetic initial-value problems are formulated around the assumption that the supplied Jacobian is approximate; this allows the Jacobian and its LU factorization to be reused for multiple internal integration time steps, accelerating the solution process.
Such solvers do not need the exact form of the Jacobian and thus the so-called ``approximate'' form is preferable.
Though this might be used as a crude form of preconditioning for such solvers, the primary purpose is merely to increase Jacobian sparsity; McNenly et al.~\cite{MCNENLY2015581} more thoroughly investigated preconditioners.
Hence, in this section we will detail the sparsity of both forms of the Jacobian for the chemical models tested.

\begin{figure}[htb]
  \centering
  \begin{subfigure}[t]{0.48\linewidth}
      \includegraphics[width=\textwidth]{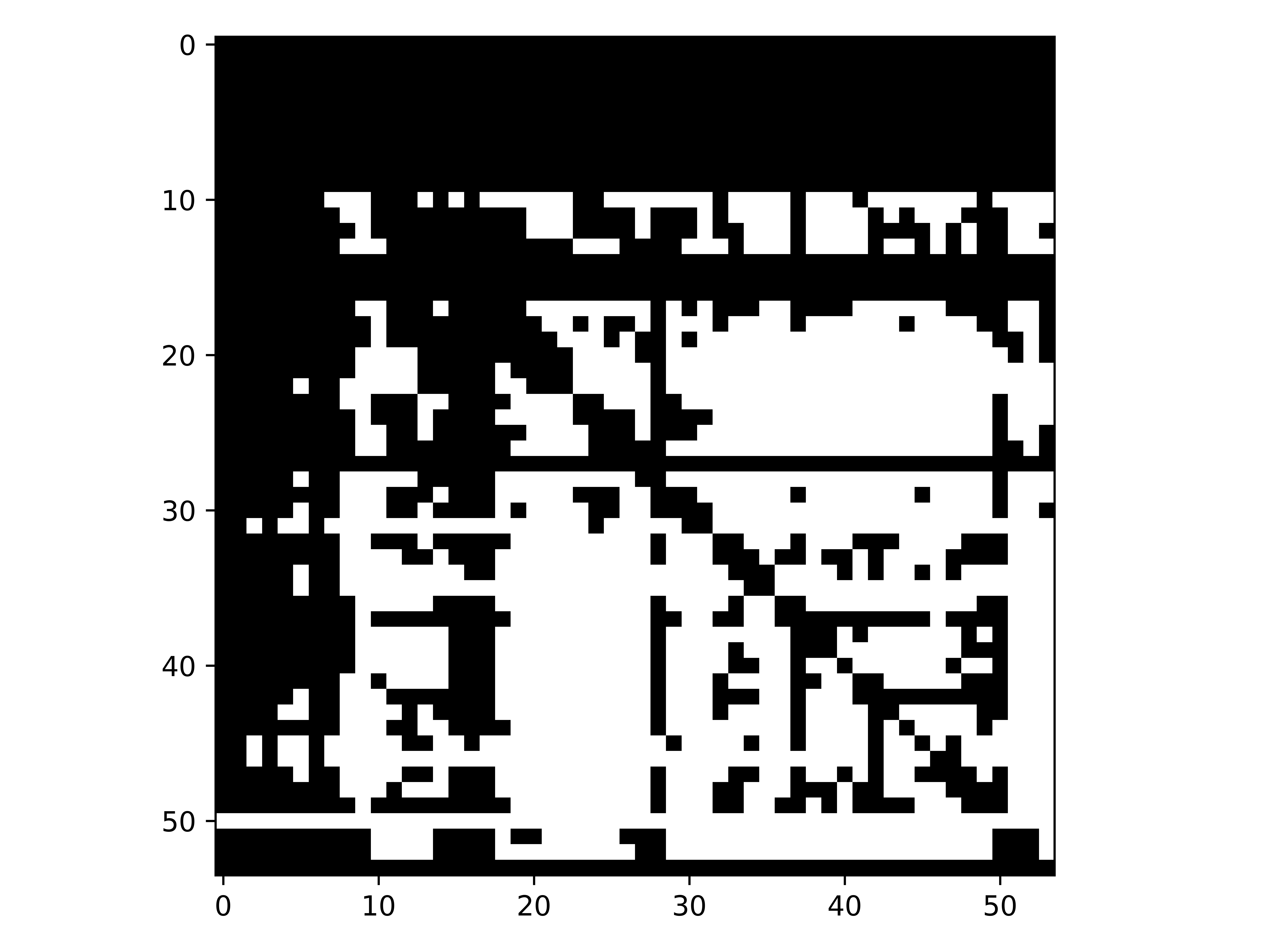}
      \caption{The ``exact'' Jacobian.}
      \label{F:jac_sparse_exact}
  \end{subfigure}
  \hfill
  \begin{subfigure}[t]{0.48\linewidth}
      \includegraphics[width=\textwidth]{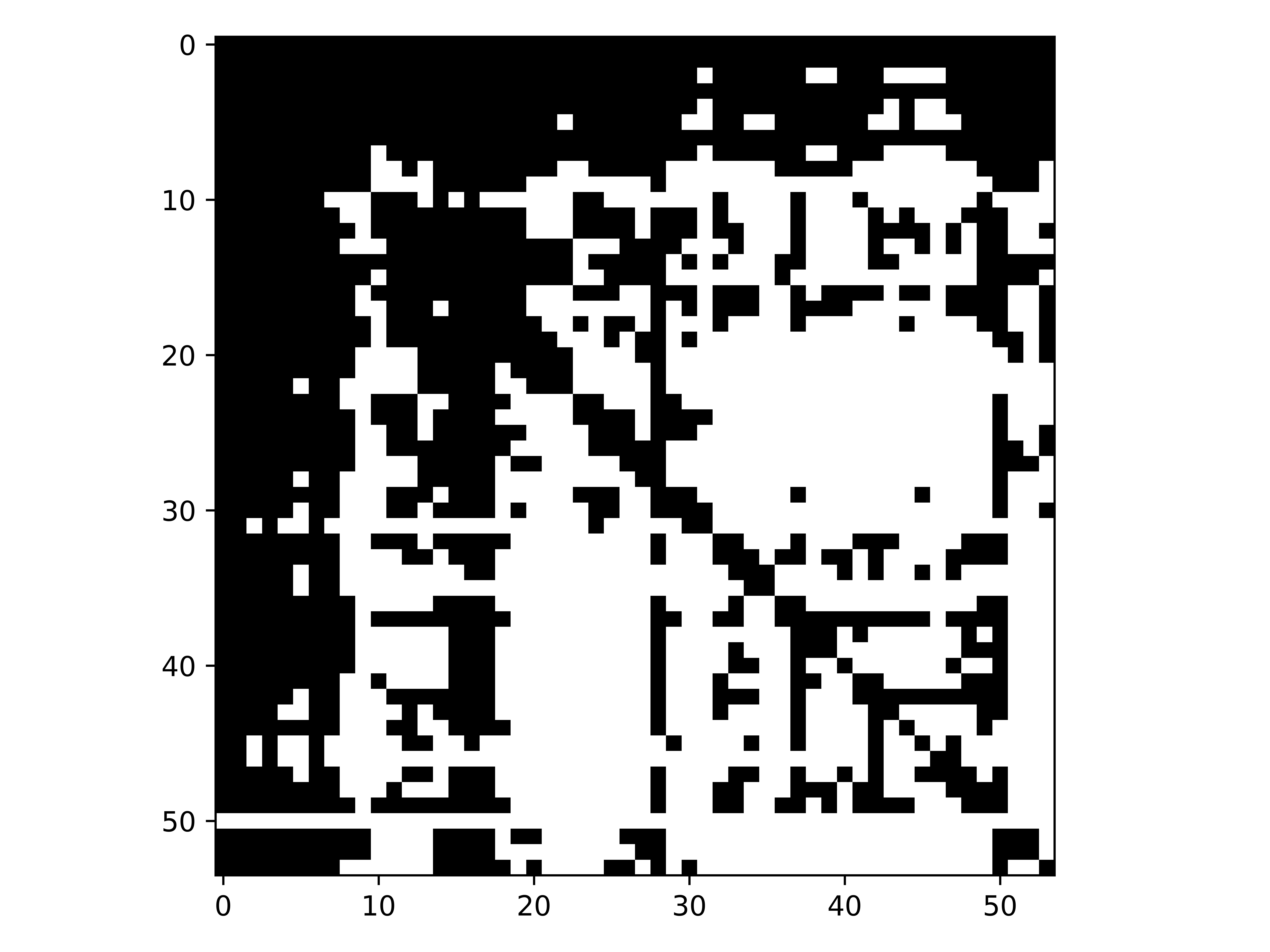}
      \caption{The ``approximate'' Jacobian.}
      \label{F:jac_sparse_approx}
  \end{subfigure}
  \caption{A graphical representation of the sparsity pattern of the chemical kinetic Jacobian generated by \texttt{pyJac} for GRI-Mech 3.0.
	   Black squares indicate a non-zero Jacobian entry, while white square correspond to an empty entry.
	   The numbers indicate the index of the entry in the state vector.}
  \label{F:jac_sparsity}
\end{figure}

\Cref{F:jac_sparsity} graphically represents the Jacobian sparsity of GRI-Mech 3.0.
In particular we note that~\cref{F:jac_sparse_approx} has several rows that are no longer fully dense, as result of its approximate form; these rows correspond to species directly interacting with \ce{N2}, largely in GRI-Mech 3.0's nitrogen chemistry reactions.
\cref{T:jac_sparsity} shows the density of the exact and approximate Jacobians for all chemical kinetic models tested in this work.
The smallest model, \ce{H2}\slash\ce{CO}, is very dense with \SI{71.4}{\percent} of the exact Jacobian entries non-zero; this drops to \SI{56.7}{\percent} for GRI-Mech 3.0, continues to decrease to \SI{28.2}{\percent} for USC-Mech II, and is just \SI{11.5}{\percent} for the isopentanol model.
The approximate Jacobian assumption drops the density of Jacobian by \textasciitilde\SIrange{3}{7}{\percent} for all models.

Currently, \texttt{pyJac} can use two common sparse-matrix storage formats~\cite{netlib_templates}: compressed row storage (CRS) and compressed column storage (CCS), used for ``C'' and ``F''-ordered data respectively.
For brevity we will outline only the CRS format but the CCS format is similar~\cite{netlib_templates}.
An $N\times N$ CRS matrix is stored using three vectors: a value vector of length $N_{NZ}$ (the number of non-zero elements in the Jacobian) that stores the elements of the Jacobian, a row pointer vector of length $N$ that stores the locations in the value vector that begin a row, and a column index vector length $N_{NZ}$ that stores the column indices of the elements in the value vector.

\begin{table}[tbp]
\centering
\begin{tabular}{@{}l S[table-format=2.1] S[table-format=2.1]@{}}
\toprule
Model                 & {Exact Jacobian Density} & {Approximate Jacobian Density} \\
\midrule
\ce{H2}\slash \ce{CO} & \SI{71.4}{\percent} & \SI{68.4}{\percent} \\
GRI-Mech 3.0          & \SI{56.7}{\percent} & \SI{49.8}{\percent} \\
USC-Mech II           & \SI{28.2}{\percent} & \SI{26.4}{\percent} \\
\ce{iC5H11OH}         & \SI{11.5}{\percent} & \SI{7.98}{\percent} \\
\bottomrule
\end{tabular}
\caption{The density of the exact and approximate Jacobians generated by \texttt{pyJac} for the various models studied.}
\label{T:jac_sparsity}
\end{table}

The Jacobian access pattern used by \texttt{pyJac} is fairly irregular; for simplicity we will only discuss looping-structure of species derivatives calculations since these form the bulk of the computation and have the most challenging Jacobian access patterns.
In general, an outer loop iterates over all reactions of a certain type (e.g., falloff reactions) and calculates the relevant Jacobian subproducts---independent of any particular species---for the reaction (e.g., the derivative of the falloff pressure modification term).
Two inner loops then iterate over the species in a reaction, updating the Jacobian entries for these species as appropriate.
This pattern leads to fairly easily vectorizable code and efficient Jacobian evaluation, since the bulk of the computation depends only on the reaction in question, as discussed in our previous work~\cite{Niemeyer:2016aa}.
Generally, this means that a lookup operation is required to find the sparse Jacobian index for any pair of state variables; in some cases this can be avoided, e.g., the rows corresponding to derivatives of the temperature and thermodynamic state parameter source-terms are fully dense in \texttt{pyJac}, and hence no lookup is necessary.
This lookup operation is currently implemented as a simple ``for'' loop, e.g., for a sparse lookup of a pair of indices $(i, j)$ in a CRS matrix, the lookup function searches the column index vector between the values $\texttt{row\_pointer}[i], \ldots, \texttt{row\_pointer}[i + 1]$ for $j$, and returns the offset from $\texttt{row\_pointer}[i]$ (or $\num{-1}$ if not found).
As will be explored in~\cref{S:jacobian_results}, this slows down sparse Jacobian evaluation, and might be improved by a static mapping of the full Jacobian indices to the sparse index (or some null value if the entry is empty).
However, this would require increased constant-data usage, a limitation for OpenCL.
Additionally, this might be an excellent usage of OpenCL's Image memory type (similar to texture memory in CUDA terminology).
Both of these sparse indexing techniques merit future investigation.

\subsection{Performance}
\label{S:results}
The performance studies in this work were run on the platforms listed in~\cref{t:cpus,t:gpus}.
Run times in each case were averaged over ten runs, each using the same set of PaSR conditions used in verification.
The OpenMP Jacobian\slash source-term kernels, as well as the OpenMP\slash OpenCL wrapping code (responsible for initializing\slash transferring memory, reading input, etc.) was compiled with \texttt{gcc v5.4.0} on the \avx/\slash\gpunew/ platforms and \texttt{gcc v4.8.5} on the \sse/\slash\gpuold/ machines.
The optimization level ``\texttt{-O3 -mtune=native}'' was used and no ``fast math'' OpenCL optimizations were enabled.
Additionally, the exact form of the Jacobian (as opposed the ``approximate'' form discussed in~\cref{S:sparsity}) was used in all cases.
Finally, unless stated otherwise: the performance results used a single CPU core, the \conp/ assumption, a vector width of \num{8}\slash\num{128}, and ``C''\slash ``F''-ordered data for the CPU\slash GPU cases, respectively; the run times reported are for the number of conditions specified in~\cref{T:error} and include data-transfer overhead to\slash from internal buffers used in \texttt{pyJac}.
The effects of choice of vector width, data ordering, and differences between \conp/ and \conv/ evaluations on the CPU\slash GPU will be explored in~\cref{S:source_results}, while parallel scaling for multiple CPU cores will be examined in~\cref{S:source_results,S:jacobian_results}.

\subsubsection{Source-term evaluation}
\label{S:source_results}

\begin{figure}[htbp]
   \centering
  \begin{subfigure}[t]{0.48\linewidth}
      \includegraphics[width=\textwidth]{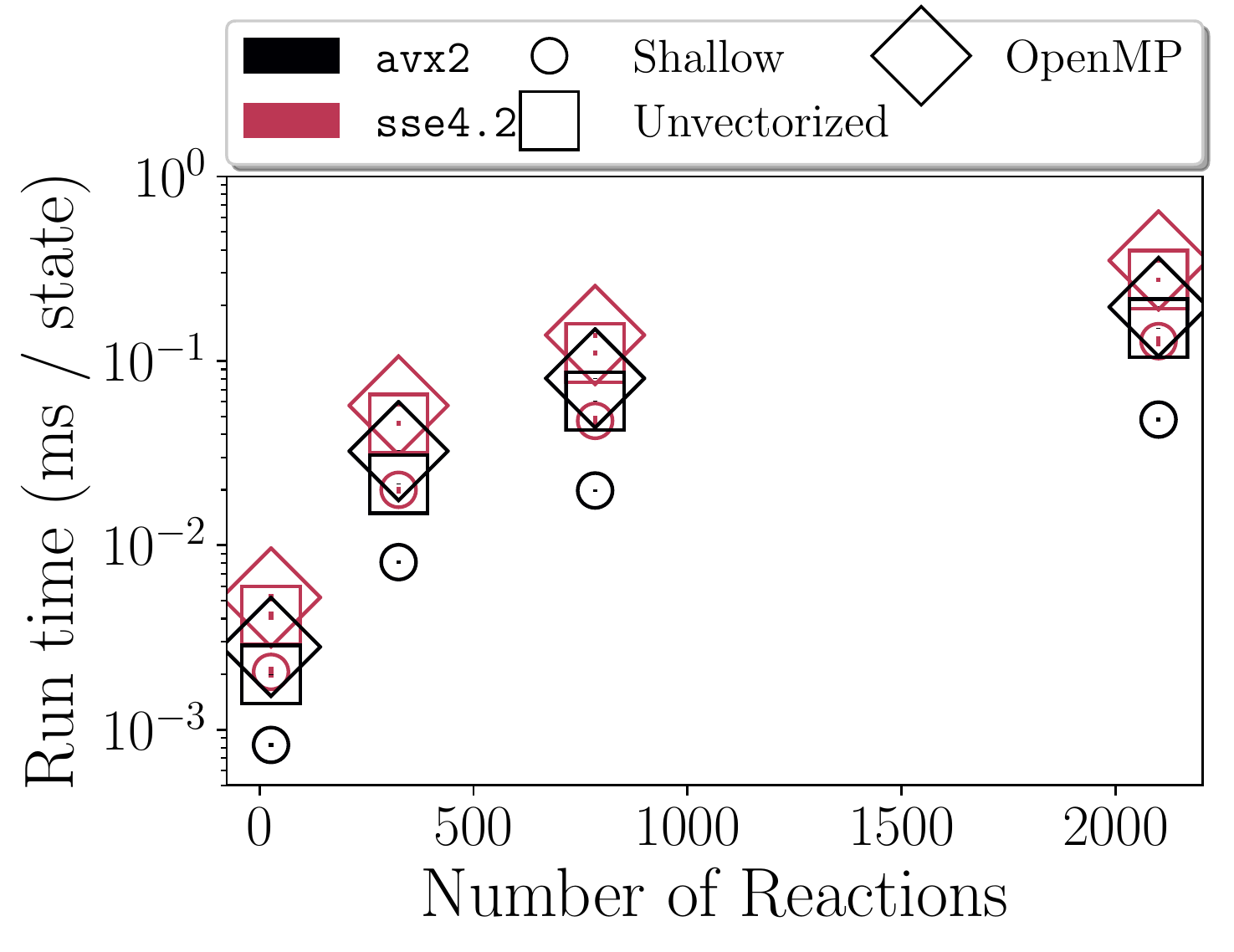}
      \caption{The mean run time per condition for each chemical model using the Intel OpenCL and OpenMP on both CPUs studied.}
      \label{F:intel_source_nonorm}
  \end{subfigure}
  \hfill
  \begin{subfigure}[t]{0.48\linewidth}
      \includegraphics[width=\textwidth]{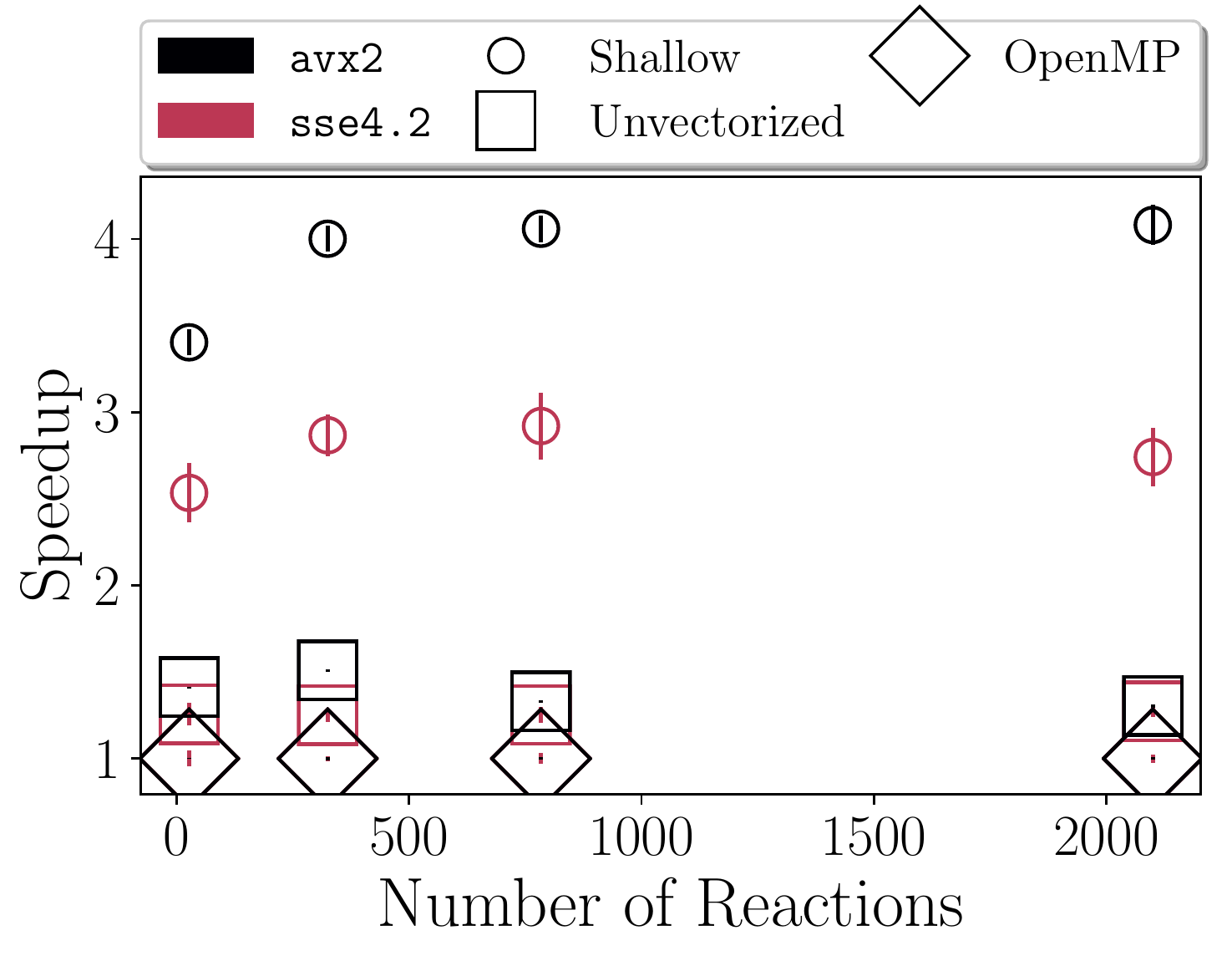}
      \caption{The speedup achieved over the baseline OpenMP parallelization by both the unvectorized and shallow-vectorized Intel OpenCL codes; the speedup is presented on per-machine basis.}
      \label{F:intel_source}
  \end{subfigure}
  \\
  \begin{subfigure}[t]{0.48\linewidth}
      \includegraphics[width=\textwidth]{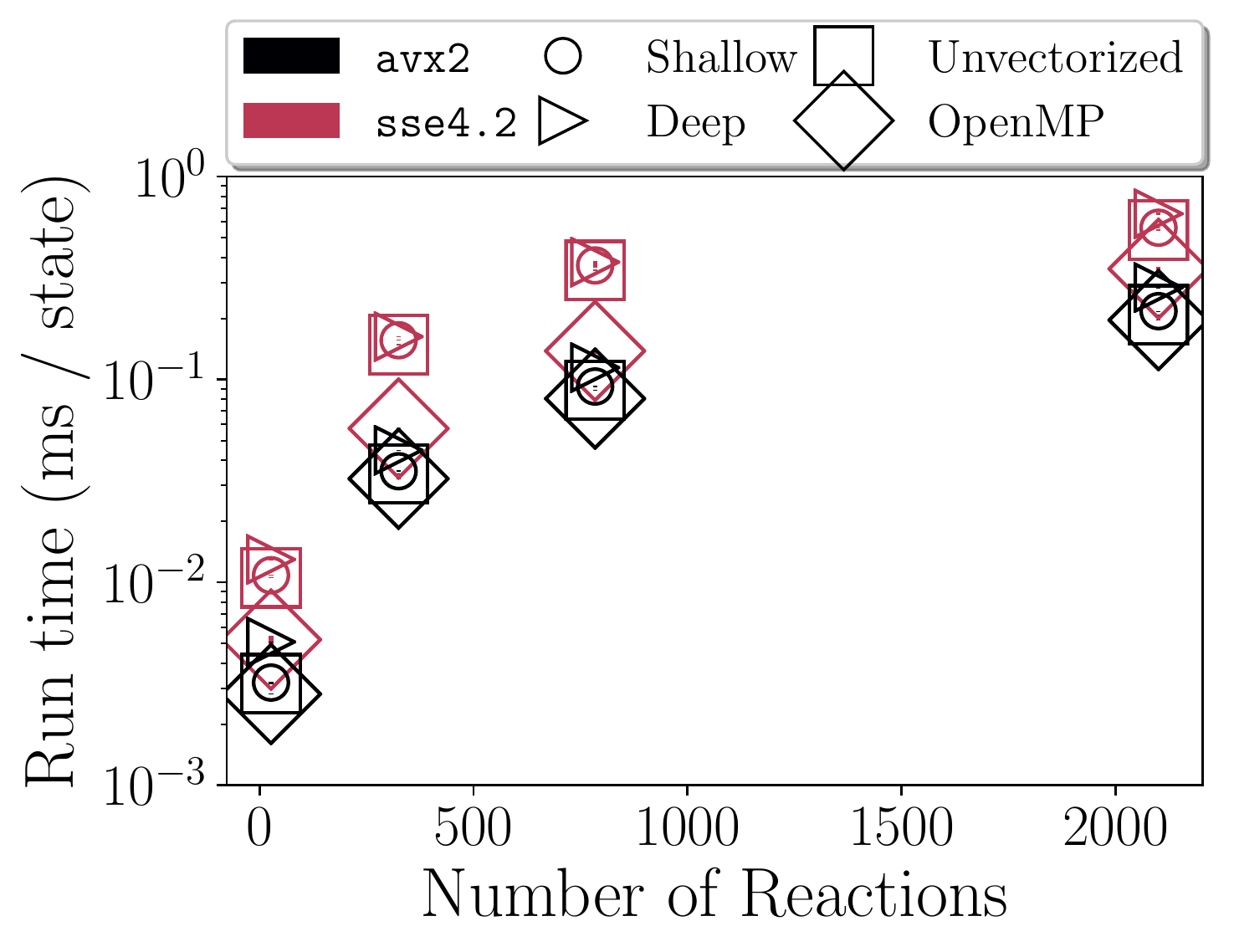}
      \caption{The mean run time per condition of the Portable OpenCL runtime compared to OpenMP parallelization.}
      \label{F:pocl_source}
  \end{subfigure}
 \caption{Mean run times per condition and speedups achieved by the various CPU OpenCL runtimes compared to OpenMP parallelization for each chemical model studied. The names in the legends correspond to the identifiers listed in~\cref{t:cpus}.}
 \label{F:cpu_source}
\end{figure}

\Cref{F:cpu_source} explores the performance of the source-term evaluations generated by \texttt{pyJac} on the CPU test platforms listed in~\cref{t:cpus}.
Source-term evaluations---critical in their own right for direct numerical simulations of reactive-flows~\cite{Spafford:2010aa}, among other applications---also provide a convenient platform to detail the effects of various code configuration options before investigating the more involved Jacobian evaluation performance.

\Cref{F:intel_source_nonorm} shows the mean run time per initial condition for both the~\avx/ and~\sse/ CPUs, using Intel OpenCL and OpenMP.
This normalization of the run time by the number of initial conditions tested is chosen to account for the varying numbers of conditions in the PaSR databases for each model (\cref{t:models}).
For both CPUs, the OpenMP implementation is the slowest for all models; interestingly, the unvectorized (i.e., strictly parallel) Intel OpenCL code is slightly faster than OpenMP in all cases.
As expected, the \avx/ machine is faster than the \sse/ CPU for the strictly parallel cases, performing \SIrange{1.82}{2.13}{$\times$} and \SIrange{1.72}{1.85}{$\times$} faster for the unvectorized OpenCL case and OpenMP, respectively.
Additionally, the shallow-vectorized OpenCL code performs significantly faster than either the OpenMP or unvectorized OpenCL codes on both processors.

\Cref{F:intel_source} details the extent of this speedup; the speedup displayed is calculated per-machine, e.g., the \avx/ shallow-vectorized code speedup is relative to OpenMP on the same CPU.
On both machines, the unvectorized OpenCL code is faster than the baseline parallel OpenMP code, by \SIrange{1.30}{1.51}{$\times$} on the \avx/ CPU and \SIrange{1.25}{1.27}{$\times$} on the \sse/ machine.
Additionally, the shallow-vectorized OpenCL code is \SIrange{2.53}{2.92}{$\times$} and \SIrange{3.40}{4.08}{$\times$} faster than the OpenMP code for the \sse/ and \avx/ machines, respectively.

In contrast,~\cref{F:pocl_source} shows the mean run time per condition of deep, shallow, and unvectorized OpenCL codes using the POCL runtime, as compared with OpenMP parallelization.
No speedup is achieved for either vectorization type on either CPU---indeed, the OpenMP case is faster on both CPUs, though we stress that Intel OpenCL runtime achieves vectorization using the same code and hardware.
The reasons for this lack of vectorization with POCL are quite technical; however, personal communication with the developers of POCL revealed that to achieve vectorization, changes are required to both the way POCL prepares LLVM intermediate representation code, as well as improvements to LLVM's loop-vectorizer itself~\cite{pocl_communication}.\footnote{Specifically, improvements such as ensuring LLVM recognizes uniform vectorization loop bounds (even if said bounds \textit{are} uniform in practice), proving vector instructions' ability to handle all edge cases identically to the corresponding scalar instruction, handing of branches and conditionals (potentially within POCL instructions themselves), and handling of memory access\slash vector-element extraction patterns.}
As will be discussed in~\cref{S:opencl}, we hope that using explicit vector types (to lessen demands on the LLVM-vectorization module) in combination with some of these changes might solve this issue, but for the moment POCL is still quite useful as a verification tool.

\begin{figure}[htbp]
   \centering
  \begin{subfigure}[t]{0.48\linewidth}
      \includegraphics[width=\textwidth]{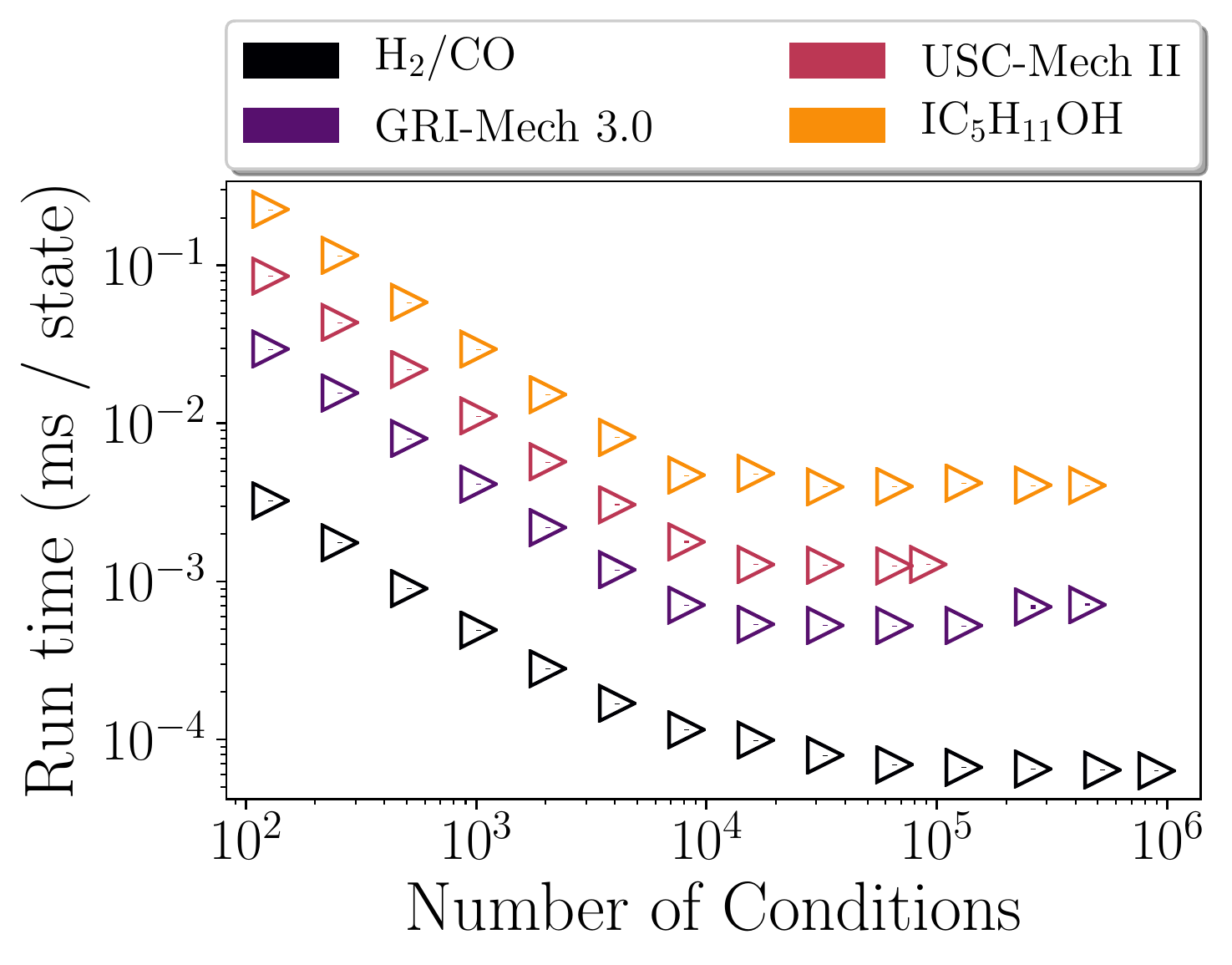}
      \caption{The mean run time per condition for each chemical model on the \gpunew/ GPU as a function of the number of initial conditions tested.}
      \label{F:gpu_source_scaling}
  \end{subfigure}
  \hfill
  \begin{subfigure}[t]{0.48\linewidth}
      \includegraphics[width=\textwidth]{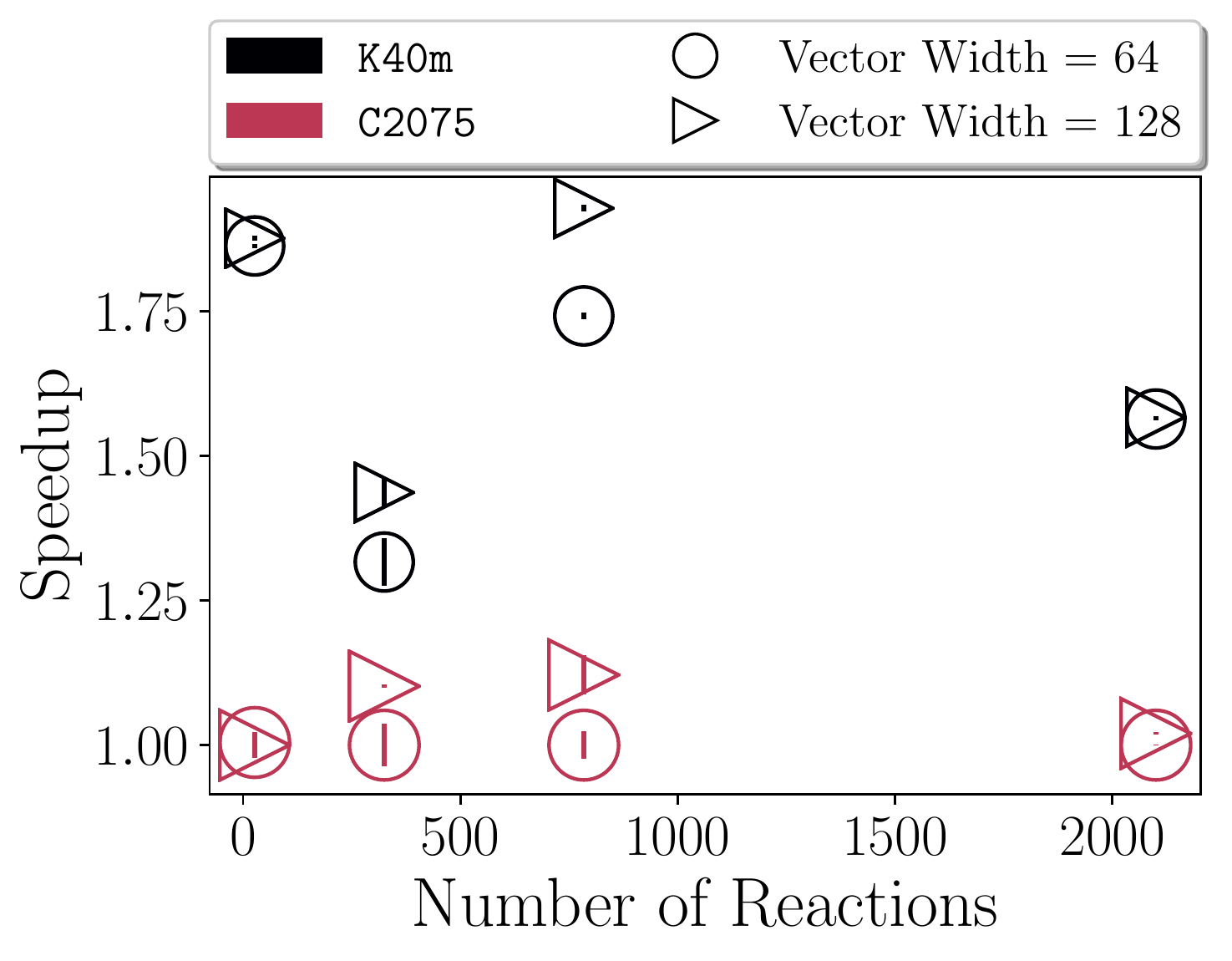}
      \caption{The speedup achieved on the \gpunew/ versus the \gpuold/ GPU; the names correspond to the identifiers listed in~\cref{t:gpus}}
      \label{F:gpu_source_speedup}
  \end{subfigure}
  \caption{\texttt{pyJac} source-term evaluation performance on the Nvidia GPUs}
  \label{F:gpu_source}
\end{figure}

\Cref{F:gpu_source} shows the performance of evaluating source terms on the GPUs listed in~\cref{t:gpus}.
\Cref{F:gpu_source_scaling} investigates how the number of initial conditions evaluated affects the mean run time per condition on the \gpunew/ GPU;
the run time decreases until around \textasciitilde\num{e4} conditions for all chemical models, at which point the GPU becomes saturated and performance levels off.
The performance plateaus slightly later for the \ce{H2}\slash\ce{CO} model compared with the others.
\Cref{F:gpu_source_speedup} shows the speedup in source-term evaluation that the \gpunew/ GPU achieves over the \gpuold/ GPU for the maximum number of conditions in~\cref{F:gpu_source_scaling} with two vector widths (i.e., GPU block size), \num{64} and \num{128}.
The best \gpunew/ case with a vector width of 128 is \SIrange{1.40}{1.88}{$\times$} faster than the slowest case (\gpuold/ with a vector-width of 64) depending on the chemical model in question.
\Cref{F:gpu_source_speedup} also shows that varying the vector-width minimally affects performance for most of the \gpunew/ and \gpuold/ cases; the GRI-Mech 3.0 and USC-Mech II models show the largest improvements with a vector width of 128: \textasciitilde\SIrange{10}{18}{\percent} for both GPUs.
This is likely caused by higher occupancy on the GPUs, but it is unclear exactly how Nvidia's OpenCL runtime balances the registers\slash warps per streaming-multiprocessor, as controls occupancy in CUDA~\cite{occupancy}.

\begin{figure}[htbp]
   \centering
  \begin{subfigure}[t]{0.48\linewidth}
      \includegraphics[width=\textwidth]{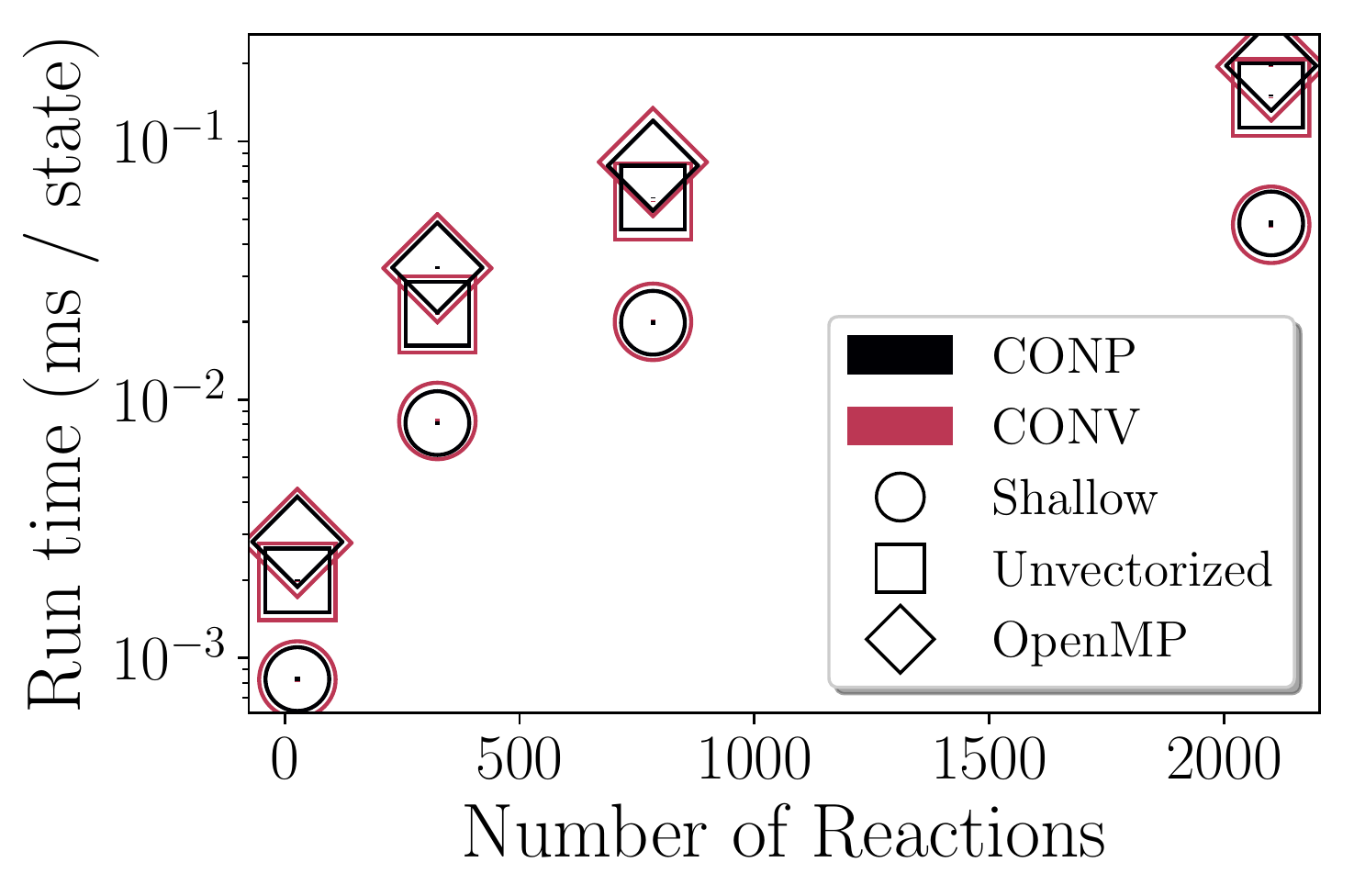}
      \caption{The mean run time per condition for each chemical model using both the \conp/ and \conv/ formulations on Intel OpenCL and OpenMP on the \avx/ CPU.}
      \label{F:source_conpvsconv}
  \end{subfigure}
  \hfill
  \begin{subfigure}[t]{0.48\linewidth}
      \includegraphics[width=\textwidth]{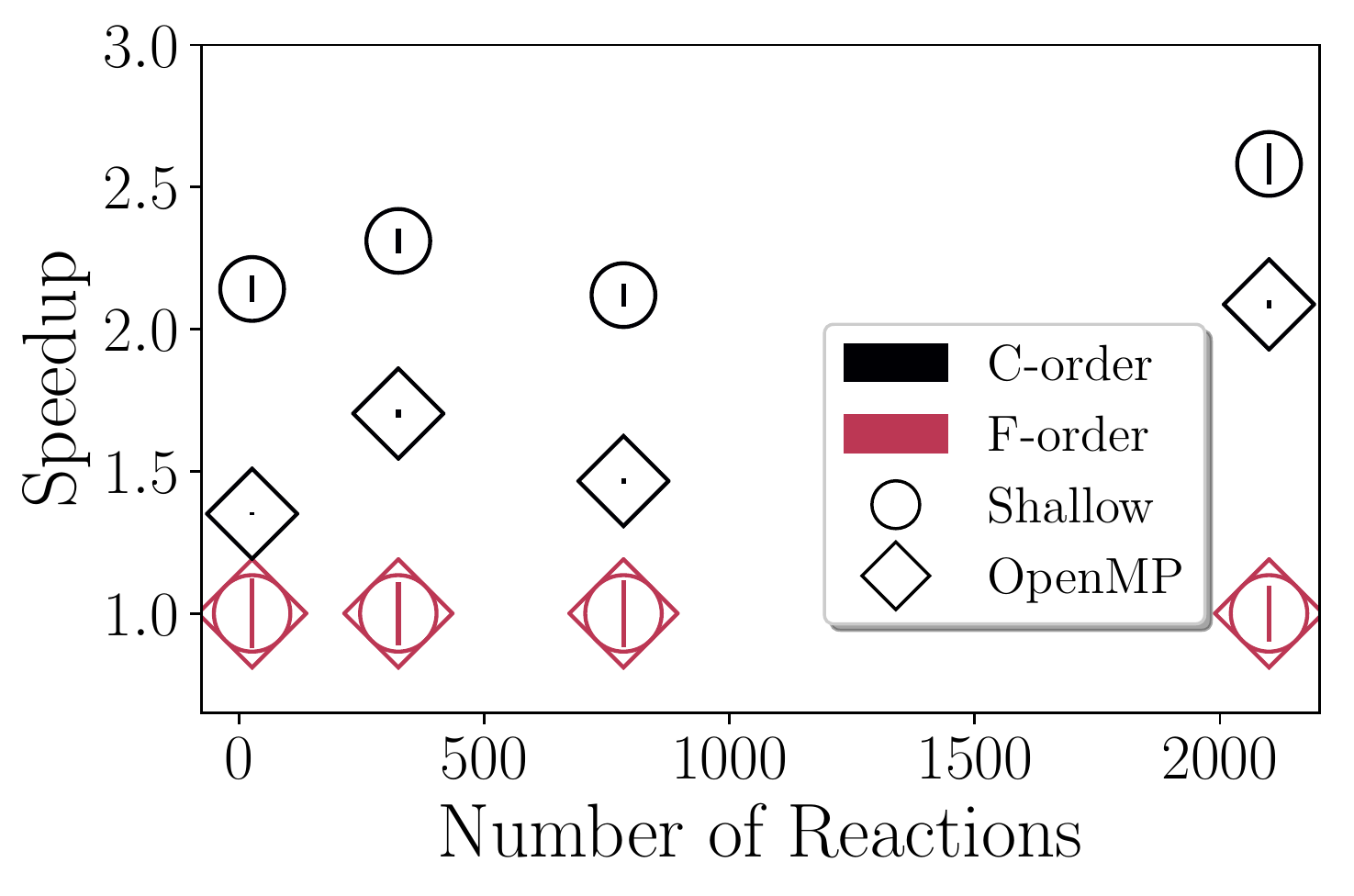}
      \caption{The speedup achieved by ``C'' ordering over ``F'' ordering for Intel OpenCL and OpenMP on the \avx/ CPU.  The speedup presented is calculated per-language (OpenMP and OpenCL) to better assess the effect of the data ordering.}
      \label{F:source_cvsf}
  \end{subfigure}
  \\
  \begin{subfigure}[t]{0.48\linewidth}
      \includegraphics[width=\textwidth]{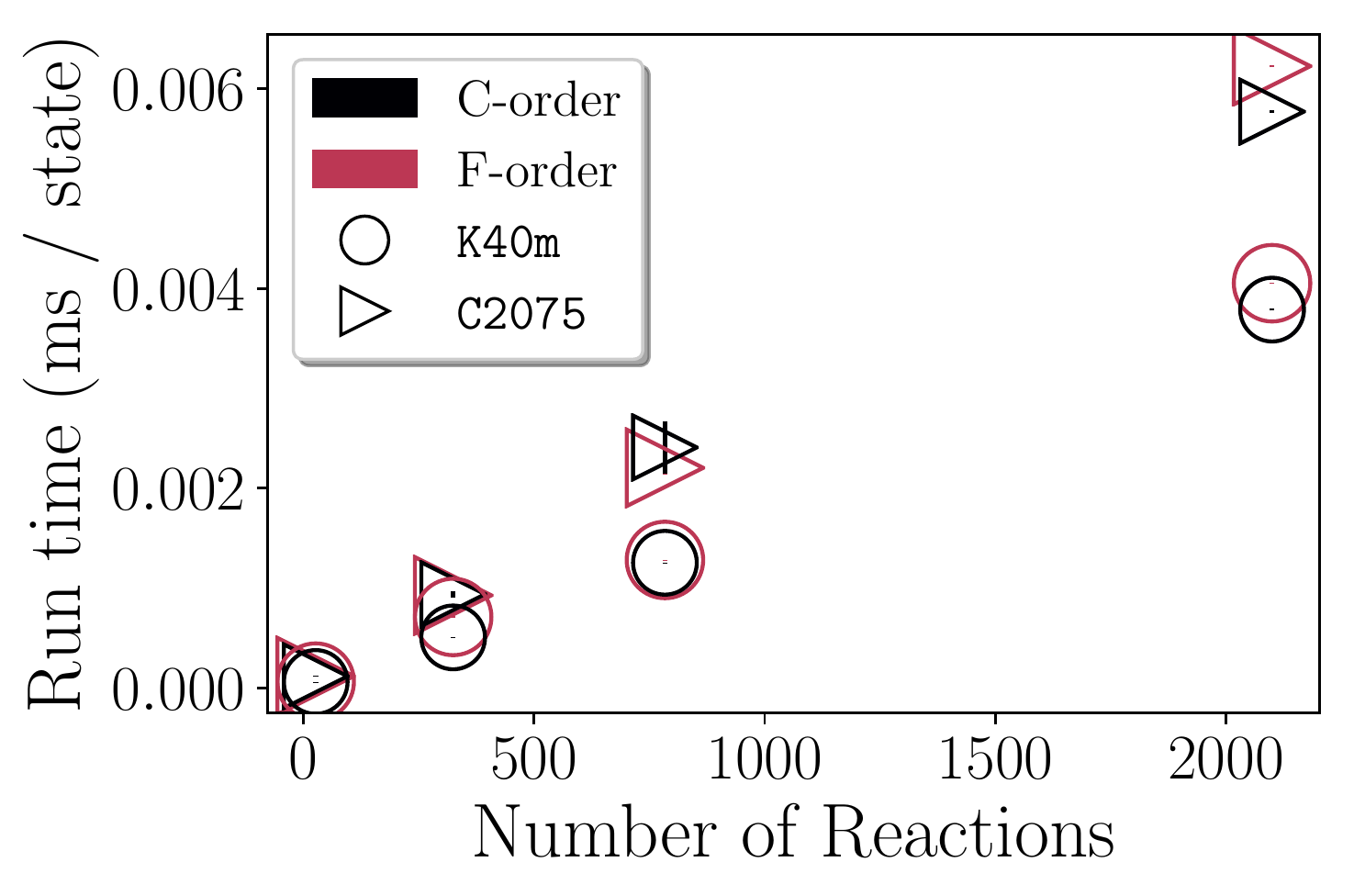}
      \caption{The affect on performance of ``C'' vs ``F''-ordering for the shallow-vectorized Nvidia OpenCL code on both GPUs with a vector-width of \num{128}.}
      \label{F:source_gpu_cvsf}
  \end{subfigure}
  \hfill
  \begin{subfigure}[t]{0.48\linewidth}
      \includegraphics[width=\textwidth]{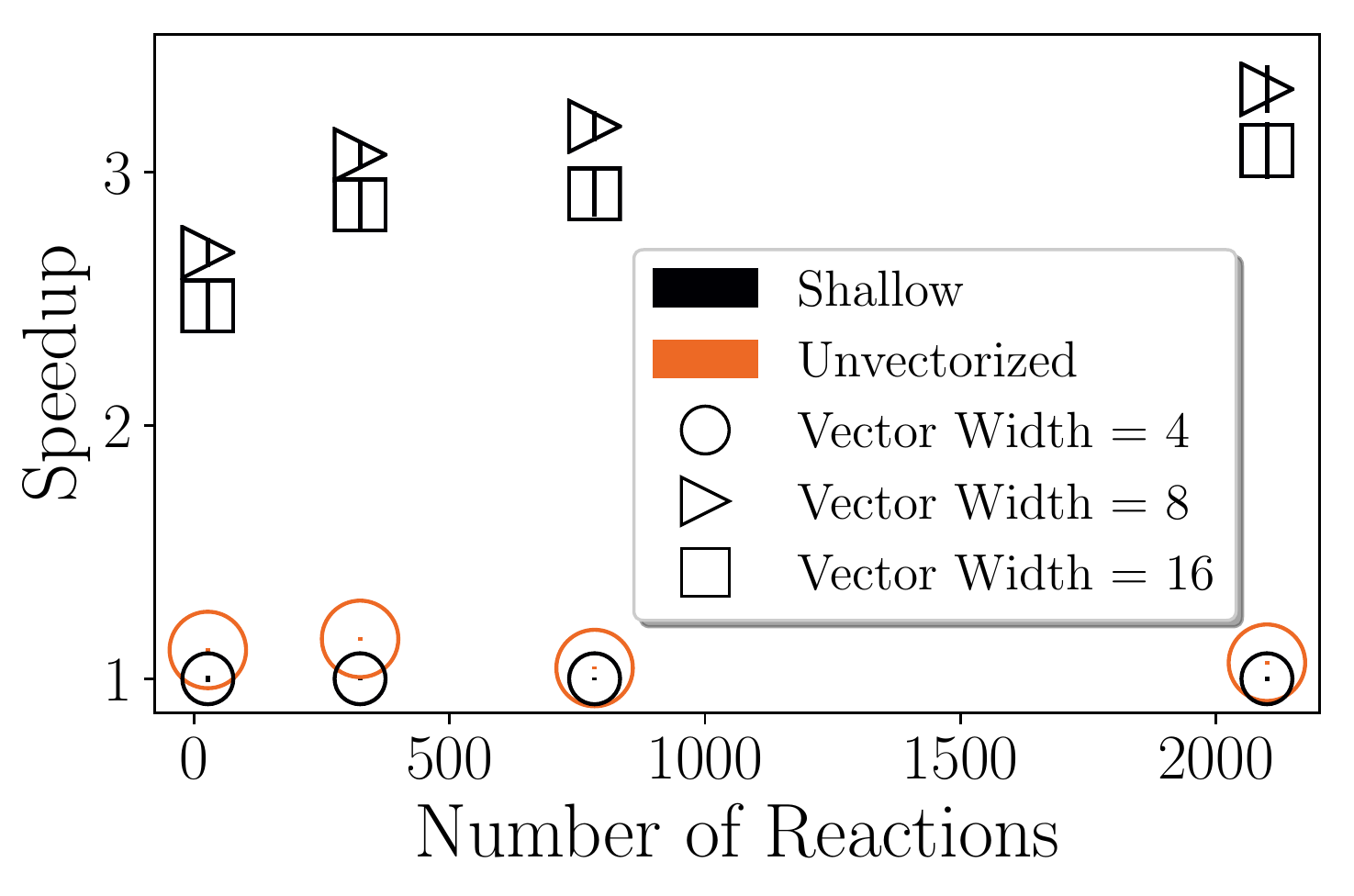}
      \caption{The effect of vector-width on ``C''-ordered shallow-vectorized Intel OpenCL source-term evaluations on the \avx/ CPU.}
      \label{F:source_vector_width}
  \end{subfigure}
  \caption{The effect of the \conp/ and \conv/ formulations, ``C'' and ``F'' data-ordering, and CPU vector-width on source-term evaluation performance in \texttt{pyJac}.
  The shallow-vectorized ``C''-ordered OpenCL cases correspond to the vectorized data ordering described in~\cref{S:data}.}
  \label{f:source_permutate}
\end{figure}

\Cref{f:source_permutate} shows how changing data-ordering patterns, the \conp/ or \conv/ formulation, and the CPU vector width affect the performance of source-term evaluation in \texttt{pyJac}.
Per~\cref{F:source_conpvsconv}, we see that the choice of \conp/ or \conv/ formulation has little to no effect on run time for OpenMP as well as the shallow-vectorized\slash unvectorized Intel OpenCL codes on the \avx/ machine.
Generally speaking, the difference between the \conp/ and \conv/ formulations only marginally affects performance regardless of CPU\slash GPU choice.

In contrast, \cref{F:source_cvsf} shows significant speedups of ``C''-ordered data over ``F''-ordered data on the \avx/ machine; the speedup presented is calculated per language, e.g., the \SIrange{1.35}{2.09}{$\times$} speedup of the ``C''-ordered OpenMP implementation is relative to the ``F''-ordered OpenMP baseline.
Additionally, the ``C''-ordered shallow-vectorization in~\cref{F:source_cvsf} and the other shallow-vectorized CPU data shown in \cref{S:source_results,S:jacobian_results} use the vectorized-data ordering described in~\cref{S:data}; this case achieves speedups of \SIrange{2.14}{2.58}{$\times$} over the ``F''-ordered shallow-vectorization, demonstrating the value of the vectorized-data ordering for CPU execution.

\Cref{F:source_gpu_cvsf} shows how the ``C''- and ``F''-ordering affect the performance of source-term evaluation on both GPUs, with the speedup presented per-GPU.
The ``C''- and ``F''-ordered shallow-vectorizations perform almost equivalently on both GPUs, with less than a \SI{10}{\percent} difference in run time between data orderings.
For the isopentanol model, ``C''-ordered data is \textasciitilde\SI{1.08}{$\times$} faster on both GPUs (while the trend is less clear for the other models).
The roughly equivalent performance between the ``C'' and ``F''-ordered approaches on GPUs counters what one might expect: typically speaking, coalesced memory access in a shallow vectorization is easier to achieve with ``F''-ordering (see~\cref{S:data}).
However, the vectorized-data ordering here ensures that memory storage is aligned to the vector width and, thus, encourages coalesced accesses.

\Cref{F:source_vector_width} shows how changing vector width affects source-term evaluation performance on the \avx/ CPU.
The vector width of 8 performs the fastest (out of those tested), while the larger vector width of 16 is slightly slower due to increased register pressure~\cite{intel_vectypes}.
It is unclear why the vector width of 4 results in no speedup at all (in fact, it is the slowest case).
Intel's vectorization guide~\cite{intel_vecknobs} mentions that a heuristic determines the optimal vector width (in this case, it appears from compiler output to be 8), so it is possible that using a vector width smaller than the heuristic breaks the implicit vectorizer.
This issue does not occur for a vector width of 4 on the \sse/ CPU.

\begin{figure}[htbp]
   \centering
  \begin{subfigure}[t]{0.48\linewidth}
      \includegraphics[width=\textwidth]{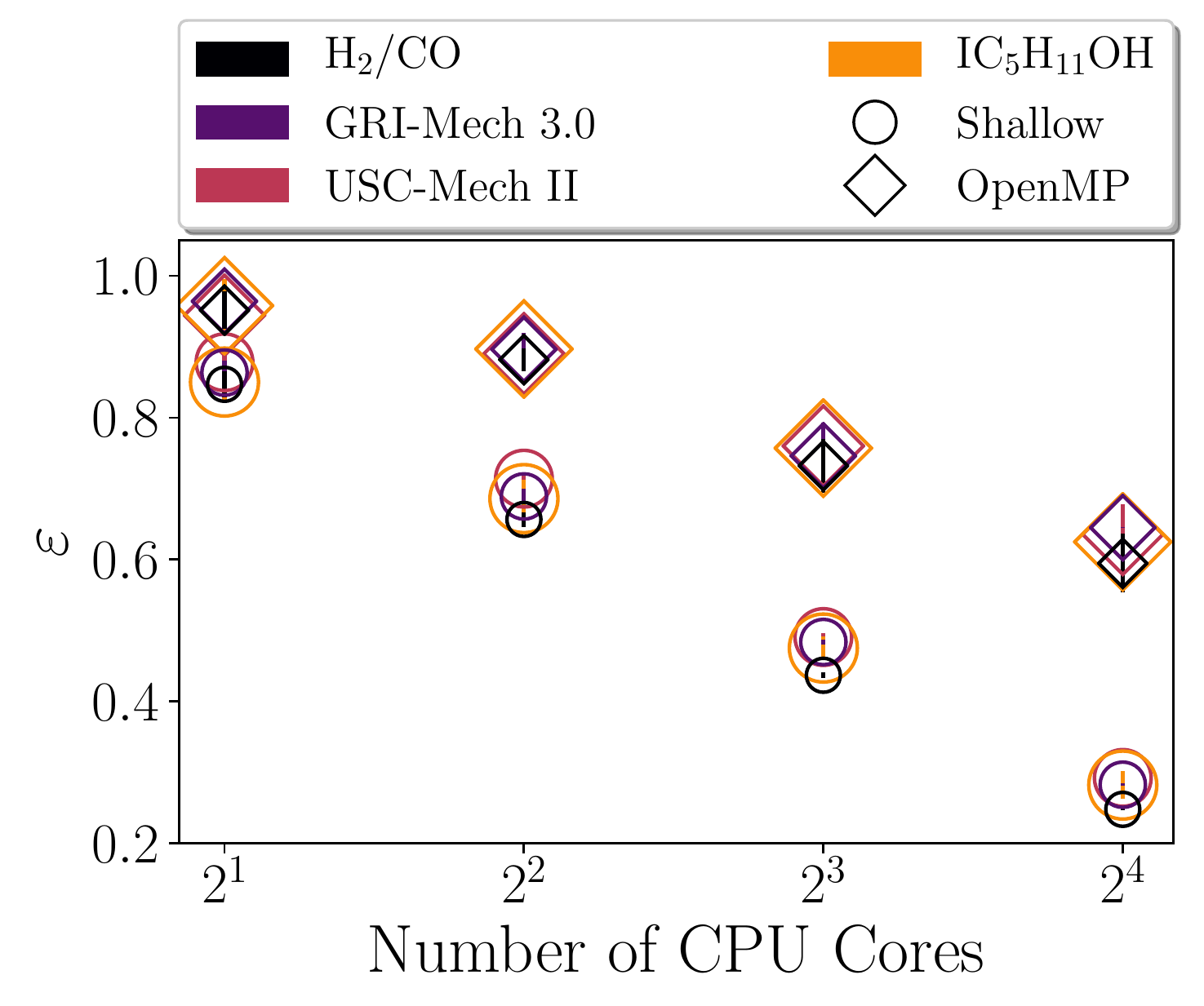}
      \caption{The strong parallel scaling efficiency (as defined in~\cref{e:strong_scaling}) of source-term evaluation for Intel OpenCL\slash OpenMP on the \avx/ machine.}
      \label{F:source_parallel_scaling}
  \end{subfigure}
  \hfill
  \begin{subfigure}[t]{0.48\linewidth}
      \includegraphics[width=\textwidth]{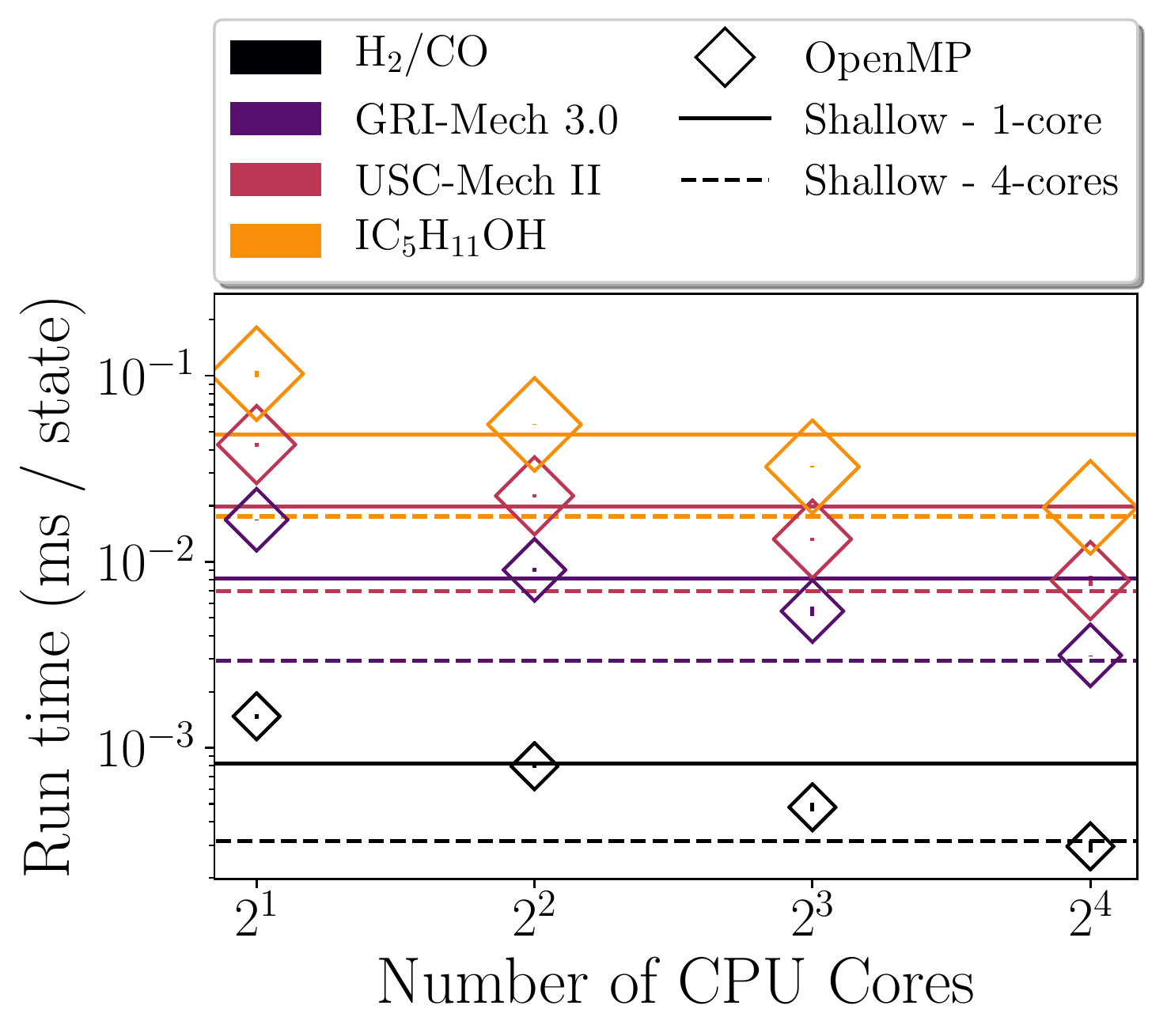}
      \caption{The mean run time per condition of OpenMP source-term evaluation on different cores compared to a shallow-vectorized Intel OpenCL evaluation on \num{1} (solid lines) and \num{4} (dashed lines) cores on \avx/ machine.}
      \label{F:source_crossover}
  \end{subfigure}
  \\
  \begin{subfigure}[t]{0.48\linewidth}
      \includegraphics[width=\textwidth]{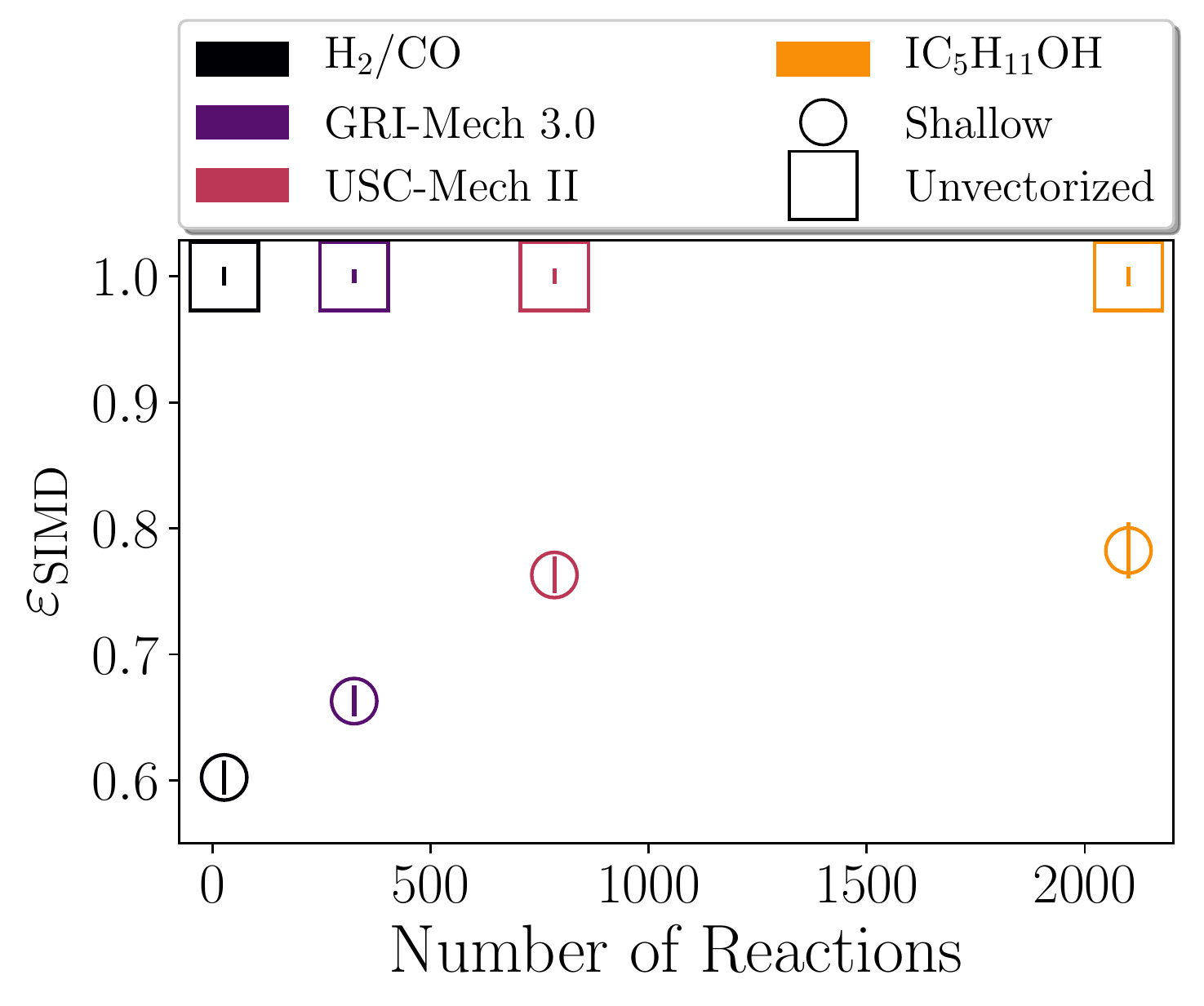}
      \caption{The SIMD efficiency (\cref{e:simd_efficiency}) of source-term evaluation for the Intel OpenCL runtime on a single core of the \avx/ CPU.}
      \label{F:source_simd_scaling}
  \end{subfigure}
  \hfill
  \begin{subfigure}[t]{0.48\linewidth}
      \includegraphics[width=\textwidth]{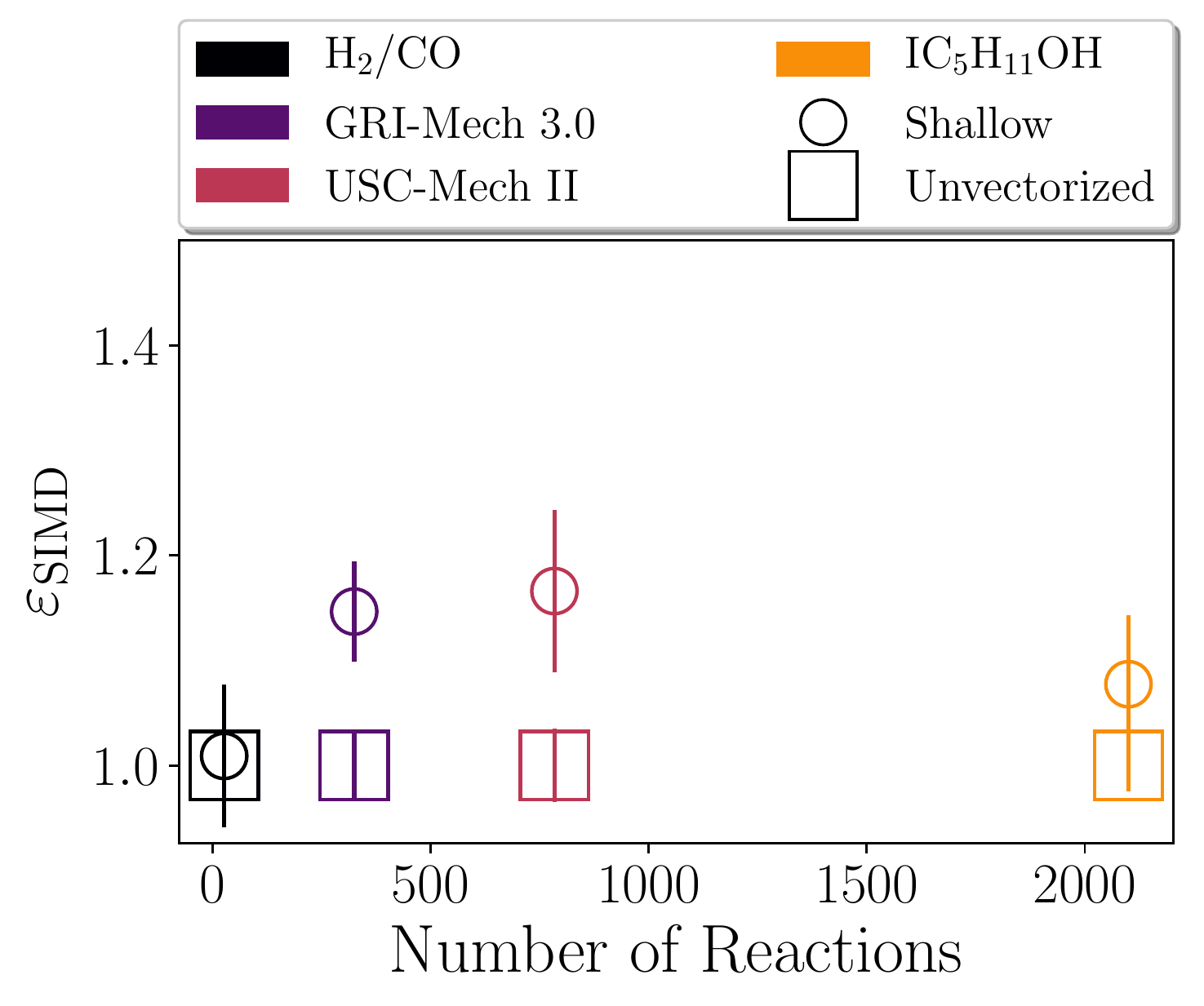}
      \caption{The SIMD efficiency (\cref{e:simd_efficiency}) of source-term evaluation for the Intel OpenCL runtime on a single core of the \sse/ CPU.}
      \label{F:source_sse_simd_scaling}
  \end{subfigure}
  \caption{The parallel scaling efficiency and SIMD efficiency of source-term evaluation for Intel OpenCL on the \avx/ and \sse/ CPUs.}
  \label{F:source_scaling}
\end{figure}

Finally,~\cref{F:source_scaling} displays the (strong) parallel scaling efficiency and SIMD efficiency for the CPU platforms.
The strong parallel scaling efficiency is defined as
\begin{equation}
 \label{e:strong_scaling}
 \varepsilon = \frac{\bar{t}_{1}}{N \bar{t}_{N}} \;,
\end{equation}
where $\bar{t}_{N}$ is the mean run time per condition on $N$ CPU cores and $\bar{t}_{1}$ the same on a single CPU core.
The strong parallel scaling efficiency measures the speedup due to the use of additional CPU cores as a fraction of linear speedup; strong scaling tends to decrease with the number of processors used due to memory-bandwidth limitations and decreasing computation work allocated per CPU core~\cite{strong_scaling}.

\Cref{F:source_parallel_scaling} shows the strong parallel scaling efficiency of source-term evaluation in \texttt{pyJac} on the \avx/ machine for both the shallow-vectorized Intel OpenCL and OpenMP codes.
In general, the \ce{H2}\slash\ce{CO} mechanism has the worst scaling efficiency for both Intel OpenCL and OpenMP, likely resulting from both its relatively small size and few falloff\slash chemically activated reaction (in particular, the additional expensive logarithm and exponential evaluations that accompany them).
As demonstrated in~\cref{S:SIMD_scaling}, the amount of computational work required per thermochemical state plays a critical role in fully utilizing SIMD instructions\slash multiple threads.
Additionally, OpenMP tends to scale better than the shallow-vectorized OpenCL code, e.g., \textasciitilde\num{0.9} and \textasciitilde\num{0.75} for four and eight CPU cores, respectively, compared to just \numrange{0.66}{0.72} and \numrange{0.44}{0.48} for OpenCL.
Though not pictured (to keep the figure readable), the unvectorized Intel OpenCL code scales only slightly worse than the OpenMP code, hence the poorer scaling is unique to the shallow-vectorized code.
This is due in large part to the superior performance of the shallow-vectorized OpenCL code, coupled with the relatively small amounts of work associated with source-term evaluation.
To illustrate this,~\cref{F:source_crossover} shows the mean run time per-condition of the OpenMP source-term evaluations for \numrange{2}{16} cores, compared to the shallow-vectorized OpenCL code on one (solid line) and four (dashed line) cores on the \avx/ machine.
For all chemical models, the mean run time per-condition (and hence the computational work allocated per-core, one of the key-drivers of parallel scaling efficiency~\cite{strong_scaling}) of OpenMP running on four cores is roughly equal to that of the shallow-vectorized OpenCL code on a single core.
Similarly, OpenMP running on \num{16} cores is roughly equivalent to the OpenCL code on \num{4} cores.
Therefore, a more fair comparison of parallel scaling efficiencies is to compare OpenCL running on \num{4} cores with OpenMP on \num{16}; the OpenMP code's parallel efficiency drops to \textasciitilde\num{0.64} for \num{16} cores, similar to OpenCL's parallel scaling efficiency of \numrange{0.66}{0.72} at \num{4} cores.
Indeed, as will be seen in~\cref{S:jacobian_results}, sparse Jacobian evaluation---the most computationally intensive task in this work---exhibits similar strong-scaling efficiency on Intel OpenCL and OpenMP.

The SIMD efficiency is defined as
\begin{equation}
 \label{e:simd_efficiency}
 \varepsilon_{\text{SIMD}} = \frac{\bar{t}_{\text{unvec}}}{W \bar{t}_{\text{shallow}}} \;,
\end{equation}
where $\bar{t}_{\text{unvec}}$ is the mean run time per condition of the unvectorized OpenCL code, $\bar{t}_{\text{shallow}}$ the same for the shallow-vectorized OpenCL code, and $W$ is the vector width reported in number of double operations (see~\cref{t:cpus}).
This measure compares the actual speedup due to shallow vectorization with the ideal speedup based on the nominal vector width of the machine.
\Cref{F:source_simd_scaling} shows the SIMD efficiency of source-term evaluation in \texttt{pyJac} on a single core of the \avx/ machine; the larger models (isopentanol and USC-Mech II) have higher SIMD efficiencies of \numrange{0.76}{0.78}, and the smaller models (\ce{H2}\slash\ce{CO}, GRI-Mech \num{3.0}) have lower SIMD efficiencies of \numrange{0.6}{0.66}.
This again demonstrates that the SIMD vectorization becomes more efficient with increasing amounts of work to perform (i.e., with increasing model size).
Interestingly,~\cref{F:source_sse_simd_scaling} shows the SIMD efficiency on the \sse/ machine as greater than one.
This is likely caused by a combination of using an OpenCL vector width greater than the native CPU vector width (i.e., eight versus two) and improved data locality for the vectorized-data ordering as discussed in~\cref{S:data}, and results in a modest \SIrange{7}{17}{\percent} improvement over the nominal vector width.

\subsubsection{Jacobian evaluation}
\label{S:jacobian_results}

\begin{figure}[htbp]
   \centering
  \begin{subfigure}[t]{0.48\linewidth}
      \includegraphics[width=\textwidth]{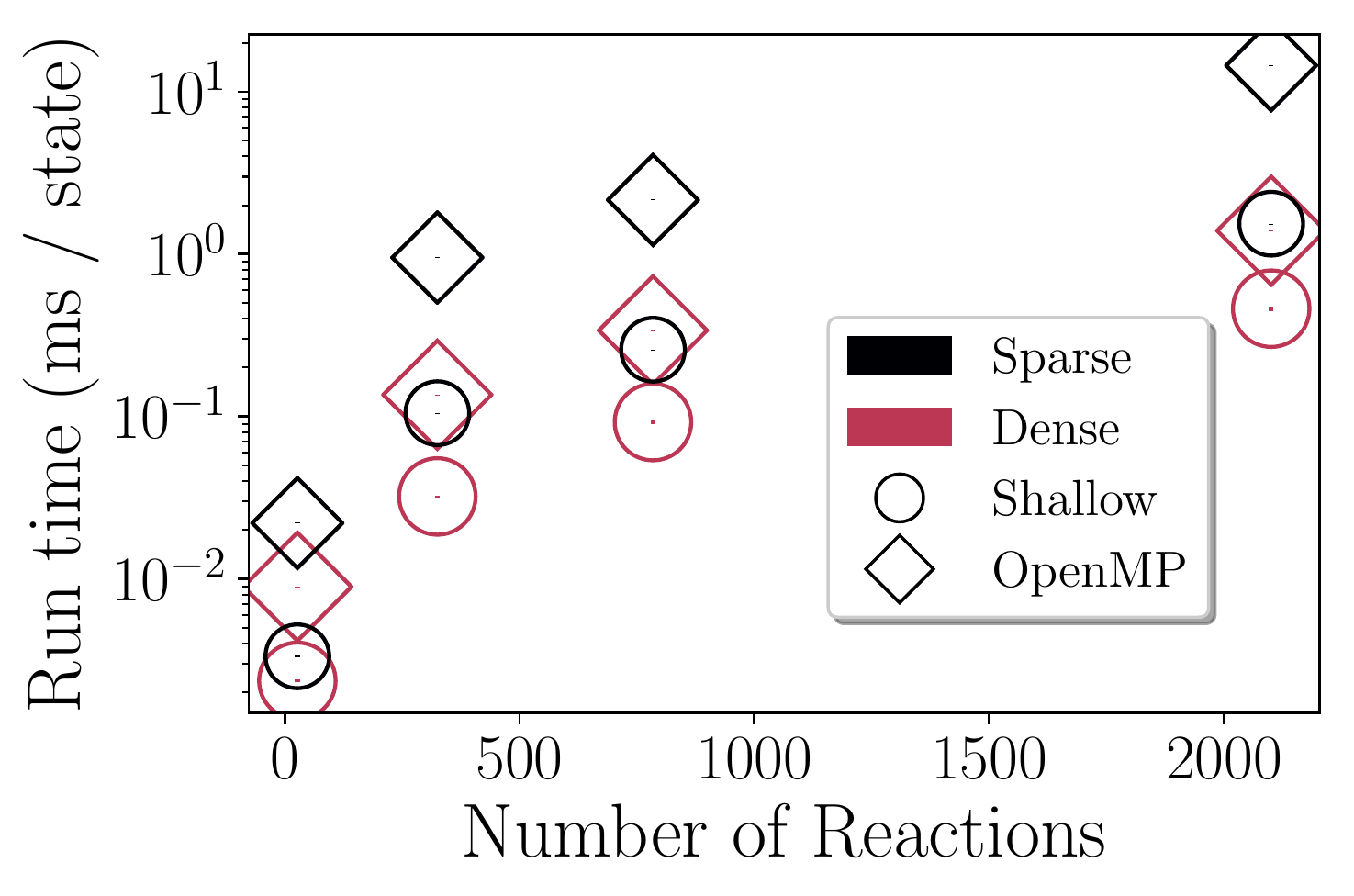}
      \caption{The mean run time per condition of sparse and dense Jacobian evaluations for Intel OpenCL\slash OpenMP on the \avx/ machine.}
      \label{F:sparse_vs_dense}
  \end{subfigure}
  \hfill
  \begin{subfigure}[t]{0.48\linewidth}
      \includegraphics[width=\textwidth]{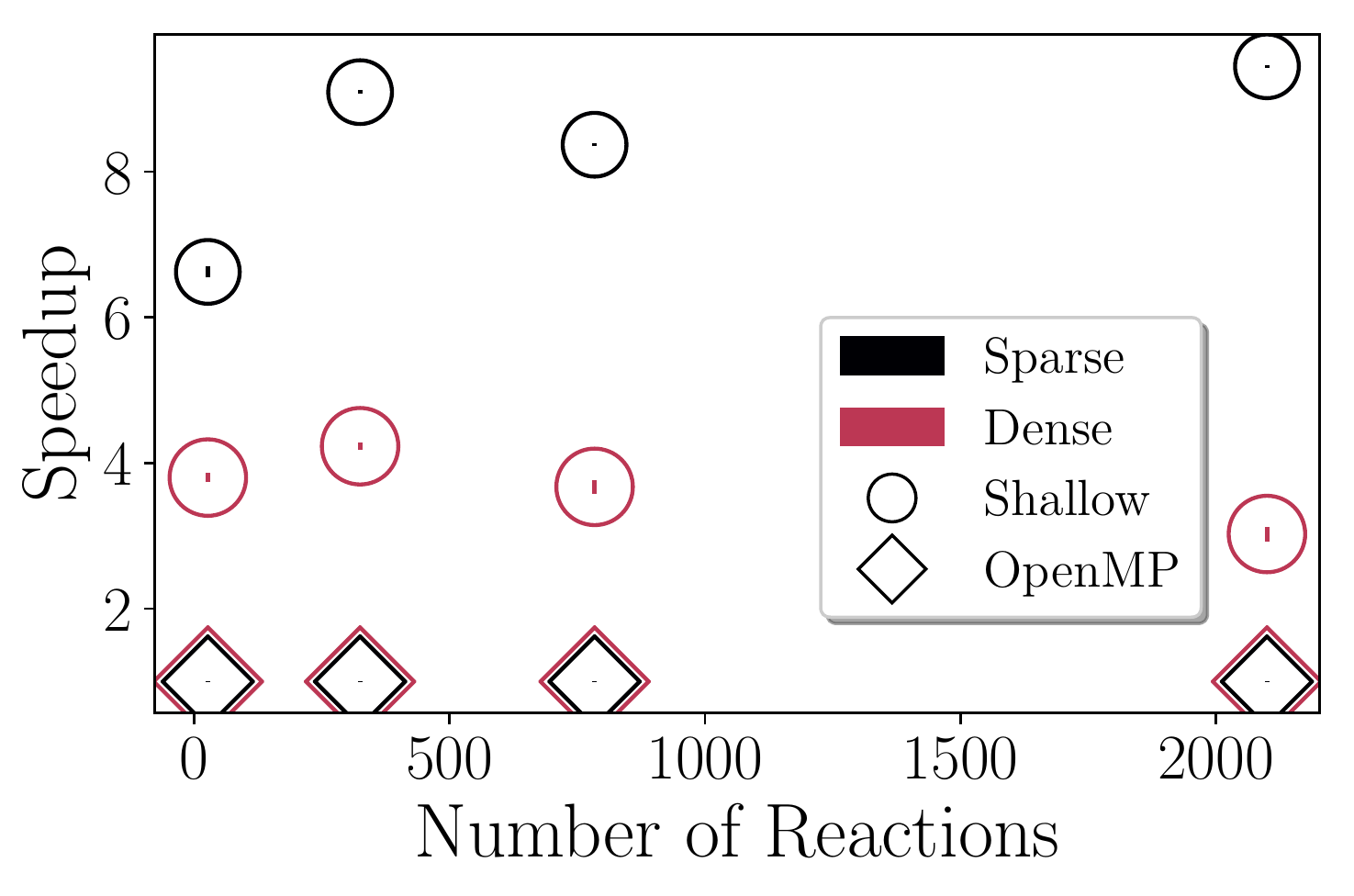}
      \caption{The speedup of the sparse and dense Jacobian evaluations for Intel OpenCL over the sparse\slash dense OpenMP baseline on the \avx/ machine.}
      \label{F:sparse_vs_dense_speedup}
  \end{subfigure}
  \\
  \begin{subfigure}[t]{0.48\linewidth}
      \includegraphics[width=\textwidth]{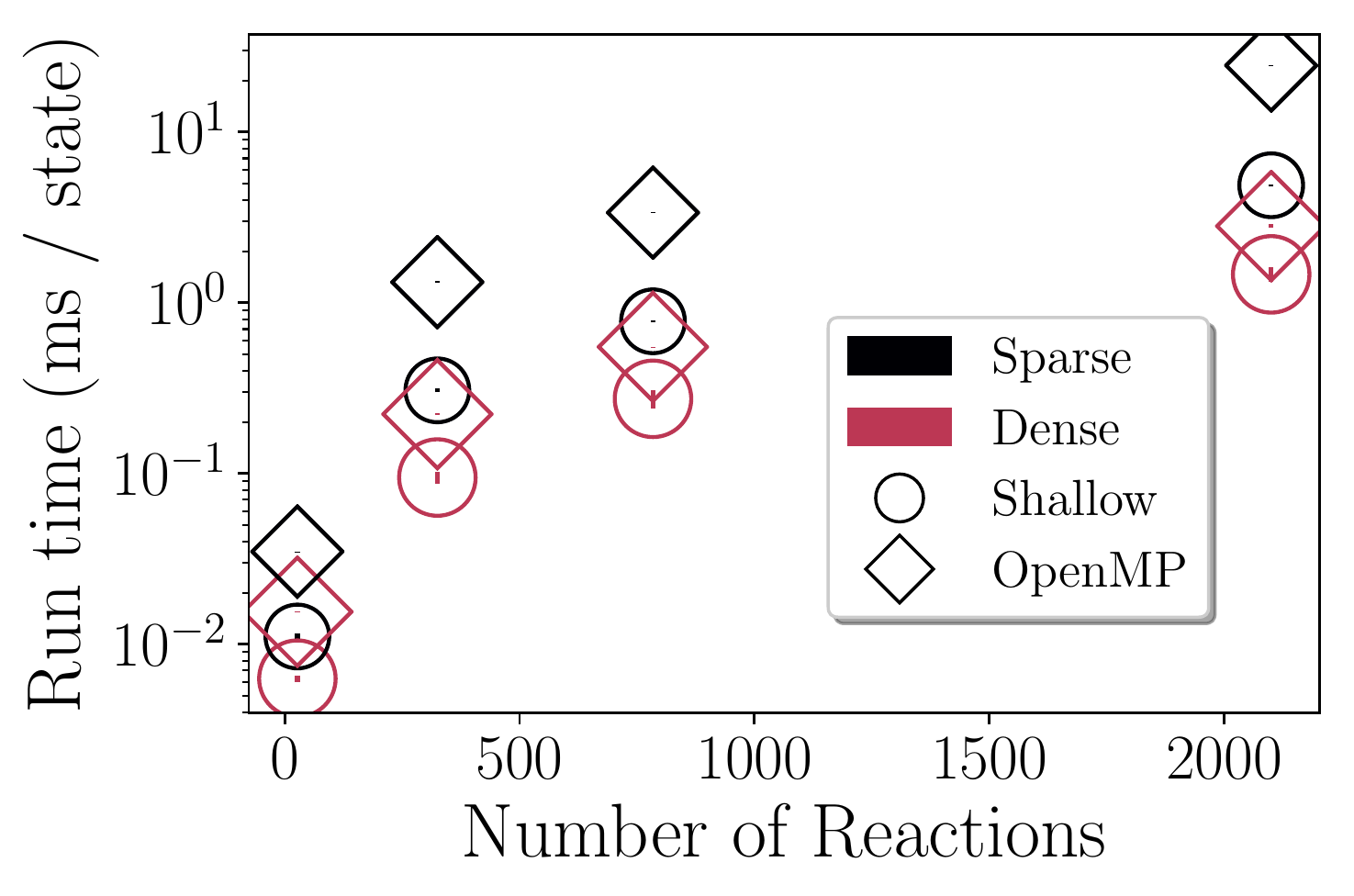}
      \caption{The mean run time per condition of sparse and dense Jacobian evaluations for Intel OpenCL\slash OpenMP on the \sse/ machine.}
      \label{F:sparse_vs_dense_sse}
  \end{subfigure}
  \hfill
  \begin{subfigure}[t]{0.48\linewidth}
      \includegraphics[width=\textwidth]{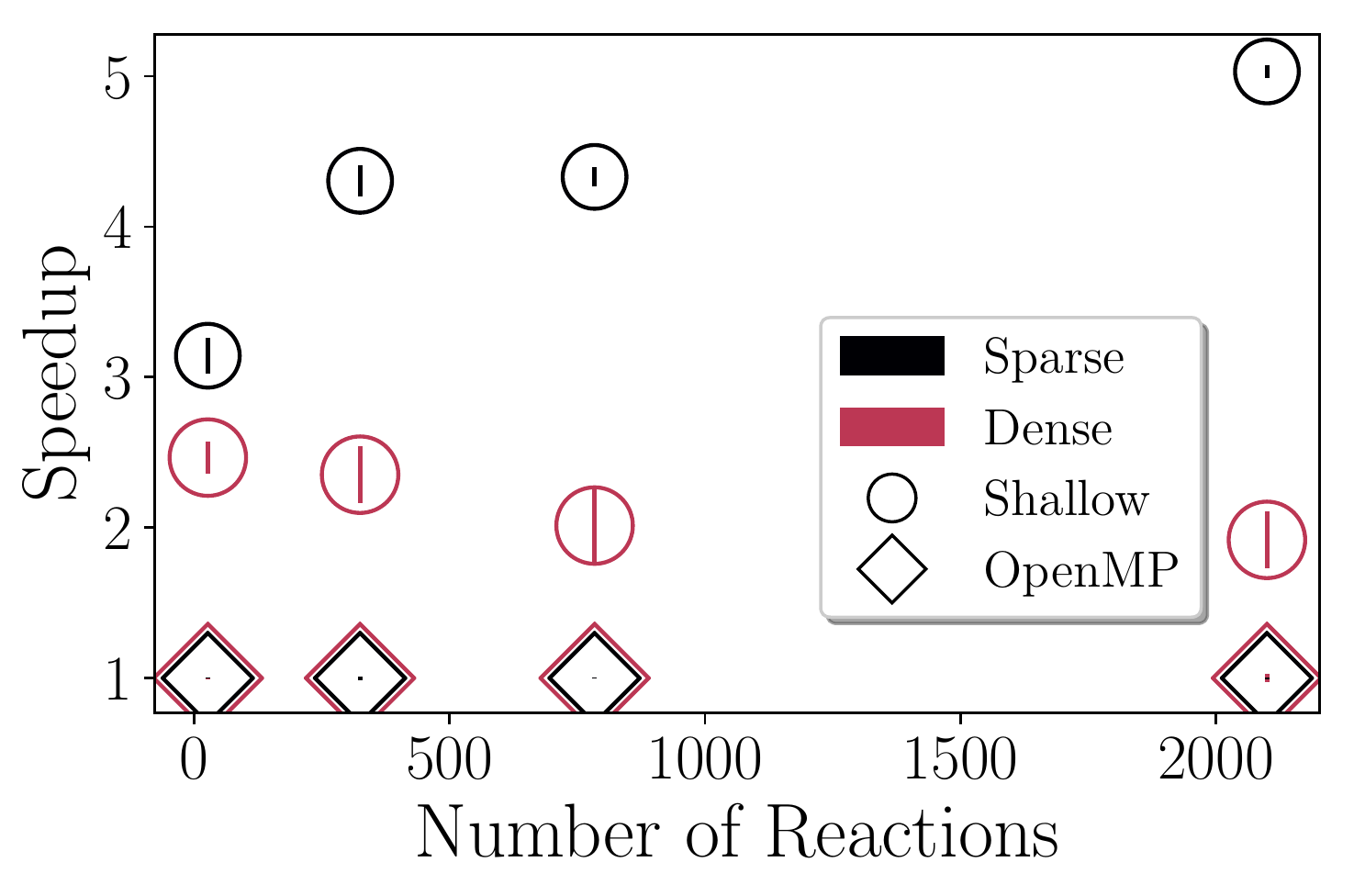}
      \caption{The speedup of the sparse and dense Jacobian evaluations for Intel OpenCL over the sparse\slash dense OpenMP baseline on the \sse/ machine.}
      \label{F:sparse_vs_dense_sse_speedup}
  \end{subfigure}
  \caption{Performance of sparse and dense Jacobian evaluations on the CPU platforms in \texttt{pyJac}.}
  \label{F:jacobian_perfomance}
\end{figure}

\Cref{F:jacobian_perfomance} shows the performance of the sparse and dense Jacobian evaluations in \texttt{pyJac} on the CPU platforms.
In~\cref{F:sparse_vs_dense}, the mean run time per condition is presented for the shallow-vectorized Intel OpenCL and OpenMP codes on the \avx/ CPU.
The sparse Jacobian evaluates slower on both Intel OpenCL and OpenMP due to indirect lookup indexing requirements, as discussed in~\cref{S:sparsity}.
Interestingly, indirect lookup less-negatively impacts the shallow-vectorized OpenCL code: the sparse OpenMP code is \SIrange{2.47}{10.42}{$\times$} slower than the dense OpenMP evaluation, while the sparse shallow-vectorized OpenCL code is just \SIrange{1.41}{3.34}{$\times$} slower than its dense counterpart.
As a result, the shallow-vectorized sparse OpenCL code performs as fast or faster than the dense OpenMP code in all cases on the \avx/ machine (\cref{F:sparse_vs_dense}).
\Cref{F:sparse_vs_dense_speedup} shows the speedup of the shallow-vectorized OpenCL, sparse and dense Jacobian evaluations over the same on OpenMP; the dense OpenCL code is \SIrange{3.03}{4.23}{$\times$} faster than the corresponding dense OpenMP code.
This speedup increases to \SIrange{6.63}{9.44}{$\times$} for the sparse Jacobian.

On the \sse/ machine,~\cref{F:sparse_vs_dense_sse} shows similar results: the sparse OpenMP code is the slowest in all cases, and the shallow-vectorized OpenCL code is nearly as fast as the dense OpenMP code.
Once again, indirect lookup less-negatively impacts the sparse OpenCL code, which is only \SIrange{1.76}{3.33}{$\times$} slower than its dense counterpart, while the sparse OpenMP code is significantly (\SIrange{2.25}{8.72}{$\times$}) slower than the dense version.
In~\cref{F:sparse_vs_dense_sse_speedup}, the speedup of the sparse and dense shallow-vectorized OpenCL codes are compared with their OpenMP versions; the dense OpenCL code is \SIrange{1.92}{2.47}{$\times$} faster while the sparse shallow-vectorization achieves a speedup of \SIrange{3.14}{5.03}{$\times$}.

\begin{figure}[htbp]
   \centering
  \begin{subfigure}[t]{0.48\linewidth}
      \includegraphics[width=\textwidth]{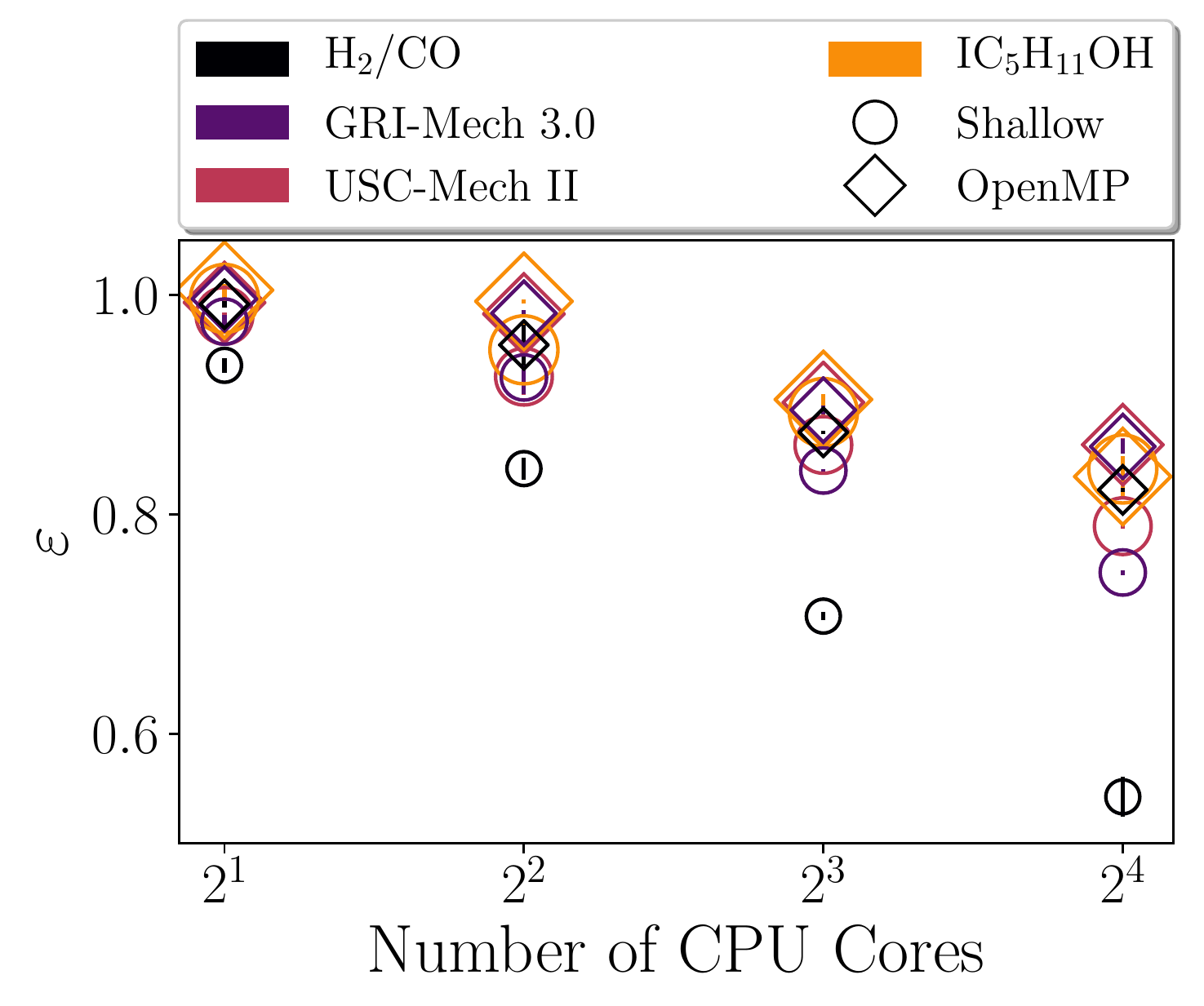}
      \caption{Sparse Jacobian evaluation parallel scaling efficiency for Intel OpenCL\slash OpenMP on the \avx/ machine.}
      \label{F:sparse_jac_scaling}
  \end{subfigure}
  \hfill
  \begin{subfigure}[t]{0.48\linewidth}
      \includegraphics[width=\textwidth]{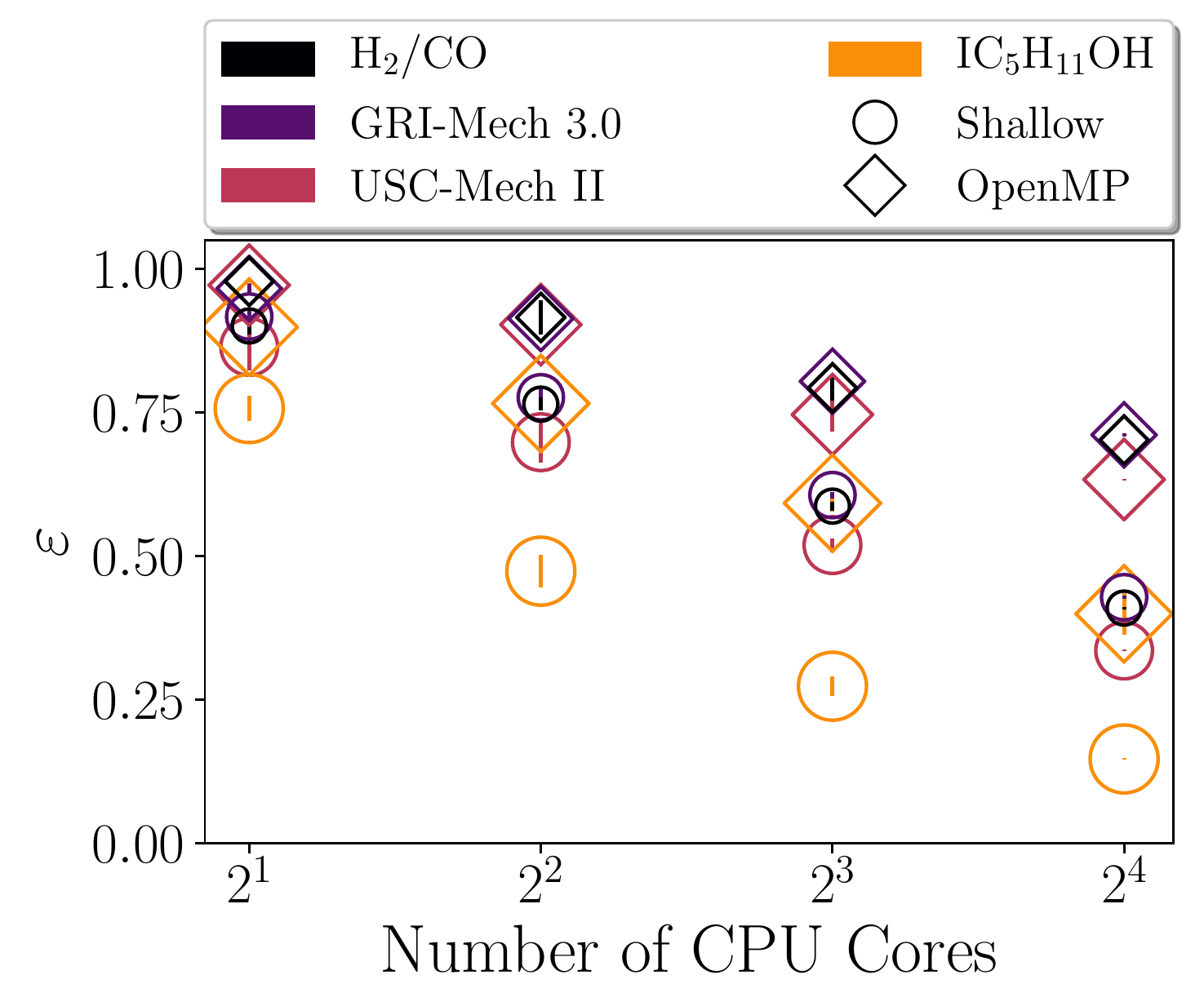}
      \caption{Dense Jacobian evaluation parallel scaling efficiency for Intel OpenCL\slash OpenMP on the \avx/ machine.}
      \label{F:dense_jac_scaling}
  \end{subfigure}
  \caption{Strong parallel scaling efficiencies of sparse and dense Jacobian evaluations for the shallow-vectorized Intel OpenCL and OpenMP codes on the \avx/ CPU.}
\end{figure}

\Cref{F:sparse_jac_scaling} compares the strong parallel scaling efficiency of the sparse shallow-vectorized OpenCL with the sparse OpenMP code.
Although the plot is challenging to read since most of the data are clustered together, it shows that the shallow-vectorized OpenCL code scales similarly to the OpenMP code, in contrast to the parallel scaling efficiency of source-term evaluation (\cref{F:source_parallel_scaling}).
The \ce{H2}\slash\ce{CO} model scales the worst for both codes, ranging from \numrange{0.94}{0.54} and \numrange{0.99}{0.82} efficiency for OpenCL and OpenMP respectively on \numrange{2}{16} cores.
As the model size increases, the efficiency of the OpenCL code improves dramatically, reaching \numrange{0.997}{0.84} for the isopentanol model.

\Cref{F:dense_jac_scaling} shows scaling for the OpenMP and shallow-vectorized dense Jacobian OpenCL codes.
In this case, the isopentanol model scales the worst for both cases.
The sheer size of the dense isopentanol Jacobian limited the total number of states for the dense isopentanol Jacobian evaluation to \num{50000}---storing the dense matrix for a single thermochemical state takes over \SI{1}{\mega\byte} of data, so \num{50000} states requires over \SI{50}{\giga\byte} of memory); this greatly drops the computation cost for this case, and adversely affects the scaling efficiency as discussed in~\cref{S:source_results}.
Excluding the isopentanol model, the dense shallow-vectorized OpenCL code scales slightly better than for source-term evaluation: \numrange{0.70}{0.78} and \numrange{0.52}{0.61} for four and eight cores, respectively (compared with \numrange{0.66}{0.72} and \numrange{0.44}{0.48} for shallow-vectorized OpenCL source-term evaluations).
This results from the higher computational cost of Jacobian evaluation, and hence more available work per CPU core.
As in~\cref{S:source_results}, the shallow-vectorized dense Jacobian code running on \num{1} and \num{4} cores of the \avx/ machine performs roughly as fast as the OpenMP code on \num{4} and \num{16} cores, respectively.
The parallel scaling efficiency of OpenMP on \num{16} cores (excluding isopentanol) is \numrange{0.63}{0.71}, similar to the shallow vectorization's efficiency of \numrange{0.70}{0.78} for \num{4} cores.

\begin{figure}[htbp]
   \centering
  \begin{subfigure}[t]{0.48\linewidth}
      \includegraphics[width=\textwidth]{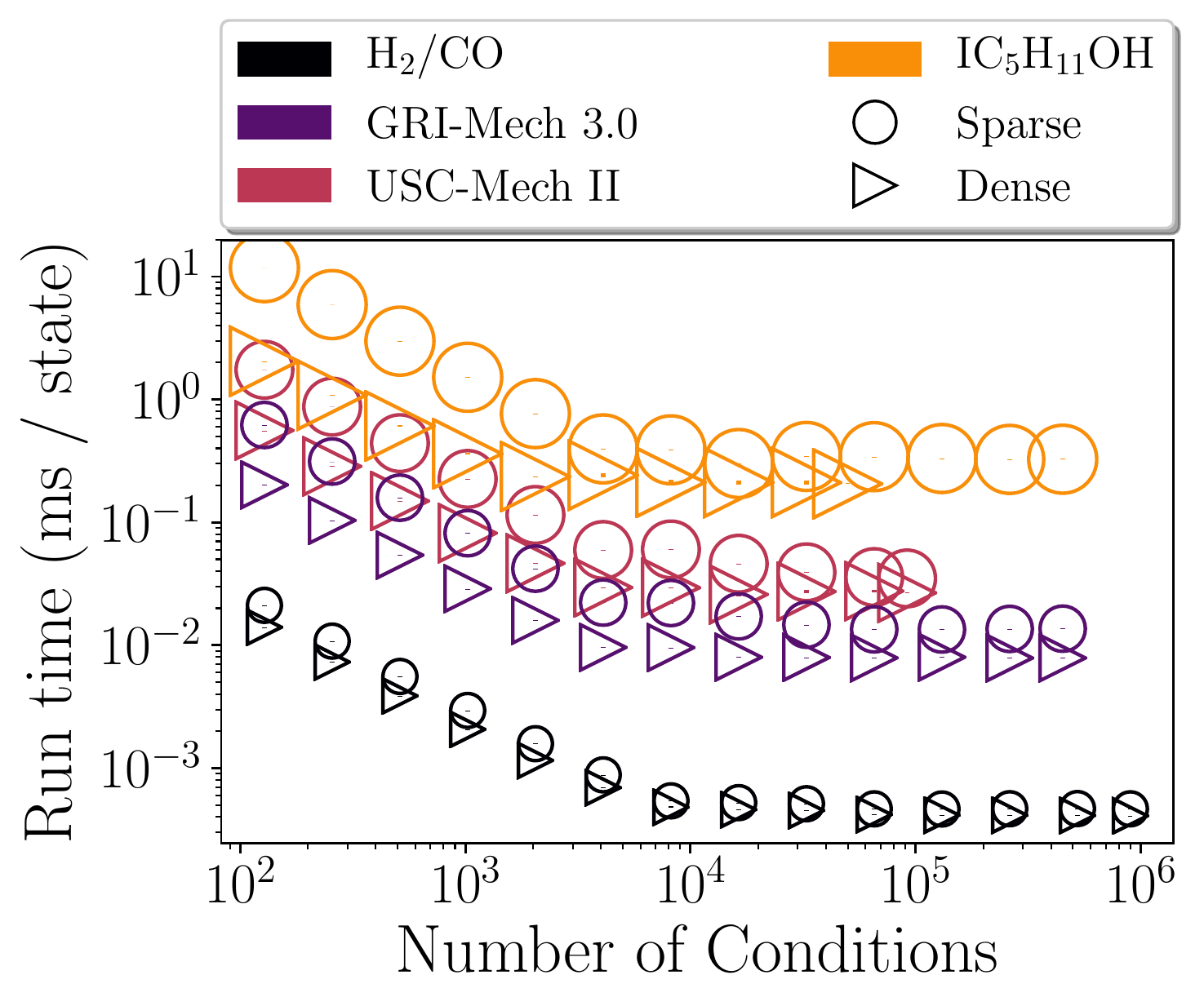}
      \caption{Mean run time per condition of sparse and dense Jacobian evaluations on the \gpunew/ GPU.}
      \label{F:gpu_sparse_vs_dense}
  \end{subfigure}
  \hfill
  \begin{subfigure}[t]{0.48\linewidth}
      \includegraphics[width=\textwidth]{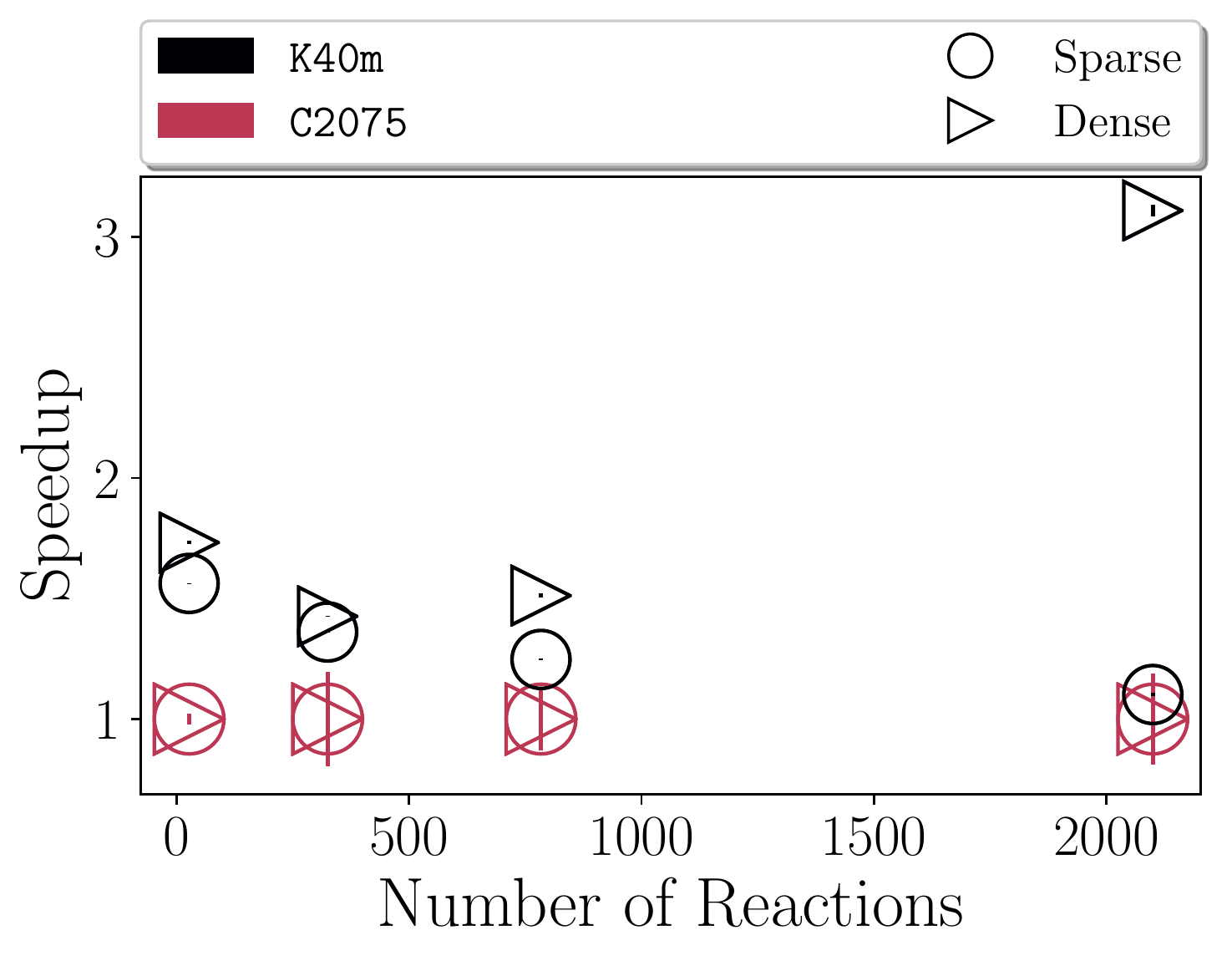}
      \caption{Speedup of the \gpunew/ over \gpuold/ GPUs for sparse and dense Jacobian evaluations, normalized per-Jacobian type (i.e., sparse versus dense).}
      \label{F:gpu_jacobian_speedup}
  \end{subfigure}
  \caption{The performance of sparse\slash dense Jacobian evaluation on the \gpunew/ and \gpuold/ GPUs.}
\end{figure}

\Cref{F:gpu_sparse_vs_dense} plots the mean run time per condition for the sparse and dense Jacobian evaluations on the \gpunew/ GPU.
As in the species evaluation case, the mean run time per condition drops steadily, for both cases becoming roughly constant after just \num{3e3} states (except for \ce{H2}\slash\ce{CO}, which levels off near \num{e4} conditions).
Sparse Jacobian evaluation is significantly slower for all models before the GPU becomes saturated (due to the indirect indexing lookup), but the performance gap between sparse and dense evaluations narrows past the saturation point.
This is likely due to the ability to fit many more sparse Jacobian matrices in the \gpunew/'s memory, as well as improved data locality\slash caching due to the smaller size of the sparse Jacobian.
\Cref{F:gpu_jacobian_speedup} presents the speedup of the \gpunew/ over the \gpuold/ GPU for sparse\slash dense Jacobian evaluation; sparse evaluation on the \gpunew/ is \SIrange{1.10}{1.59}{$\times$} faster than on the \gpuold/, while dense evaluation is \SIrange{1.36}{3.0}{$\times$} faster.
The speedup on the \gpunew/ decreases with increasing model size for the sparse formulation, but increases for larger models when dense; this likely results from the larger available memory of the \gpunew/, and hence fewer data-transfer operations to\slash from the GPU.

\begin{figure}[htbp]
   \centering
  \begin{subfigure}[t]{0.48\linewidth}
      \includegraphics[width=\textwidth]{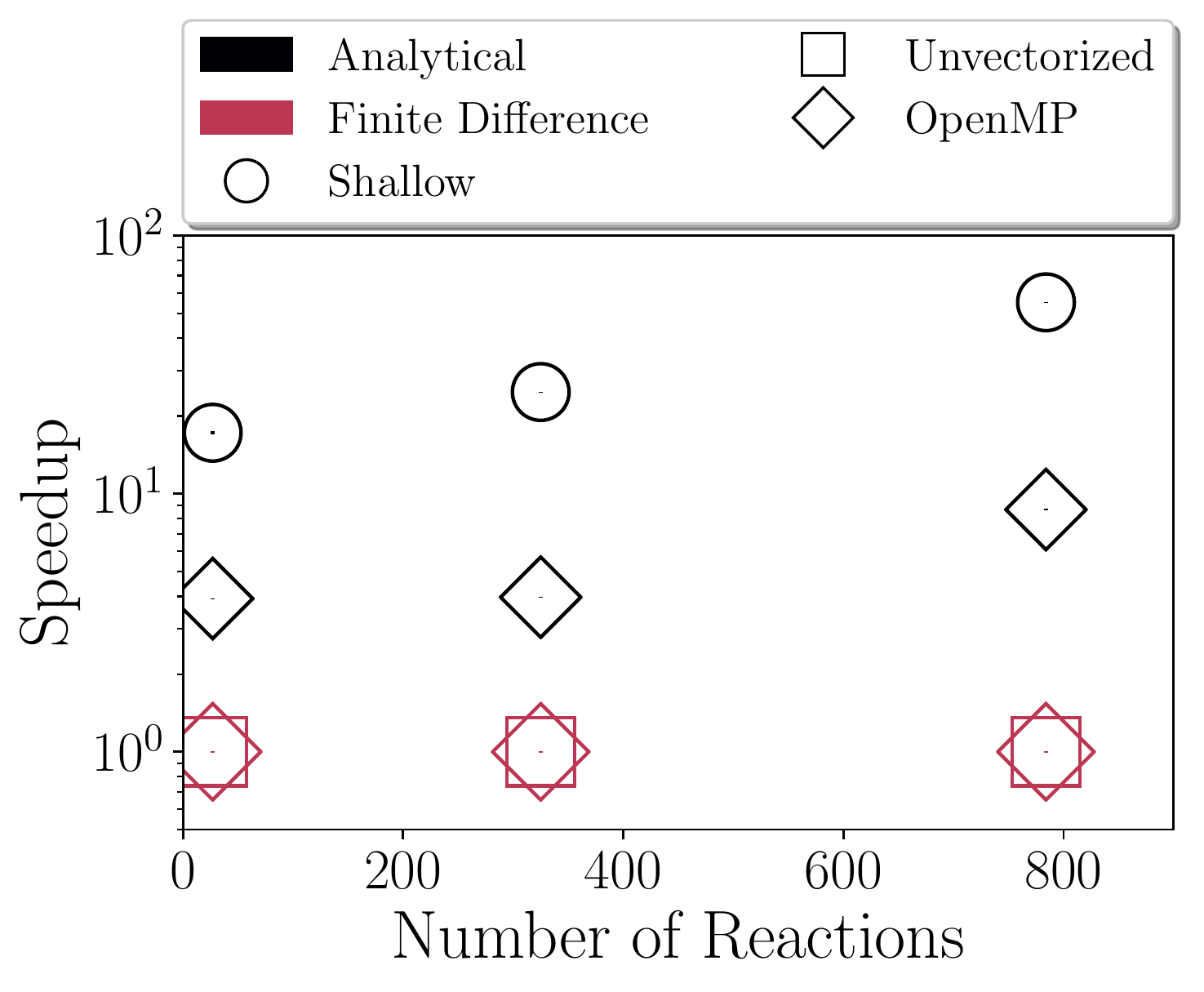}
      \caption{Speedup of the sparse analytical Jacobian versus finite-difference Jacobian evaluation on the \avx/ CPU, normalized per-language.}
      \label{F:fd_vs_analytical_cpu}
  \end{subfigure}
  \hfill
  \begin{subfigure}[t]{0.48\linewidth}
      \includegraphics[width=\textwidth]{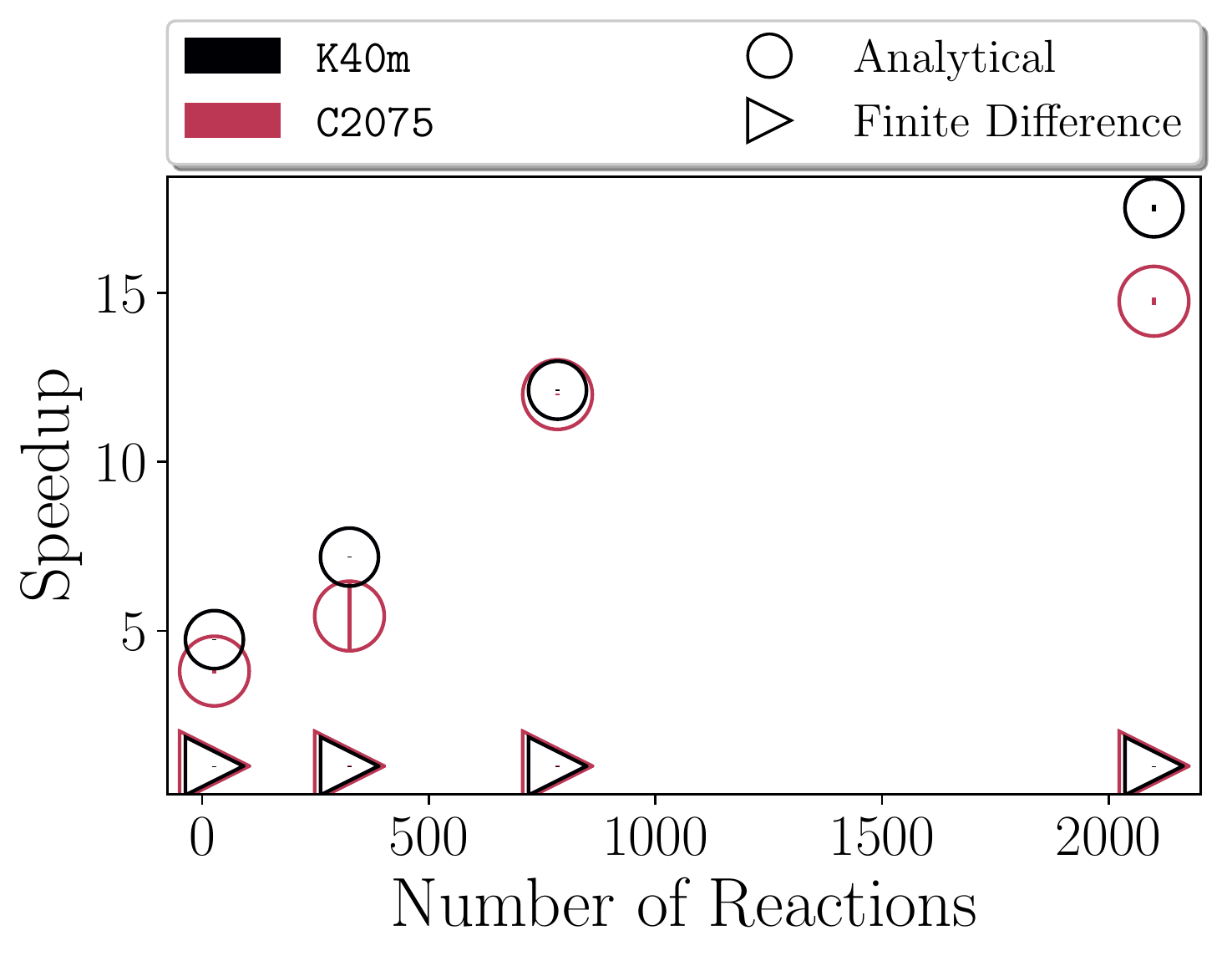}
      \caption{Speedup of the analytical versus finite-difference Jacobian evaluation on both the \gpunew/ and \gpuold/ GPUs, normalized per-GPU.}
      \label{F:fd_vs_analytical_gpu}
  \end{subfigure}
  \caption{The performance of a sparse, first-order forward finite-difference Jacobian compared to the analytical Jacobian on the \avx/ CPU and both GPUs.}
  \label{F:fd_vs_analytical}
\end{figure}

\Cref{F:fd_vs_analytical} compares the performance of the sparse analytical Jacobian with a sparse first-order finite-difference Jacobian on both the \avx/ CPU and \gpuold/\slash \gpunew/ GPUs.
\Cref{F:fd_vs_analytical_cpu} shows large speedups for both OpenMP and shallow-vectorized OpenCL; the analytical OpenMP Jacobian is \SIrange{3.92}{8.67}{$\times$} faster, while the analytical OpenCL Jacobian achieves speedups of \SIrange{17.22}{55.11}{$\times$}.
We excluded the isopentanol case here, since a single run of the sparse finite-difference Jacobian using either OpenCL or OpenMP took over 12 hours of run time.
In addition the current finite-difference formulation breaks Intel's auto-vectorizer, hence we compared OpenCL against the unvectorized OpenCL code (we did not prioritize fixing this issue, since we implemented the finite-difference Jacobian for comparison purposes only).
Although we do not display the dense finite-difference Jacobian speedup in~\cref{F:fd_vs_analytical_cpu}, the dense OpenCL and OpenMP analytical Jacobian codes outperform the finite-difference variants by even larger margins: \SIrange{24.44}{245.63}{$\times$} for OpenCL and \SIrange{9.68}{112.73}{$\times$} for OpenMP (these data do include the isopentanol model, though limited to \num{50000} conditions as discussed previously).
\Cref{F:fd_vs_analytical_gpu} compares the sparse analytical and finite-difference Jacobians on the GPUs.
The analytical Jacobian on the \gpunew/ and \gpuold/ shows speedups of \SIrange{3.81}{17.60}{$\times$} that increase with chemical model size; the \gpunew/ has a larger speedup than the \gpuold/ for the isopentanol model (\SI{17.60}{$\times$} vs.\ \SI{14.75}{$\times$}) due to its larger available memory.
Although not pictured, the dense analytical Jacobian on the \gpunew/ GPU has larger speedups compared with the dense finite-difference Jacobian: \SIrange{4.04}{45.13}{$\times$}.
The \gpunew/ GPU again shows significantly larger speedups over the \gpuold/ for the isopentanol model (\SI{45.13}{$\times$} vs.\ \SI{23.85}{$\times$}), further underscoring the effect of more available memory on the \gpunew/.

\begin{figure}[htbp]
   \centering
  \begin{subfigure}[t]{0.48\linewidth}
      \includegraphics[width=\textwidth]{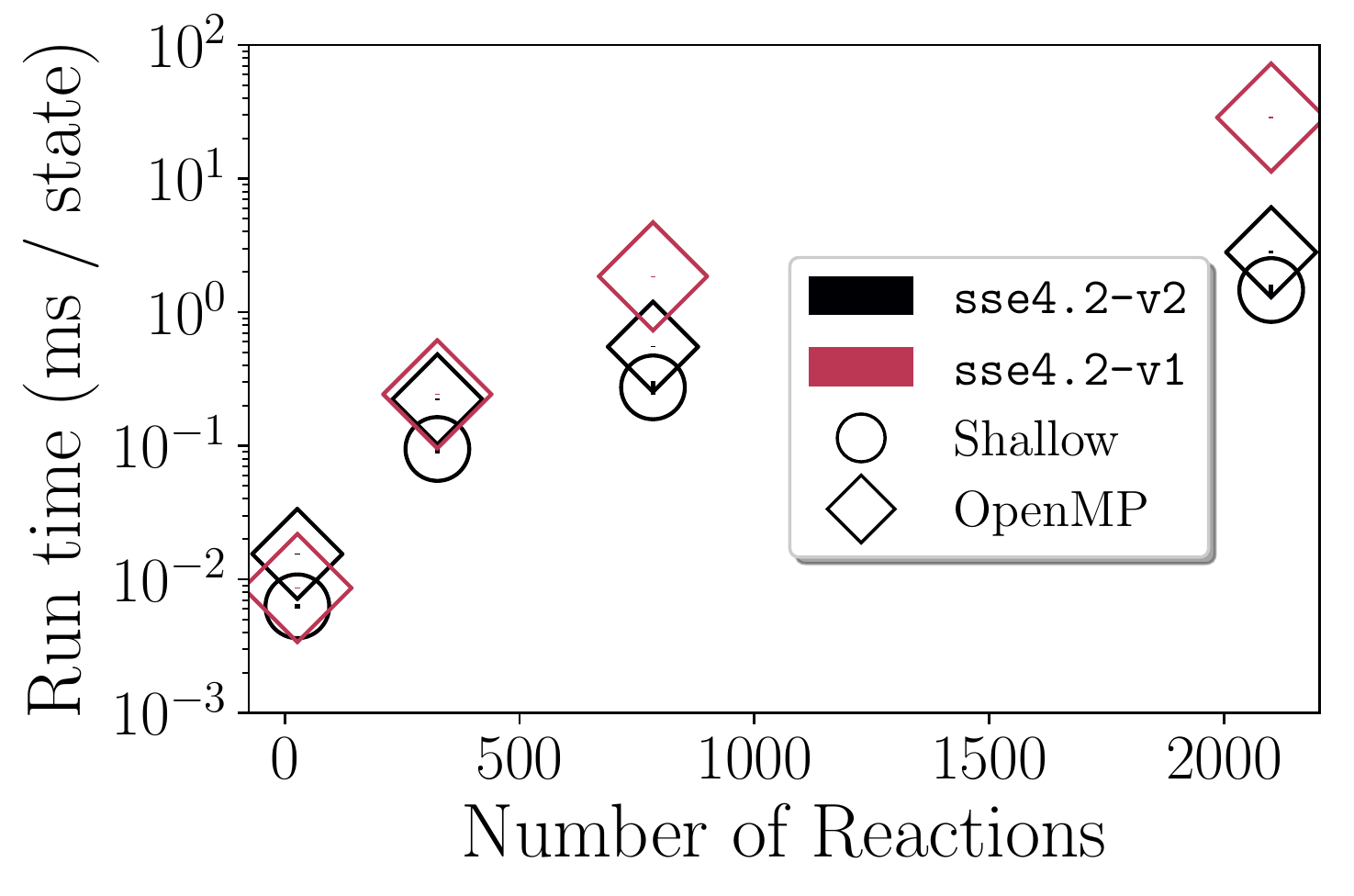}
      \caption{\sse/ CPU}
      \label{F:v1_vs_v2_cpu}
  \end{subfigure}
  \hfill
  \begin{subfigure}[t]{0.48\linewidth}
      \includegraphics[width=\textwidth]{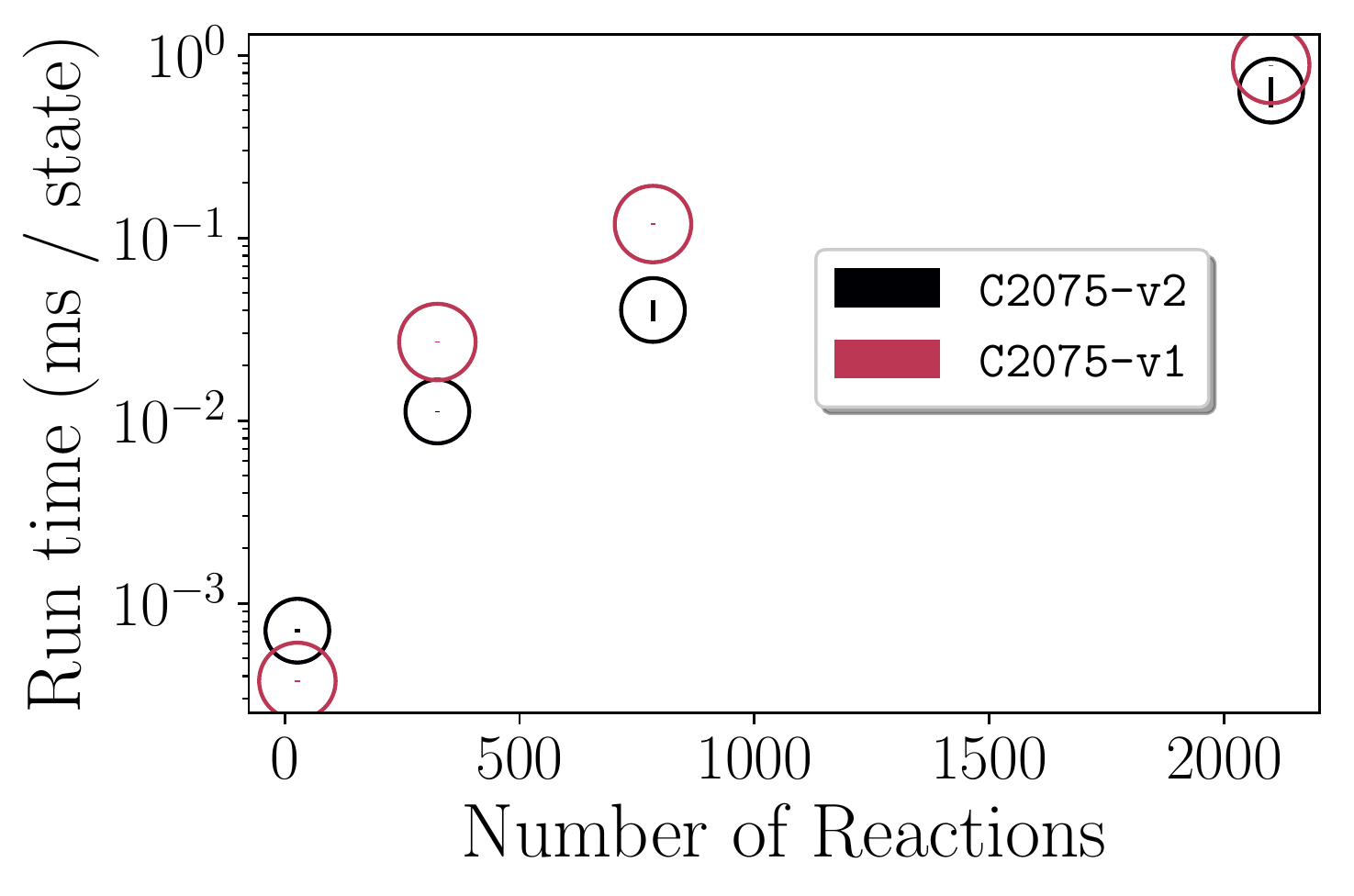}
      \caption{\gpuold/ GPU}
      \label{F:v1_vs_v2_gpu}
  \end{subfigure}
  \caption{Performance comparison of dense Jacobian evaluation using \texttt{pyJac-v2} and \texttt{pyJac-v1}~\cite{pyjac16}.}
  \label{F:v1_vs_v2}
\end{figure}

\Cref{F:v1_vs_v2} compares the performance of evaluating the dense analytical Jacobian of \texttt{pyJac-v2} with that of the previous version, \texttt{pyJac-v1}~\cite{pyjac16}, on the \sse/ CPU and \gpuold/ GPU.
(We selected dense Jacobian evaluation for this comparison since it was the only type implemented in the previous version of \texttt{pyJac}.)
On the \sse/ CPU, the \texttt{pyJac-v2} evaluates faster than \texttt{pyJac-v1} for OpenMP for the larger chemical models; the static OpenMP code generated by \texttt{pyJac-v1} (see~\cref{s:unittest}) is \SI{1.79}{$\times$} faster for the \ce{H2}\slash\ce{CO} model, and only \SI{1.09}{$\times$} slower for the GRI-Mech 3.0 model.
In contrast, the loop-based OpenMP code of \texttt{pyJac-v2} is \SIrange{3.37}{10.19}{$\times$} faster than \texttt{pyJac-v1} for the USC-Mech II and isopentanol models.
The shallow-vectorized OpenCL \texttt{pyJac-v2} code is faster than \texttt{pyJac-v1} in all cases, achieving speedups of \SIrange{1.37}{19.56}{$\times$} that increase with model size.
\Cref{F:v1_vs_v2_gpu} compares the performance of the \texttt{pyJac-v2} with \texttt{pyJac-v1} for evaluating dense analytical Jacobians on the \gpuold/ GPU.
As with the CPU, the static code of \texttt{pyJac-v1} is slightly faster for the \ce{H2}\slash\ce{CO} model, but \texttt{pyJac-v2} outperforms \texttt{pyJac-v1} by \SIrange{1.25}{2.84}{$\times$} for the other models.
Performance likely drops for the isopentanol model due to the limited number of conditions in dense evaluation for \texttt{pyJac-v2}, as noted earlier.

\section{Practical notes on OpenCL use}
\label{S:opencl}

While OpenCL provides a simple interface to enable cross-platform execution, and significant speedups were achieved via shallow-vectorized OpenCL code in this work, there are some serious potential pitfalls in its use.
The closed-source OpenCL runtimes tested in this work (Intel and Nvidia) contain several bugs that result in compilation failures, simply incorrect vectorized machine code, or even segmentation faults.
Further, these runtimes (in our experience) tend to be less responsive to fixing said bugs, with relatively infrequent new releases (or in Nvidia's case, lack of changelogs\slash public records of bug-fixes).
On the other hand, the open-source OpenCL runtime used in this work, POCL, has far fewer implementation bugs, and when issues arise the community is very responsive to bug reports and user outreach; however, POCL fails to achieve vectorization as noted in~\cref{S:source_results}.

To demonstrate the type of issue discussed, we created a minimum working example~\cite{Nvidia_mwe} that shows a failure of Nvidia's OpenCL runtime corresponding to the GPU driver version \num{375.26} on a Tesla \gpunew/ GPU; simply changing to another runtime (e.g., Intel) produces the correct result---with no code changes or recompilation.
This provides a particularly vexing problem for the programmer, since little can often be done to resolve the issue; thankfully, in this case we were able to upgrade the driver version to resolve the problem.
Further, as noted throughout this work, certain code-generation patterns can break OpenCL execution\slash vectorization, e.g., the failure of POCL to achieve vectorization or Intel OpenCL's failure to vectorize the finite-difference Jacobian.
Indeed the vectorization process attempted by most OpenCL runtimes (and thus, reasons for incorrect\slash unvectorized code output) is obscured from the user, making reasoning about errors or performance trends quite difficult.
Thankfully, \texttt{loo.py} allows relatively easy switching between output languages; the most significant code change requires building of the wrapper that initializes\slash transfers memory and calls the source-rate\slash Jacobian kernel.
These issues make it critical to provide adequate implementation details (e.g., runtime version, platform, etc., as given in~\cref{s:test_platforms}) for codes utilizing OpenCL in order to enable reproducible results.

\section{Conclusions}
\label{S:future}

In this work, we developed automatically generated OpenCL codes for SIMD and SIMT-vectorized evaluations of thermochemical source-term and sparse\slash dense chemical kinetic Jacobian matrices.
The main contributions of this work are:
\begin{itemize}
 \item Deriving and verifying a new Jacobian formulation that greatly increases sparsity;
 \item Enabling vectorized execution on the CPU, GPU, and other accelerators; and
 \item Achieving significant speedups over a strictly parallel Jacobian and source-term evaluation on SIMD-enabled processors (e.g., CPUs).
\end{itemize}

These efforts are made publicly available (see~\cref{A:availability}) via the open-source, high-performance, chemical kinetics code \texttt{pyJac}.
The new molar-based formulation resulted in highly sparse chemical kinetic Jacobians, and allows selection of either the constant-volume or constant-pressure approximation.
In addition, sparsity can be increased further by eliminating components associated with the bath gas, as discussed in \cref{S:sparsity}; this approximation is not a key feature of this work, and future efforts to incorporate more sophisticated Jacobian approximations would be a worthwhile endeavor.

We also demonstrated source-term and Jacobian evaluation for a range of chemical kinetic models~\cite{Burke:2011fh,smith_gri-mech_30,Wang:2007,Sarathy:2013jr} and multiple CPUs\slash GPUs (\cref{t:cpus,t:gpus}).
In addition to parallel OpenMP evaluation on the CPU, this work enabled the shallow-vectorized evaluation of the chemical-kinetic source terms and analytical Jacobian on both the CPU and GPU via OpenCL.
Deep vectorization is possible on the Portable OpenCL (POCL) platform~\cite{poclIJPP}, but it yields no performance benefit as POCL did not achieve vectorizations for any execution pattern studied.
Deep vectorization deserves further study with other platforms (e.g., CUDA).

We demonstrated significant speedups in shallow SIMD-vectorized execution over a parallel OpenMP code for evaluating the chemical-kinetic source terms and sparse\slash dense Jacobian: up to \SI{4.09}{$\times$}, \SI{9.44}{$\times$}, and \SI{4.23}{$\times$}, respectively, on an \avx/-capable CPU.
Sparse Jacobians evaluate more slowly than dense Jacobians on all CPU\slash GPU platforms due to indirect lookup requirements in array indexing, but this adversely affects the shallow-vectorized OpenCL code less than OpenMP.
Further, analytical Jacobians evaluate significantly faster than a first-order finite-difference-based approach on all platforms.
Finally, we compared the performance of evaluating dense, analytical Jacobians in this new version of \texttt{pyJac} with the previous version~\cite{pyjac16}.
The OpenMP version evaluates moderately slower on the CPU for the smallest chemical model (e.g., \SI{1.79}{$\times$} on the \avx/ CPU), but significantly faster for the larger models---up to \SI{10.19}{$\times$}.
The shallow-vectorized OpenCL code runs faster than the previous version over all chemical models, reaching speedups of \SI{19.56}{$\times$}.

The OpenMP code-generation is currently only capable of parallel execution, but extending this platform to shallow\slash deep-vectorizations (via \texttt{loo.py} and compiler \texttt{\#pragma}s) is a key priority going forwards since OpenMP is a standard library on most machines.
In addition, CUDA~\cite{Nvidia:2018} has been significantly more reliable in previous works~\cite{Niemeyer:2016aa,CurtisGPU:2017}, while Intel's open-source OpenCL alternative, ISPC~\cite{pharr2012ispc}, has been relatively stable and easy to work with during preliminary efforts with the unit-testing discussed in~\cref{s:unittest}.
The current deep-vectorization formulation would be executable for both CUDA and ISPC targets, as these languages implement double-precision atomic operations, further recommending their use.
It is also possible, particularly for the Intel OpenCL\slash POCL runtimes, that better performance\slash stability might be achieved using so-called ``explicit'' vectorization, i.e., through use of built-in vector types such as the \texttt{double8}.
Specifically, this change could enable vectorization on the POCL runtime and might also enable deep vectorization on the Intel OpenCL runtime.
Using OpenCL Image\slash CUDA Texture memory to accelerate the indirect lookup for sparse matrix evaluation should be investigated as well.
Finally, future extensions of this work will include extending to additional target languages (e.g., vectorized OpenMP, CUDA, ISPC) to improve ease of use and reliability, improving the existing OpenCL targets (e.g., to enable meaningful deep-vectorized evaluation), and implementing reaction sorting~\cite{Sewerin20151375} to improve SIMD efficiency (\cref{S:SIMD_scaling}).

One key component missing in this work is vectorized sparse\slash dense linear-algebra subroutines to maximize the performance of LU-factorization and matrix-vector multiplication (commonly used in implicit-integration techniques).
Third-party\slash open-source options exist for some target languages, e.g., cuBLAS~\cite{cublas}, clBLAS\slash clSPARSE~\cite{clmath} or SuperLU~\cite{superlu99}, but these do not necessarily cover all targets\slash required linear-algebra operations and, in the case of the CUDA\slash OpenCL libraries, are often optimized operation on one large matrix instead of many (relatively) smaller matrices.
The extent to which these existing programs may be used needs to be assessed and missing operations should be implemented in \texttt{loo.py} to ensure easy switching between target languages and vectorization types.

\section{Acknowledgments}
This material is based upon work supported by the National Science Foundation
under grants ACI-1534688 (Curtis and Sung) and ACI-1535065 (Niemeyer).

\pagebreak


\appendix
\setcounter{figure}{0}
\setcounter{table}{0}

\renewcommand*{\thesection}{\appendixname~\Alph{section}}

\section{Availability of material}
\label{A:availability}

The results for this article were obtained using \texttt{pyJac v2.0.0b0}~\cite{pyjac2}.
The most recent version of \texttt{pyJac} can be found at its GitHub repository:
\url{https://github.com/SLACKHA/pyJac}.
All figures, and the data and plotting scripts necessary to reproduce them,
are available openly under the CC-BY license~\cite{data}.

\section{Jacobian error statistics per test platform}
\label{A:per_platform}

This section gives more detail on the results presented in~\cref{S:jac_valid}, breaking down the reported error statistics per test platform\slash language.
The error of the Intel OpenCL runtime is presented in~\cref{T:intel_error}, the Portable OpenCL (POCL) runtime in~\cref{T:pocl_error}, OpenMP in~\cref{T:omp_error}, and the Nvidia OpenCL runtime in~\cref{T:nv_error}.
POCL and OpenMP tend to have the smallest error norms, while Nvidia tends to have the largest; in particular the stringent filtered error norm $E_{\mathcal{C} = 10^{20}}$ is two orders of magnitude larger for the Nvidia runtime than the other test platforms with the \ce{H2}\slash\ce{CO} and USC-Mech II models.

\begin{table}[htbp]
\sisetup{retain-zero-exponent=true}
\centering
\begin{tabular}{@{}l S[table-format=.0] S[table-format=1.3e1] S[table-format=1.3e1] S[table-format=1.3e1] S[table-format=1.3e1] @{}}
\toprule
Model                 & \multicolumn{1}{c}{$E_{\mathcal{L}}$} & \multicolumn{1}{c}{$E_{\mathcal{C} = 10^{20}}$}   & \multicolumn{1}{c}{$E_{\mathcal{C} = 10^{15}}$} \\
\midrule
\ce{H2}\slash \ce{CO} & \num{1.455e-14}      & \num{8.084e-01}  & \num{1.907e-05} \\
GRI-Mech 3.0          & \num{1.567e-14}      & \num{1.469e-07}  & \num{1.316e-07} \\
USC-Mech II           & \num{9.632e-15}      & \num{5.567e-03}  & \num{1.704e-07} \\
\ce{iC5H11OH}         & \num{1.227e-10}      & \num{1.363e-03}  & \num{2.864e-05} \\
\bottomrule
\end{tabular}
\caption{Summary of Jacobian matrix verification results for the Intel OpenCL runtime.
The reported error statistics are the maximum filtered relative error $E_\mathcal{C}$ and LAPACK error $E_{\mathcal{L}}$ over all vectorization patterns (\cref{t:platforms}), \conp/\slash \conv/, and sparse\slash dense Jacobians.
The threshold for the filtered relative error is the same as reported in~\cref{S:jac_valid}.
}
\label{T:intel_error}
\end{table}

\begin{table}[htbp]
\sisetup{retain-zero-exponent=true}
\centering
\begin{tabular}{@{}l S[table-format=.0] S[table-format=1.3e1] S[table-format=1.3e1] S[table-format=1.3e1] S[table-format=1.3e1] @{}}
\toprule
Model                 & \multicolumn{1}{c}{$E_{\mathcal{L}}$} & \multicolumn{1}{c}{$E_{\mathcal{C} = 10^{20}}$}   & \multicolumn{1}{c}{$E_{\mathcal{C} = 10^{15}}$} \\
\midrule
\ce{H2}\slash \ce{CO} & \num{1.456e-14}      & \num{1.230e-01}  & \num{3.951e-06} \\
GRI-Mech 3.0          & \num{1.014e-14}      & \num{1.890e-07}  & \num{1.877e-07} \\
USC-Mech II           & \num{9.632e-15}      & \num{8.998e-04}  & \num{1.201e-08} \\
\ce{iC5H11OH}         & \num{9.133e-15}      & \num{1.723e-05}  & \num{5.108e-07} \\
\bottomrule
\end{tabular}
\caption{Summary of Jacobian matrix verification results for the Portable OpenCL (POCL) runtime.
The reported error statistics are the maximum filtered relative error $E_\mathcal{C}$ and LAPACK error $E_{\mathcal{L}}$ over all vectorization patterns (\cref{t:platforms}), \conp/\slash \conv/, and sparse\slash dense Jacobians.
The threshold for the filtered relative error is the same as reported in~\cref{S:jac_valid}.
}
\label{T:pocl_error}
\end{table}

\begin{table}[htbp]
\sisetup{retain-zero-exponent=true}
\centering
\begin{tabular}{@{}l S[table-format=.0] S[table-format=1.3e1] S[table-format=1.3e1] S[table-format=1.3e1] S[table-format=1.3e1] @{}}
\toprule
Model                 & \multicolumn{1}{c}{$E_{\mathcal{L}}$} & \multicolumn{1}{c}{$E_{\mathcal{C} = 10^{20}}$}   & \multicolumn{1}{c}{$E_{\mathcal{C} = 10^{15}}$} \\
\midrule
\ce{H2}\slash \ce{CO} & \num{5.962e-15}      & \num{3.614e-02}  & \num{1.657e-06} \\
GRI-Mech 3.0          & \num{1.297e-15}      & \num{1.321e-07}  & \num{1.316e-07} \\
USC-Mech II           & \num{9.630e-15}      & \num{4.185e-04}  & \num{6.746e-09} \\
\ce{iC5H11OH}         & \num{6.131e-15}      & \num{1.721e-05}  & \num{5.108e-07} \\
\bottomrule
\end{tabular}
\caption{Summary of Jacobian matrix verification results for OpenMP execution.
The reported error statistics are the maximum filtered relative error $E_\mathcal{C}$ and LAPACK error $E_{\mathcal{L}}$ over all vectorization patterns (\cref{t:platforms}),  \conp/\slash \conv/, and sparse\slash dense Jacobians.
The threshold for the filtered relative error is the same as reported in~\cref{S:jac_valid}.
}
\label{T:omp_error}
\end{table}

\begin{table}[htbp]
\sisetup{retain-zero-exponent=true}
\centering
\begin{tabular}{@{}l S[table-format=.0] S[table-format=1.3e1] S[table-format=1.3e1] S[table-format=1.3e1] S[table-format=1.3e1] @{}}
\toprule
Model                 & \multicolumn{1}{c}{$E_{\mathcal{L}}$} & \multicolumn{1}{c}{$E_{\mathcal{C} = 10^{20}}$}   & \multicolumn{1}{c}{$E_{\mathcal{C} = 10^{15}}$} \\
\midrule
\ce{H2}\slash \ce{CO} & \num{1.862e-14}      & \num{1.741e+00}  & \num{4.508e-05} \\
GRI-Mech 3.0          & \num{1.489e-14}      & \num{3.842e-07}  & \num{3.687e-07} \\
USC-Mech II           & \num{1.174e-14}      & \num{1.119e-02}  & \num{1.983e-07} \\
\ce{iC5H11OH}         & \num{8.602e-15}      & \num{1.748e-05}  & \num{5.109e-07} \\
\bottomrule
\end{tabular}
\caption{Summary of Jacobian matrix verification results for Nvidia OpenCL execution.
The reported error statistics are the maximum filtered relative error $E_\mathcal{C}$ and LAPACK error $E_{\mathcal{L}}$ over all vectorization patterns (\cref{t:platforms}),  \conp/\slash \conv/, and sparse\slash dense Jacobians.
The threshold for the filtered relative error is the same as reported in~\cref{S:jac_valid}.
}
\label{T:nv_error}
\end{table}

\FloatBarrier

\section{SIMD efficiency Scaling Example}
\label{S:SIMD_scaling}

This simple example demonstrates how the SIMD efficiency of shallow-vectorized OpenCL source-term evaluation depends on the size of the chemical model in question, i.e., the amount of computational work per source-term evaluation.
The base chemical model for this example was the isopentanol model~\cite{Sarathy:2013jr} used throughout this article, and the same thermochemical state database described in~\cref{s:test_platforms} was used for source-term evaluation.
As in~\cref{S:results} all reported results are based on \num{10} individual runs and in this example all cases were run on the \avx/ machine using the Intel OpenCL runtime.

\begin{algorithm}[htbp]
\algnewcommand\algorithmicinput{\textbf{Input:}}
\algnewcommand\INPUT{\item[\algorithmicinput]}
\begin{algorithmic}[0]
  \caption{A greedy selection algorithm to remove reactions from a base chemical model $M$, while preserving the number of active species.}
  \label{a:model_gen}
  \INPUT{Base chemical model $M$ with reactions $R$ and species $S$}
  \Function{Determine Species Count}{\text{active}}
    \For {Species $S_k$ in model $M$}
      \State  $\text{species\smallunderscore{}rxn\smallunderscore{}count}\left[k\right] \gets 0$
      \For {Reaction $R_j$ in model $M$}
	\If{$\text{active}[j]$ and $\left(\left\lvert \nu_{k, j}^{\prime}\right\rvert + \left\lvert \nu_{k, j}^{\prime\prime} \right\rvert \right) > 0$}
	  \State $\text{species\smallunderscore{}rxn\smallunderscore{}count}[k] \gets \text{species\smallunderscore{}rxn\smallunderscore{}count}[k] + 1$
	\EndIf
      \EndFor
    \EndFor
    \Return \text{species\smallunderscore{}rxn\smallunderscore{}count}
  \EndFunction
  \Procedure{Model Generation}{$M$}
  \State $\text{active}[j] \gets \text{True}$ for all reactions $R_j$ in $M$
  \State $\text{species\smallunderscore{}rxn\smallunderscore{}count}\gets \Call{Determine Species Count}{\text{active}}$
  \While {$\operatorname{min}\left(\text{species\smallunderscore{}rxn\smallunderscore{}count}\right) \ge 1$}
     \State $\text{species\smallunderscore{}rxn\smallunderscore{}count}\gets \Call{Determine Species Count}{\text{active}}$
     \For {Reaction $R_j$ in model $M$}
	\State $\text{rxn\smallunderscore{}count}[j] \gets \min_{\forall S_k \in R_j}{\left(\text{species\smallunderscore{}rxn\smallunderscore{}count}\left[k\right]\right)}$
     \EndFor
     \State $\text{remove\smallunderscore{}at} \gets \operatorname{argmax}\left(\text{rxn\smallunderscore{}count}\right)$
     \State $\text{active}[\text{remove\smallunderscore{}at}] \gets \text{False}$
  \EndWhile
  \EndProcedure
\end{algorithmic}
\end{algorithm}

First, the reactions in the isopentanol model were converted to simple reversible Arrhenius reactions by either simply dropping third-body efficiency calculations (third-body enhanced reactions), using the high-pressure-limit coefficients (falloff\slash chemically-activated and P-Log reactions) or fitting Arrhenius parameters to the calculated rate constant at a fixed pressure (Chebyshev reactions).
This conversion made the cost of reaction rate evaluation roughly equivalent between all reactions in model, separating the effect of chemical model size from computational intensities of different reaction types on the SIMD efficiency.
Next, a greedy reaction removal algorithm (\cref{a:model_gen}) generated a number of models ranging from \numrange{2100}{186} reactions, in increments of 200 reactions (except the final increment from \num{200} to \num{186} reactions).

\begin{figure}[htbp]
\centering
\includegraphics[width=0.5\linewidth]{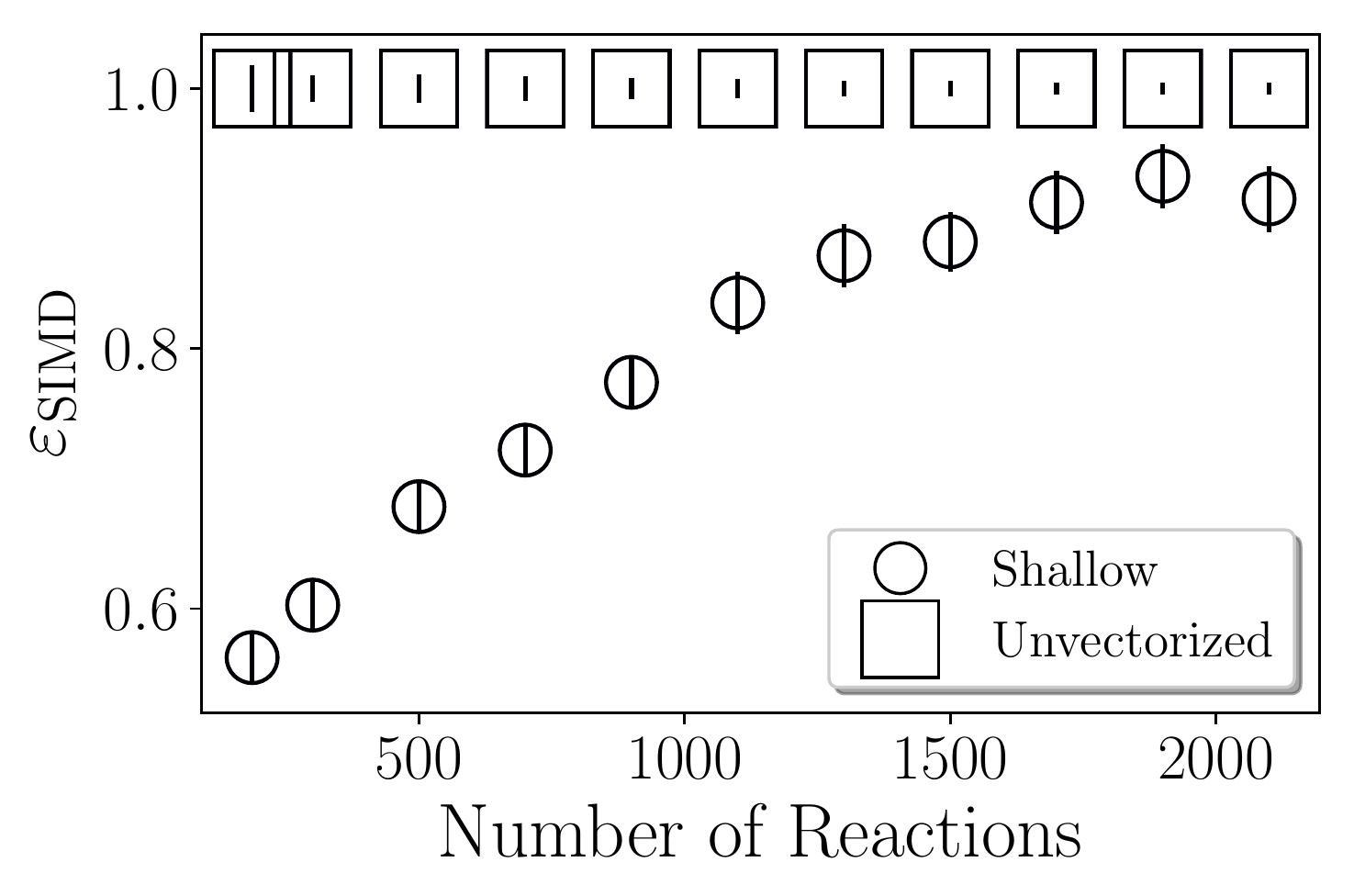}
\caption{The effect on SIMD efficiency of varying chemical model size.}
\label{F:simd_vs_rxns}
\end{figure}

To discern the effect of varying chemical model size on the SIMD efficiency, shallow-vectorized and unvectorized source-term evaluation performance tests were run.
As demonstrated in~\cref{F:simd_vs_rxns} the SIMD efficiency strongly depends on the generated model size and thus the amount of computational work per thermochemical state.
In addition, the range of SIMD efficiency in this example (\numrange{0.56}{0.91}) is larger than the range of SIMD efficiencies calculated for real chemical models, as seen in~\cref{F:source_simd_scaling}.
For smaller models---e.g., \ce{H2}\slash\ce{CO} which had a SIMD efficiency of \num{0.6} on the \avx/ CPU---this suggests that the presence of more computationally intensive fall-off\slash chemically activated reactions in model can increase the SIMD efficiency.
However, the base isopentanol model achieved a SIMD efficiency of only \num{0.78} on the \avx/ machine in~\cref{S:source_results}, suggesting that more work could be done to optimize the source-term evaluations.
In particular, it is likely that a reaction sorting method, such as suggested by Sewerin and Rigopoulos~\cite{Sewerin20151375}, would be particularly beneficial to reduce the number of vector gather\slash scatter\slash masking operations incurred during source-term evaluation.

\pagebreak
\bibliography{paper}

\begin{thebibliography}{10}
\expandafter\ifx\csname url\endcsname\relax
  \def\url#1{\texttt{#1}}\fi
\expandafter\ifx\csname urlprefix\endcsname\relax\def\urlprefix{URL }\fi
\expandafter\ifx\csname href\endcsname\relax
  \def\href#1#2{#2} \def\path#1{#1}\fi

\bibitem{LU2009192}
T.~Lu, C.~K. Law, Toward accommodating realistic fuel chemistry in large-scale
  computations, Prog. Energy Combust. Sci. 35~(2) (2009) 192 -- 215.
\newblock \href {http://dx.doi.org/10.1016/j.pecs.2008.10.002}
  {\path{doi:10.1016/j.pecs.2008.10.002}}.

\bibitem{WESTBROOK2011742}
C.~Westbrook, C.~Naik, O.~Herbinet, W.~Pitz, M.~Mehl, S.~Sarathy, H.~Curran,
  Detailed chemical kinetic reaction mechanisms for soy and rapeseed biodiesel
  fuels, Combust. Flame 158~(4) (2011) 742 -- 755, {Special} Issue on Kinetics.
\newblock \href {http://dx.doi.org/10.1016/j.combustflame.2010.10.020}
  {\path{doi:10.1016/j.combustflame.2010.10.020}}.

\bibitem{Spafford:2010aa}
K.~Spafford, J.~Meredith, J.~Vetter, J.~Chen, R.~Grout, R.~Sankaran,
  Accelerating {S3D}: A {GPGPU} case study, in: Euro-Par 2009 Parallel Process.
  Workshops, LNCS 6043, Springer-Verlag, Berlin, Heidelberg, 2010, pp.
  122--131.
\newblock \href {http://dx.doi.org/10.1007/978-3-642-14122-5_16}
  {\path{doi:10.1007/978-3-642-14122-5_16}}.

\bibitem{Lu:2006bb}
T.~Lu, C.~K. Law, Linear time reduction of large kinetic mechanisms with
  directed relation graph: \textit{n}-heptane and iso-octane, Combust. Flame
  144~(1-2) (2006) 24--36.
\newblock \href {http://dx.doi.org/10.1016/j.combustflame.2005.02.015}
  {\path{doi:10.1016/j.combustflame.2005.02.015}}.

\bibitem{Pepiot-Desjardins:2008}
P.~Pepiot-Desjardins, H.~Pitsch, An efficient error-propagation-based reduction
  method for large chemical kinetic mechanisms, Combust. Flame 154~(1--2)
  (2008) 67--81.
\newblock \href {http://dx.doi.org/10.1016/j.combustflame.2007.10.020}
  {\path{doi:10.1016/j.combustflame.2007.10.020}}.

\bibitem{Hiremath:2010jw}
V.~Hiremath, Z.~Ren, S.~B. Pope, A greedy algorithm for species selection in
  dimension reduction of combustion chemistry, Combust. Theor. Model. 14~(5)
  (2010) 619--652.
\newblock \href {http://dx.doi.org/10.1080/13647830.2010.499964}
  {\path{doi:10.1080/13647830.2010.499964}}.

\bibitem{Niemeyer:2010bt}
K.~E. Niemeyer, C.-J. Sung, M.~P. Raju, Skeletal mechanism generation for
  surrogate fuels using directed relation graph with error propagation and
  sensitivity analysis, Combust. Flame 157~(9) (2010) 1760--1770.
\newblock \href {http://dx.doi.org/10.1016/j.combustflame.2009.12.022}
  {\path{doi:10.1016/j.combustflame.2009.12.022}}.

\bibitem{Curtis:2015}
N.~J. Curtis, K.~E. Niemeyer, C.-J. Sung, An automated target species selection
  method for dynamic adaptive chemistry simulations, Combust. Flame 162~(4)
  (2015) 1358--1374.
\newblock \href {http://dx.doi.org/10.1016/j.combustflame.2014.11.004}
  {\path{doi:10.1016/j.combustflame.2014.11.004}}.

\bibitem{Lu:2007}
T.~Lu, C.~K. Law, Diffusion coefficient reduction through species bundling,
  Combust. Flame 148~(3) (2007) 117--126.
\newblock \href {http://dx.doi.org/10.1016/j.combustflame.2006.10.004}
  {\path{doi:10.1016/j.combustflame.2006.10.004}}.

\bibitem{Ahmed:2007fa}
S.~S. Ahmed, F.~Mau{\ss}, G.~Mor{\'e}ac, T.~Zeuch, A comprehensive and compact
  \textit{n}-heptane oxidation model derived using chemical lumping, Phys.
  Chem. Chem. Phys. 9~(9) (2007) 1107--1126.
\newblock \href {http://dx.doi.org/10.1039/b614712g}
  {\path{doi:10.1039/b614712g}}.

\bibitem{Pepiot:2008kq}
P.~Pepiot-Desjardins, H.~Pitsch, An automatic chemical lumping method for the
  reduction of large chemical kinetic mechanisms, Combust. Theor. Model. 12~(6)
  (2008) 1089--1108.
\newblock \href {http://dx.doi.org/10.1080/13647830802245177}
  {\path{doi:10.1080/13647830802245177}}.

\bibitem{Maas:1992ws}
U.~Maas, S.~B. Pope, Simplifying chemical kinetics: intrinsic low-dimensional
  manifolds in composition space, Combust. Flame 88~(3-4) (1992) 239--264.
\newblock \href {http://dx.doi.org/10.1016/0010-2180(92)90034-M}
  {\path{doi:10.1016/0010-2180(92)90034-M}}.

\bibitem{Lam:1994ws}
S.-H. Lam, D.~A. Goussis, The {CSP} method for simplying kinetics, Int. J.
  Chem. Kinet. 26~(4) (1994) 461--486.
\newblock \href {http://dx.doi.org/10.1002/kin.550260408}
  {\path{doi:10.1002/kin.550260408}}.

\bibitem{Lu:2001ve}
T.~Lu, Y.~Ju, C.~K. Law, Complex {CSP} for chemistry reduction and analysis,
  Combust. Flame 126~(1--2) (2001) 1445--1455.
\newblock \href {http://dx.doi.org/10.1016/S0010-2180(01)00252-8}
  {\path{doi:10.1016/S0010-2180(01)00252-8}}.

\bibitem{Gou:2010}
X.~Gou, W.~Sun, Z.~Chen, Y.~Ju, A dynamic multi-timescale method for combustion
  modeling with detailed and reduced chemical kinetic mechanisms, Combust.
  Flame 157~(6) (2010) 1111--1121.
\newblock \href {http://dx.doi.org/10.1016/j.combustflame.2010.02.020}
  {\path{doi:10.1016/j.combustflame.2010.02.020}}.

\bibitem{turanyi2016analysis}
T.~Tur{\'a}nyi, A.~S. Tomlin, Analysis of kinetic reaction mechanisms,
  Springer, 2016.

\bibitem{SCHWER2002270}
D.~A. Schwer, J.~E. Tolsma, W.~H. Green, P.~I. Barton, On upgrading the
  numerics in combustion chemistry codes, Combust. Flame 128~(3) (2002)
  270--291.
\newblock \href {http://dx.doi.org/10.1016/S0010-2180(01)00352-2}
  {\path{doi:10.1016/S0010-2180(01)00352-2}}.

\bibitem{Niemeyer:2016aa}
K.~E. Niemeyer, N.~J. Curtis, C.~J. Sung, \texttt{pyJac}: analytical {Jacobian}
  generator for chemical kinetics, Comput. Phys. Comm. 215 (2017) 188--203.
\newblock \href {http://dx.doi.org/10.1016/j.cpc.2017.02.004}
  {\path{doi:10.1016/j.cpc.2017.02.004}}.

\bibitem{GAO2015287}
Y.~Gao, Y.~Liu, Z.~Ren, T.~Lu, A dynamic adaptive method for hybrid integration
  of stiff chemistry, Combust. Flame 162~(2) (2015) 287--295.
\newblock \href {http://dx.doi.org/10.1016/j.combustflame.2014.07.023}
  {\path{doi:10.1016/j.combustflame.2014.07.023}}.

\bibitem{superlu99}
J.~W. Demmel, S.~C. Eisenstat, J.~R. Gilbert, X.~S. Li, J.~W.~H. Liu, A
  supernodal approach to sparse partial pivoting, SIAM J. Matrix Analys. Appl.
  20~(3) (1999) 720--755.
\newblock \href {http://dx.doi.org/10.1137/S0895479895291765}
  {\path{doi:10.1137/S0895479895291765}}.

\bibitem{Shi:2012aa}
Y.~Shi, W.~H. Green, H.-W. Wong, O.~O. Oluwole, Accelerating multi-dimensional
  combustion simulations using {GPU} and hybrid explicit\slash implicit {ODE}
  integration, Combust. Flame 159~(7) (2012) 2388--2397.
\newblock \href {http://dx.doi.org/10.1016/j.combustflame.2012.02.016}
  {\path{doi:10.1016/j.combustflame.2012.02.016}}.

\bibitem{Niemeyer:2014aa}
K.~E. Niemeyer, C.-J. Sung, Accelerating moderately stiff chemical kinetics in
  reactive-flow simulations using {GPUs}, J. Comput. Phys. 256 (2014) 854--871.
\newblock \href {http://dx.doi.org/10.1016/j.jcp.2013.09.025}
  {\path{doi:10.1016/j.jcp.2013.09.025}}.

\bibitem{Sewerin20151375}
F.~Sewerin, S.~Rigopoulos, A methodology for the integration of stiff chemical
  kinetics on {GPUs}, Combust. Flame 162~(4) (2015) 1375--1394.
\newblock \href {http://dx.doi.org/10.1016/j.combustflame.2014.11.003}
  {\path{doi:10.1016/j.combustflame.2014.11.003}}.

\bibitem{CurtisGPU:2017}
N.~J. Curtis, K.~E. Niemeyer, C.-J. Sung, An investigation of {GPU}-based stiff
  chemical kinetics integration methods, Combust. Flame 179 (2017) 312--324.
\newblock \href {http://dx.doi.org/10.1016/j.combustflame.2017.02.005}
  {\path{doi:10.1016/j.combustflame.2017.02.005}}.

\bibitem{stone2018}
C.~P. Stone, A.~T. Alferman, K.~E. Niemeyer, Accelerating finite-rate chemical
  kinetics with coprocessors: comparing vectorization methods on {GPUs},
  {MICs}, and {CPUs}, Comput. Phys. Comm. 226 (2018) 18--29.
\newblock \href {http://dx.doi.org/10.1016/j.cpc.2018.01.015}
  {\path{doi:10.1016/j.cpc.2018.01.015}}.

\bibitem{khan2018science}
H.~N. Khan, D.~A. Hounshell, E.~R. Fuchs, Science and research policy at the
  end of {Moore}'s law, Nat. Electron. 1~(1) (2018) 14--21.
\newblock \href {http://dx.doi.org/10.1038/s41928-017-0005-9}
  {\path{doi:10.1038/s41928-017-0005-9}}.

\bibitem{stone2010opencl}
J.~E. Stone, D.~Gohara, G.~Shi, {OpenCL}: A parallel programming standard for
  heterogeneous computing systems, IEEE Des. Test 12~(3) (2010) 66--73.
\newblock \href {http://dx.doi.org/10.1109/MCSE.2010.69}
  {\path{doi:10.1109/MCSE.2010.69}}.

\bibitem{lindholm2008Nvidia}
E.~Lindholm, J.~Nickolls, S.~Oberman, J.~Montrym, {NVIDIA} {Tesla}: A unified
  graphics and computing architecture, IEEE micro 28~(2) (2008) 39--55.
\newblock \href {http://dx.doi.org/10.1109/MM.2008.31}
  {\path{doi:10.1109/MM.2008.31}}.

\bibitem{Safta:2011vn}
C.~Safta, H.~N. Najm, O.~M. Knio, {TChem} - a software toolkit for the analysis
  of complex kinetic models, Tech. Rep. SAND2011-3282, Sandia National
  Laboratories (May 2011).

\bibitem{Curtis2017:tchem}
N.~J. Curtis, K.~E. Niemeyer, Fileset for testing thread-safety of {TChem},
  figshare (Jan. 2017).
\newblock \href {http://dx.doi.org/10.6084/m9.figshare.4563982.v1}
  {\path{doi:10.6084/m9.figshare.4563982.v1}}.

\bibitem{Youssefi:2011tm}
M.~R. Youssefi, Development of analytic tools for computational flame
  diagnostics, Master's thesis, University of Connecticut,
  \url{http://digitalcommons.uconn.edu/gs_theses/145/} (Aug. 2011).

\bibitem{Bisetti:2012jw}
F.~Bisetti, Integration of large chemical kinetic mechanisms via exponential
  methods with {Krylov} approximations to {Jacobian} matrix functions, Combust.
  Theor. Model. 16~(3) (2012) 387--418.
\newblock \href {http://dx.doi.org/10.1080/13647830.2011.631032}
  {\path{doi:10.1080/13647830.2011.631032}}.

\bibitem{Perini:2012gy}
F.~Perini, E.~Galligani, R.~D. Reitz, An analytical {Jacobian} approach to
  sparse reaction kinetics for computationally efficient combustion modeling
  with large reaction mechanisms, Energy Fuels 26~(8) (2012) 4804--4822.
\newblock \href {http://dx.doi.org/10.1021/ef300747n}
  {\path{doi:10.1021/ef300747n}}.

\bibitem{HANSEN2018257}
M.~A. Hansen, J.~C. Sutherland, On the consistency of state vectors and
  {Jacobian} matrices, Combust. Flame 193 (2018) 257--271.
\newblock \href {http://dx.doi.org/j.combustflame.2018.03.017}
  {\path{doi:j.combustflame.2018.03.017}}.

\bibitem{Dijkmans:2014bb}
T.~Dijkmans, C.~M. Schietekat, K.~M. Van~Geem, G.~B. Marin, {GPU} based
  simulation of reactive mixtures with detailed chemistry in combination with
  tabulation and an analytical {Jacobian}, Comput. Chem. Eng. 71 (2014)
  521--531.
\newblock \href {http://dx.doi.org/10.1016/j.compchemeng.2014.09.016}
  {\path{doi:10.1016/j.compchemeng.2014.09.016}}.

\bibitem{Bauer:2014}
M.~Bauer, S.~Treichler, A.~Aiken, Singe: Leveraging warp specialization for
  high performance on {GPUs}, SIGPLAN Not. 49~(8) (2014) 119--130.
\newblock \href {http://dx.doi.org/10.1145/2692916.2555258}
  {\path{doi:10.1145/2692916.2555258}}.

\bibitem{lu_yoo_chen_law_2010}
T.~F. LU, C.~S. YOO, J.~H. CHEN, C.~K. LAW, Three-dimensional direct numerical
  simulation of a turbulent lifted hydrogen jet flame in heated coflow: a
  chemical explosive mode analysis, J. Fluid Mech. 652 (2010) 45---64.
\newblock \href {http://dx.doi.org/10.1017/S002211201000039X}
  {\path{doi:10.1017/S002211201000039X}}.

\bibitem{Shi:2011aa}
Y.~Shi, W.~H. Green, H.-W. Wong, O.~O. Oluwole, Redesigning combustion modeling
  algorithms for the graphics processing unit ({GPU}): Chemical kinetic rate
  evaluation and ordinary differential equation integration, Combust. Flame
  158~(5) (2011) 836--847.
\newblock \href {http://dx.doi.org/10.1016/j.combustflame.2011.01.024}
  {\path{doi:10.1016/j.combustflame.2011.01.024}}.

\bibitem{Niemeyer:2011aa}
K.~E. Niemeyer, C.-J. Sung, C.~G. Fotache, J.~C. Lee, Turbulence-chemistry
  closure method using graphics processing units: a preliminary test, in: Fall
  2011 Technical Meeting of the Eastern States Section of the Combust.
  Institute.
\newblock \href {http://dx.doi.org/10.6084/m9.figshare.3384964}
  {\path{doi:10.6084/m9.figshare.3384964}}.

\bibitem{Le2013596}
H.~P. Le, J.-L. Cambier, L.~K. Cole, {GPU}-based flow simulation with detailed
  chemical kinetics, Comput. Phys. Comm. 184~(3) (2013) 596--606.
\newblock \href {http://dx.doi.org/10.1016/j.cpc.2012.10.013}
  {\path{doi:10.1016/j.cpc.2012.10.013}}.

\bibitem{Stone:2013aa}
C.~P. Stone, R.~L. Davis, Techniques for solving stiff chemical kinetics on
  graphical processing units, J. Propul. Power 29~(4) (2013) 764--773.
\newblock \href {http://dx.doi.org/10.2514/1.B34874}
  {\path{doi:10.2514/1.B34874}}.

\bibitem{Brown:1989vl}
P.~N. Brown, G.~D. Byrne, A.~C. Hindmarsh, {VODE}: a variable-coefficient {ODE}
  solver, SIAM J. Sci. Stat. Comput. 10~(5) (1989) 1038--1051.
\newblock \href {http://dx.doi.org/10.1137/0910062}
  {\path{doi:10.1137/0910062}}.

\bibitem{Yonkee2016}
N.~Yonkee, J.~C. Sutherland, {PoKiTT}: Exposing task and data parallelism on
  heterogeneous architectures for detailed chemical kinetics, transport, and
  thermodynamics calculations, SIAM Journal on Scientific Computing 38~(5)
  (2016) S264--S281.
\newblock \href {http://dx.doi.org/10.1137/15M1026237}
  {\path{doi:10.1137/15M1026237}}.

\bibitem{wanner1991solving}
G.~Wanner, E.~Hairer, Solving Ordinary Differential Equations II: Stiff and
  Differential-Algebraic Problems, 2nd Edition, Springer-Verlag, Berlin, 1996.
\newblock \href {http://dx.doi.org/10.1007/978-3-642-05221-7}
  {\path{doi:10.1007/978-3-642-05221-7}}.

\bibitem{Hochbruck:1998}
M.~Hochbruck, C.~Lubich, H.~Selhofer, Exponential integrators for large systems
  of differential equations, SIAM J. Sci. Comput. 19~(5) (1998) 1552--1574.
\newblock \href {http://dx.doi.org/10.1137/S1064827595295337}
  {\path{doi:10.1137/S1064827595295337}}.

\bibitem{Hockbruck:2009}
M.~Hochbruck, A.~Ostermann, J.~Schweitzer, Exponential {Rosenbrock}-type
  methods, SIAM J. Numer. Anal. 47~(1) (2009) 786--803.
\newblock \href {http://dx.doi.org/10.1137/080717717}
  {\path{doi:10.1137/080717717}}.

\bibitem{Hindmarsh:2005}
A.~C. Hindmarsh, P.~N. Brown, K.~E. Grant, S.~L. Lee, R.~Serban, D.~E.
  Shumaker, C.~S. Woodward, Sundials: Suite of nonlinear and
  differential/algebraic equation solvers, ACM Trans. Math. Softw. 31~(3)
  (2005) 363--396.
\newblock \href {http://dx.doi.org/10.1145/1089014.1089020}
  {\path{doi:10.1145/1089014.1089020}}.

\bibitem{Linford:2011}
J.~C. Linford, J.~Michalakes, M.~Vachharajani, A.~Sandu, Automatic generation
  of multicore chemical kernels, IEEE Trans. Parallel Distrib. Syst. 22~(1)
  (2011) 119--131.
\newblock \href {http://dx.doi.org/10.1109/TPDS.2010.106}
  {\path{doi:10.1109/TPDS.2010.106}}.

\bibitem{kroshko2013efficient}
A.~Kroshko, R.~J. Spiteri, Efficient {SIMD} solution of multiple systems of
  stiff {IVPs}, J. Comput. Sci 4~(5) (2013) 377--385.
\newblock \href {http://dx.doi.org/10.1016/j.jocs.2012.08.017}
  {\path{doi:10.1016/j.jocs.2012.08.017}}.

\bibitem{gray2000rules}
J.~Gray, P.~Shenoy, Rules of thumb in data engineering, in: Data Engineering,
  2000. Proceedings. 16th International Conference on, IEEE, 2000, pp. 3--10.

\bibitem{Nvidia:2018}
{NVIDIA}, {CUDA C} programming guide, version 9.0,
  \href{https://docs.nvidia.com/cuda/pdf/CUDA_C_Programming_Guide.pdf}{\nolinkurl{https://docs.nvidia.com/cuda/pdf/CUDA_C_Programming_Guide.pdf}}
  (Jan 2018).

\bibitem{TurnsStephenR2012Aitc}
S.~R. Turns, An Introduction to Combustion: Concepts and Applications, 3rd
  Edition, McGraw-Hill, New York, 2012.

\bibitem{kloeckner_loopy_2014}
A.~{Kl{\"o}ckner}, in: Proc. ARRAY '14: ACM SIGPLAN Workshop Libr., Lang.,
  Compil. Array Progr., Assoc. Comput. Mach., Edinburgh, Scotland., 2014.
\newblock \href {http://dx.doi.org/10.1145/2627373.2627387}
  {\path{doi:10.1145/2627373.2627387}}.

\bibitem{pyjac16}
N.~J. Curtis, K.~E. Niemeyer, \texttt{pyJac} v1.0.6 (Feb. 2018).
\newblock \href {http://dx.doi.org/10.5281/zenodo.1182789}
  {\path{doi:10.5281/zenodo.1182789}}.

\bibitem{kloeckner_pycuda_2012}
A.~{Kl{\"o}ckner}, N.~{Pinto}, Y.~{Lee}, B.~{Catanzaro}, P.~{Ivanov},
  A.~{Fasih}, {PyCUDA} and {PyOpenCL}: A scripting-based approach to {GPU}
  run-time code generation, Parallel Comput. 38~(3) (2012) 157--174.
\newblock \href {http://dx.doi.org/10.1016/j.parco.2011.09.001}
  {\path{doi:10.1016/j.parco.2011.09.001}}.

\bibitem{Cantera}
D.~G. Goodwin, H.~K. Moffat, R.~L. Speth, {Cantera}: An object-oriented
  software toolkit for chemical kinetics, thermodynamics, and transport
  processes, \url{http://www.cantera.org}, version 2.3.0 (2017).
\newblock \href {http://dx.doi.org/10.5281/zenodo.170284}
  {\path{doi:10.5281/zenodo.170284}}.

\bibitem{hogan2014fast}
R.~J. Hogan, Fast reverse-mode automatic differentiation using expression
  templates in \texttt{C++}, ACM Trans. Math. Software 40~(4) (2014) 26.
\newblock \href {http://dx.doi.org/10.1145/2560359}
  {\path{doi:10.1145/2560359}}.

\bibitem{adept-v11}
R.~J. Hogan, \texttt{Adept} v1.1, Available at
  \url{https://github.com/rjhogan/Adept} (Jun. 2015).

\bibitem{poclIJPP}
P.~J\"a\"askel\"ainen, C.~S. de~La~Lama, E.~Schnetter, K.~Raiskila, J.~Takala,
  H.~Berg, pocl: A performance-portable {OpenCL} implementation, Int. J.
  Parallel Program. 43~(5) (2015) 752--785.
\newblock \href {http://dx.doi.org/10.1007/s10766-014-0320-y}
  {\path{doi:10.1007/s10766-014-0320-y}}.

\bibitem{dagum1998openmp}
L.~Dagum, R.~Menon, {OpenMP}: an industry standard {API} for shared-memory
  programming, Computational Sci. \& Engineering, IEEE 5~(1) (1998) 46--55.
\newblock \href {http://dx.doi.org/10.1109/99.660313}
  {\path{doi:10.1109/99.660313}}.

\bibitem{travis:2018}
G.~Travis~CI, {Travis CI - Test and Deploy Your Code with Confidence},
  \url{https://about.travis-ci.com/} (2018).

\bibitem{intelopencl:2018}
\relax{Intel}\textregistered\ {Corporation}, {OpenCL}\texttrademark\ drivers
  and runtimes for {Intel}\textregistered\ architecture,
  \url{https://software.intel.com/en-us/articles/opencl-drivers#latest_CPU_runtime}
  (2018).

\bibitem{Lattner:2004:LCF:977395.977673}
C.~Lattner, V.~Adve,
  \href{http://dl.acm.org/citation.cfm?id=977395.977673}{{LLVM}: A compilation
  framework for lifelong program analysis \& transformation}, in: Proceedings
  of the International Symposium on Code Generation and Optimization:
  Feedback-directed and Runtime Optimization, CGO '04, IEEE Computer Society,
  Washington, DC, USA, 2004, pp. 75--.
\newline\urlprefix\url{http://dl.acm.org/citation.cfm?id=977395.977673}

\bibitem{Nvidia_memory}
MichaelE1000, {Bug report on {NVIDIA} forums},
  \href{https://devtalk.nvidia.com/default/topic/1019997/cuda-programming-and-performance/how-to-handle-cl_mem_object_allocation_failure-errors-if-amount-of-useable-memory-is-not-known-/}{NVIDIA
  Devtalk Forums}, accessed 03-06-18.

\bibitem{Burke:2011fh}
M.~P. Burke, M.~Chaos, Y.~Ju, F.~L. Dryer, S.~J. Klippenstein, Comprehensive
  {H}$_2$\slash {O}$_2$ kinetic model for high-pressure combustion, Int. J.
  Chem. Kinet. 44~(7) (2011) 444--474.
\newblock \href {http://dx.doi.org/10.1002/kin.20603}
  {\path{doi:10.1002/kin.20603}}.

\bibitem{smith_gri-mech_30}
G.~P. Smith, D.~M. Golden, M.~Frenklach, N.~W. Moriarty, B.~Eiteneer,
  M.~Goldenberg, C.~T. Bowman, R.~K. Hanson, S.~Song, W.~C. Gardiner, V.~V.
  Lissianski, Z.~Qin, {GRI}-{Mech} 3.0,
  \url{http://www.me.berkeley.edu/gri_mech/}.

\bibitem{Wang:2007}
H.~Wang, X.~You, A.~V. Joshi, S.~G. Davis, A.~Laskin, F.~Egolfopoulos, C.~K.
  Law, {USC Mech Version II}.\ {High}-temperature combustion reaction model of
  {H}$_2$\slash{CO}\slash{C}$_1$--{C}$_4$ compounds,
  \url{http://ignis.usc.edu/USC_Mech_II.htm} (May 2007).

\bibitem{Sarathy:2013jr}
S.~M. Sarathy, S.~Park, B.~W. Weber, W.~Wang, P.~S. Veloo, A.~C. Davis,
  C.~Togbe, C.~K. Westbrook, O.~Park, G.~Dayma, Z.~Luo, M.~A. Oehlschlaeger,
  F.~N. Egolfopoulos, T.~Lu, W.~J. Pitz, C.-J. Sung, P.~Dagaut, A comprehensive
  experimental and modeling study of iso-pentanol combustion, Combust. Flame
  160~(12) (2013) 2712--2728.
\newblock \href {http://dx.doi.org/10.1016/j.combustflame.2013.06.022}
  {\path{doi:10.1016/j.combustflame.2013.06.022}}.

\bibitem{Anderson:1999aa}
E.~Anderson, Z.~Bai, C.~Bischof, S.~Blackford, J.~Demmel, J.~Dongarra,
  J.~Du~Croz, A.~Greenbaum, S.~Hammarling, A.~McKenney, D.~Sorensen, {LAPACK}
  Users' Guide, 3rd Edition, SIAM, Philadelphia, PA, 1999.

\bibitem{MCNENLY2015581}
M.~J. McNenly, R.~A. Whitesides, D.~L. Flowers, Faster solvers for large
  kinetic mechanisms using adaptive preconditioners, Proceedings of the
  Combustion Institute 35~(1) (2015) 581--587.
\newblock \href {http://dx.doi.org/10.1016/j.proci.2014.05.113}
  {\path{doi:10.1016/j.proci.2014.05.113}}.

\bibitem{netlib_templates}
R.~Barrett, M.~Berry, T.~F. Chan, J.~Demmel, J.~Donato, J.~Dongarra,
  V.~Eijkhout, R.~Pozo, C.~Romine, H.~V. der Vorst, Templates for the Solution
  of Linear Systems: Building Blocks for Iterative Methods, 2nd Edition, SIAM,
  Philadelphia, PA, 1994, note: the section on
  \href{http://netlib.org/linalg/html_templates/node90.html}{sparse matricies
  formats}.

\bibitem{pocl_communication}
M.~Babej, P.~J\"a\"askel\"ainen, Debugging auto vectorizer, Private
  Communication, archived on
  \href{https://sourceforge.net/p/pocl/mailman/pocl-devel/thread/CACZgdAj7A70PbFcbjfMF1xDYsaP692PkP\%2BZa8AgMYhmoiDdgdQ\%40mail.gmail.com/\#msg36218233}{POCL
  mailing list} (Feb. 2018).

\bibitem{occupancy}
NVIDIA, Achieved occupancy,
  \href{https://docs.nvidia.com/gameworks/content/developertools/desktop/analysis/report/cudaexperiments/kernellevel/achievedoccupancy.htm}{Achieved
  Occupancy} (Mar 2018).

\bibitem{intel_vectypes}
\relax{Intel}\textregistered\ {Corporation}, Using vector data types,
  \url{https://software.intel.com/en-us/node/540561}, accessed on 02/19/18.

\bibitem{intel_vecknobs}
\relax{Intel}\textregistered\ {Corporation}, Vectorizer knobs,
  \url{https://software.intel.com/en-us/node/540560}, accessed on 02/19/18.

\bibitem{strong_scaling}
G.~G. Howes, Parallel performance and optimization,
  \url{http://homepage.physics.uiowa.edu/~ghowes/teach/ihpc10/lec/ihpc10Lec_PerformanceHPC10.pdf},
  slides from Iowa High Performance Computing Summer School, University of
  Iowa, 08/2010 - Accessed on 02/19/18.

\bibitem{Nvidia_mwe}
N.~J. Curtis, {A minimum working example showing the failure of simple OpenCL
  code on the NVIDIA Linux x64 Tesla 375.26 Driver},
  \url{https://figshare.com/s/03aa9064aa6fe3508d3d} (jun 2018).
\newblock \href {http://dx.doi.org/10.6084/m9.figshare.6533915}
  {\path{doi:10.6084/m9.figshare.6533915}}.

\bibitem{pharr2012ispc}
M.~Pharr, W.~R. Mark, ispc: A {SPMD} compiler for high-performance {CPU}
  programming, in: Innovat. Parallel Comput. (InPar), 2012, 2012, pp. 1--13.
\newblock \href {http://dx.doi.org/10.1109/InPar.2012.6339601}
  {\path{doi:10.1109/InPar.2012.6339601}}.

\bibitem{cublas}
\relax{NVIDIA} Corporation, Dense linear algebra on gpus,
  \href{cuBLAS}{https://developer.nvidia.com/cublas}, accessed: 03-12-18.

\bibitem{clmath}
clMathLibraries, clmathlibraries, \href{clBLAS\slash
  clSPARSE}{https://github.com/clMathLibraries}, accessed: 03-12-18.

\bibitem{pyjac2}
N.~J. Curtis, K.~E. Niemeyer, \texttt{pyJac} v2.0.0-beta.0 (Jun. 2018).
\newblock \href {http://dx.doi.org/10.5281/zenodo.1289979}
  {\path{doi:10.5281/zenodo.1289979}}.

\bibitem{data}
N.~J. Curtis, K.~E. Niemeyer, C.-J. Sung, Data, plotting scripts, and figures
  for ``using simd and simt vectorization to evaluate sparse chemical kinetic
  jacobian matrices and thermochemical source terms ''.
\newblock \href {http://dx.doi.org/10.6084/m9.figshare.6534146}
  {\path{doi:10.6084/m9.figshare.6534146}}.

\end{thebibliography}

\end{document}